\documentclass[a4paper,12pt]{article}
\usepackage{epsfig}
\usepackage{array}
\usepackage{amsmath}
\usepackage{amssymb}
\newcommand\iindent{{}~\indent}
\newcommand{\sect}[1]{\setcounter{equation}{0}\section{#1}}

\newcommand\phup{^{\phantom{|}}}
\def\rf#1{(\ref{eq:#1})}
\def\lab#1{\label{eq:#1}}
\def\nn{\nonumber \\}
\newcommand\fract[2]{{\textstyle\frac{#1}{#2}}}
\newcommand{\beano}{\begin{eqnarray*}}
\newcommand{\enano}{\end{eqnarray*}}
\def\bea{\begin{eqnarray}}
\def\ena{\end{eqnarray}}
\def\foot#1{\footnotemark\footnotetext{#1}}
\newcommand{\rg}{r_{\!g}}
\newcommand{\hg}{h_g}
\newcommand{\hgv}{h_g^{\vee}}

\relax

\font\fld=msbm10 at 12 pt

\newcommand{\fl}[1]{\mbox{\fld #1}}     

\def\ra{\rightarrow}

\newcommand{\height}{\mbox{\rm ht}}

\newcommand{\one}{1 \hspace{-3pt}{\rm I}}
\def\PRL#1#2#3{{\sl Phys. Rev. Lett.} {\bf#1} (#2) #3}
\def\NPB#1#2#3{{\sl Nucl. Phys.} {\bf B#1} (#2) #3}

\def\CMP#1#2#3{{\sl Commun. Math. Phys.} {\bf #1} (#2) #3}
\def\PRD#1#2#3{{\sl Phys. Rev.} {\bf D#1} (#2) #3}

\def\PLB#1#2#3{{\sl Phys. Lett.} {\bf #1B} (#2) #3}

\def\IJMPA#1#2#3{{\sl Int. J. Mod. Phys.} {\bf A#1} (#2) #3}

\def\MPLA#1#2#3{{\sl Mod. Phys. Lett.} {\bf A#1} (#2) #3}

\def\JETPL#1#2#3{{\sl  Sov. Phys. JETP Lett.} {\bf #1} (#2) #3}

\def\EPJC#1#2#3{{\sl Eur. Phys. J.} {\bf C#1} (#2) #3}
\def\JHEP#1#2#3{{\sl JHEP} {\bf #1} (#2) #3}
\def\ex#1{{\rm e\>}^{#1}}
\def\Buildrel#1\over#2\under#3{\mathrel{\mathop{\kern0pt
#2}\limits^{#1}_{#3}}}
\def\bfm#1{\boldsymbol{#1}}
\def\alv{\bfm{\alpha}}
\def\bev{\bfm{\beta}}

\def\lav{\bfm{\lambda}}
\def\ArrFl#1#2{\Buildrel {m}_{#1}\over{\hbox to
33pt{\rightarrowfill}}\under{{#2}}\>}
\def\ArrFls#1#2{\Buildrel {m}_{#1}\over{\hbox to
25pt{\rightarrowfill}}\under{\ex{#2}}\>}
\def\ArrFract#1#2{\fract{#1}{#2}}
\def\bil#1#2{\langle #1,#2\rangle}
\begin{document}

\begin{titlepage}
\vspace{-1cm}
\noindent
\rightline{DCPT-02/81}\\
\rightline{SPhT-T04/048}\\
\rightline{hep-th/0405275}\\
\rightline{May 2004}\\
\phantom{bla}
\vfill
\begin{center}
{\large\bf Mass scales and crossover phenomena in the\\[0.4cm]
Homogeneous Sine-Gordon Models}
\end{center}

\vspace{0.3cm}
\begin{center}
Patrick Dorey$^{1,2}$ and
J.~Luis Miramontes$^{3,2}$
\par \vskip .2in \noindent
{\em 
${}^1$ Department of Mathematical Sciences\\
University of Durham\\
Durham DH1 3LE, UK}\\
\par \vskip .1in \noindent
{\em 
${}^2$ Service de Physique Th{\'e}orique\\
CEA-Saclay\\
F-91191 Gif-sur-Yvette Cedex, France}\\
\par \vskip .1in \noindent
{\em 
${}^3$ Departamento de F\'\i sica de Part\'\i culas,\\
Facultad de F\'\i sica\\
Universidad de Santiago de Compostela\\
15782 Santiago de Compostela, Spain}\\
\par \vskip .1in \noindent
e-mails: 
{\tt p.e.dorey@durham.ac.uk},
{\tt miramont@usc.es}
\normalsize
\end{center}
\vspace{.2in}
\begin{abstract}
The finite-size behaviours of the homogeneous sine-Gordon models
are analysed in detail, using the thermodynamic Bethe ansatz.
Crossovers are observed which allow scales associated
with both stable and unstable quantum particles to be picked up.
By introducing the
concept of shielding, we show that these match precisely with the
mass scales found classically, supporting the idea that the
full set of unstable particle states persists even far from the
semiclassical regime. General rules for the effective TBA systems
governing individual crossovers are given, and we also comment on the
Lagrangian treatment of the theories, novel issues which arise in the
form-factor approach for theories with unstable particles, and the
role of heterotic cosets in the
staircase flows exhibited by the HSG models.

\end{abstract}

\vfill
\end{titlepage}
\sect{Introduction}

\iindent
The homogeneous sine-Gordon (HSG) models~\cite{HSG}
are two-dimensional quantum field theories
with a number of remarkable properties.
They are integrable perturbations of
level-$k$ $G$-parafermions~\cite{QHSG}, that is
of coset
conformal field theories of the form $G_k/U(1)^{\rg}$, where $G$ is a simple
compact Lie group with Lie algebra $g$, $k>1$ is an integer, and $\rg$ is
the rank of $g$\,.
In the limit
$k\to\infty$  they can be analysed using semiclassical
techniques, and the resulting data
used to make conjectures for the
mass spectra and exact S-matrices
at arbitrary values of $k$ \cite{SMAT}%
\foot{We restrict ourselves to simply-laced $G$ in this
paper, but we note that S-matrices for the non simply-laced cases
have also been proposed, in \cite{KORFF}.}.
Checks of the S-matrices using both thermodynamic Bethe ansatz
(TBA)~\cite{HSGTBA} and
form-factor~\cite{FFother,FFRG,FF2} approaches leave
little doubt that they
describe the perturbed parafermionic theories correctly, even for small
values of $k$ far from the semiclassical regime.

An interesting feature to emerge from the semiclassical
studies was the presence of unstable particles~\cite{HSGSOL}.
If~$\hg$ is the Coxeter number of~$g$, a total
of $(k{-}1)\,\rg
\hg/2$ particle-like states (in the simply-laced cases)
were identified, $k{-}1$
for each positive root of $g$\,. Those corresponding
to the simple roots are stable; all the rest are semiclassically
unstable.  On the other hand, in the
S-matrix treatment of the full quantum theory, only stable
particles are seen directly.
Evidence for the unstable particles must come by more indirect
routes, such as the appearance of resonance
poles at complex rapidities in the scattering amplitudes of
pairs of stable asymptotic particle states. However, for
the HSG models only the $\rg{-}1$ sets of unstable particles
associated with the roots of height two can
be picked up in this way\foot{Recall that
if $\bev=\sum_1^{\rg}c_i\alv_i$ is a root of $g$,
with $\{\alv_i\}\equiv\Delta$ a set of simple roots,
then the {\em height}\/ of $\bev$ with respect to $\Delta$
is $\height(\bev)=
\sum_1^{\rg}c_i$\,.}.
And at small values of $k$, far from the semiclassical regime,
the resonance peaks for the height-two particles
are very broad, and their interpretation
becomes delicate.
This raises an immediate question --
how should we verify the existence of the remaining particles in the
quantum theory,
and what influence do they have on the physical properties
of the models?
Are the extra unstable particles
even present in the models with $k$ small, or
do they merely emerge when the limit $k\to\infty$ is taken?

In this paper we shall address these and related issues by
returning to the study of
finite-size effects using the TBA technique. At mass scales
where new stable or
unstable particle states become important, we predict analytically (and
confirm numerically)
a change in the behaviour of the finite-size
scaling function, even for those unstable
particles which are not seen directly in the two-particle S-matrices.

The influence of the extra unstable particles
gives the HSG models a much richer structure
of renormalisation group
flows than was initially thought,
unifying and generalising
the simplest flows between conformal field theories within a common
structure. 
Moreover, as a result of parity breaking, some of these
flows turn out to involve heterotic coset models.

Our main observations were first presented in
\cite{Ltalk,Dorey:2002sc}, and \cite{Dorey:2002sc} can be consulted for
further background material. The rest of the present
paper is organised as follows. In section~2 some key features of the HSG
models and their resonances are recalled, and our strategy for detecting
the unstable resonances is described. This requires us to be able to
separate the relevant mass scales suitably, and section~3 is devoted to
an analysis of this problem in the classical theory. The idea of
`shielding' is introduced, and it is shown how certain
mass scales one might na\"\i vely expect to find in a classical HSG model
may be missing. Then in section~4 we show,  through a detailed analysis
of the TBA equations which we illustrate with a number of examples,
that these classical effects are precisely matched in the
finite-size behaviour of the quantum theory. The maximal number of
separated steps that a general HSG model can exhibit, a question
rendered non-trivial by shielding, is discussed in section~5.
We then move on to other ways to understand the crossovers, first
discussing predictions from the Lagrangian approach in section~6 before
making some comments on form factor calculations in section~7. Finally,
section~8 contains our conclusions and there are two appendices.

\sect{The HSG models and their resonances}
\label{PARTsec}

\iindent
Throughout this paper, we shall be considering the HSG models
corresponding to perturbations of the level-$k$ parafermionic
coset conformal field theories $G_k/U(1)^{\rg}$,
where $G$ is a simple compact Lie group, with
Lie algebra $g$ of rank $\rg$, Coxeter number $\hg$, and dual Coxeter number
$\hg^\vee$. Since we will only discuss the cases where $g$ is
simply-laced, $\hg^\vee=\hg$, but we shall
preserve the distinction in formulae of
more general applicability.

Unlike the cases
first studied by Zamolodchikov~\cite{ZamPCFT} and others,
the HSG models are multiparameter deformations of
conformal field theories\foot{Other possibilities for
constructing multiparameter integrable perturbations of
conformal field theories are discussed in, for example,
\cite{TATEO,FATEEV,SALEUR}.}. The basic operators of the unperturbed theory
lie in multiplets
$\Phi{}^{\bfm{\Lambda},\>\bfm{\overline{\Lambda}}}_{\;\bfm{\omega},
\>\bfm{\overline{\omega}}}$ labelled by two representations of $G$ with
highest weights
$(\bfm{\Lambda},\bfm{\overline{\Lambda}})$, and two weights
$(\bfm{\omega},\bfm{\overline{\omega}})$ in those
representations~\cite{GEPNER}, and the HSG
perturbing operators are certain spinless primary fields
$\phi\in\Phi{}^{\rm adj,\>adj}_{\;\bfm{0},\>\bfm{0}}$, with conformal
dimensions
$\Delta_{\Phi}= \bar\Delta_{\Phi}= \hgv/(k+\hgv)$.
Since the multiplicity of the weight~$\bfm{0}$ in the adjoint
representation is
$\rg$\,, $\Phi{}^{\rm
adj,\>adj}_{\;\bfm{0},\>\bfm{0}}$ is $\rg^{\>2}$-dimensional, and has
a basis $\{\phi{}^{\rm
adj,\>adj}_{\;p\>,\;q}\mid p,q=1\ldots\rg\}$. The perturbations within
this multiplet which
lead to HSG models are
conveniently parametrised by a pair of
$\rg$-dimensional vectors
$\bfm\lambda$ and $\overline{\bfm\lambda}$ as
\begin{equation}
\phi_{\bfm\lambda,\overline{\bfm\lambda}}=
\sum_{p,q=1}^{\rg}\lambda_p\> \overline\lambda_q \>\phi{}^{\rm
adj,\>adj}_{\;p\>,\;q} \>,
\lab{PertOp}
\end{equation}
where $\lambda_1\ldots \lambda_{\rg}$ and $\overline{\lambda}_1\ldots
\overline{\lambda}_{\rg}$ are the components of the vectors
$\bfm\lambda$ and $\overline{\bfm\lambda}$\,.
{}From this perspective the HSG actions have the form
\begin{equation}
S_{HSG}=
S_{CFT}+\mu\int d^2x\>
\phi_{\bfm\lambda,\overline{\bfm\lambda}}\,,
\lab{PCFTaction}
\end{equation}
where $S_{CFT}$ denotes an action for the conformal field theory of
level-$k$ $G$-parafermions,
and $\mu$ is a dimensionful coupling which can be related to the
overall mass scale, once
the combined normalisation of
$\bfm\lambda$ and $\overline{\bfm\lambda}$ has been fixed by
demanding the standard short-distance behaviour of two-point functions
involving
$\phi_{\bfm\lambda,\overline{\bfm\lambda}}$\,.
Note that $\phi_{\bfm\lambda,\overline{\bfm\lambda}}$
is trivially invariant under a joint rescaling
\begin{equation}
\bfm\lambda\to\alpha\bfm\lambda\,,\quad\
\overline{\bfm\lambda}\to\alpha^{-1}
\overline{\bfm\lambda}\,,
\end{equation}
which together with the normalisation condition
leaves $2\rg-2$
dimensionless parameters in $\bfm\lambda$ and
$\overline{\bfm\lambda}$. Thus the
theory is determined by a total of $2\rg-1$ parameters, one of which can
be mapped onto the overall scale.

A more explicit
construction of the HSG actions~\rf{PCFTaction} is provided by the
identification of $S_{CFT}$ with the gauged Wess-Zumino-Novikov-Witten
(WZW) action associated with a coset of the form
$G/H$ at level $k$, where $H\subset G$ is a maximal
abelian torus (see~\cite{HSG,QHSG} and section~\ref{LAGsec}
for details). Then, $\phi{}^{\rm adj,\>adj}_{\;p\>,\;q}$ are the Cartan
matrix elements of the spinless primary field corresponding to the WZW field
in the adjoint representation~\cite{KZ}.
One of the nicest features of this formulation is that it
simplifies the analysis of these models in the large-$k$ limit,
which corresponds to both the weak-coupling
(perturbative) and semiclassical regimes of the perturbed
gauged WZW action. 

The HSG models can also be characterised by their long-distance,
infrared, behaviour.
The exact $S$-matrices proposed in~\cite{SMAT}
describe the scattering of a set of stable
solitonic particles labelled by two quantum
numbers, $(i,a)$, where $i=1\ldots\rg$ labels a simple root of $g$,
and $a=1\ldots k{-}1$.
The mass of the particle $(i,a)$ is
\begin{equation}
M_a^i = M m_i\> {\mu}_a\>,
\lab{Perron}
\end{equation}
where $M$ is a dimensionful overall mass scale,
$m_1,\ldots, m_{\rg}$ are $\rg$ arbitrary (non-vanishing)
relative masses,
one for each simple root of $g$,
and the numbers
$\mu_a=\sin (\pi a/k)/\sin(\pi/k)$
are the components of the Perron-Frobenius
eigenvector of the $a_{k-1}$ Cartan matrix.
The S-matrix elements of these
particles depend on a
further $\rg-1$ 
real `resonance parameters' $\sigma_{ij}=-\sigma_{ji}$,
defined for each pair
$\{ i,j\}$ of neighbouring
nodes on the Dynkin diagram of $g$. The resonance
parameters are most conveniently
specified by assigning a variable $\sigma_i$ to each node
of $g$ and setting $\sigma_{ij}=\sigma_i-\sigma_j$.
The resulting set of infrared parameters $M$, $\{m_i\}$, and
$\{\sigma_i\}$ is redundant, but the obvious symmetries
$M\rightarrow\alpha M$, $\{m_i\}\rightarrow\{\alpha^{-1} m_i\}$, and
$\{\sigma_i\}\rightarrow \{\sigma_i+\beta\}$, ensure that there
are only $2\rg-1$ independent parameters, just as for
the ultraviolet description of the models.

Classically, the theory exhibits further
solitonic particle-like
solutions associated with all of the other positive roots
$\bfm\beta\in \Phi^+_g$~\cite{HSGSOL}.
Their masses can be specified in a
concise way via 
\begin{equation}
\bfm\lambda_{\pm}=\sum_1^{\rg}m_i\> e^{\pm\sigma_i}\>
\bfm\lambda_i
\lab{LambdaW}
\end{equation}
where the $\bfm\lambda_i$\,, $i=1\dots \rg$\,, are the fundamental
weights of $g$ and satisfy $\bfm\lambda_i\cdot \bfm\alpha_j=\delta_{ij}$.
The
relative mass scale for
the solitonic particles associated with the positive root
$\bfm\beta$\, is then
\begin{equation}
m^2_{\bfm\beta}=(\bfm\lambda_+\cdot
\bfm\beta)\,(\bfm\lambda_-\cdot\bfm\beta)~,
\lab{MassRoot}
\end{equation}
which reduces to $m_i^2$ for $\bfm\beta=\bfm\alpha_i$.
The semiclassical analysis performed using the gauged WZW formulation
in~\cite{SMAT} shows that the classical particles associated with
non-simple roots decay, and so do not appear directly
in the spectrum of asymptotic quantum
states\foot{As recalled in section~\ref{LAGsec} below, this
formulation also shows that $\bfm\lambda_+$ and $\bfm\lambda_-$
correspond in the semiclassical
limit to the two vectors $\bfm\lambda$ and $\overline{\bfm\lambda}$
specifying the action~\rf{PCFTaction}.}.
In particular, if $\alv_i+\alv_j$ is a root of $g$ or, equivalently,
$\{i,j\}$ is a pair of neighbouring nodes on the Dynkin diagram of $g$, the
relative mass scale and decay width of the soliton particles associated with
that root are $m_{\alv_i+\alv_j}^2 =  m_i^2 +m_j^2 + 2m_i m_j
\cosh(\sigma_i-\sigma_j)$ and
$\Gamma_{\alv_i,\alv_j}= {\pi\over k} {2m_i m_j\over m_{\alv_i+\alv_j}}
\sinh |\sigma_i-\sigma_j|$.

In the full quantum theory, such
long-lived unstable particles
should correspond to {\em resonances} in interactions among the
stable particles~\cite{EDEN}.
For example, if two stable particles
scatter at a center-of-mass energy
$\sqrt{s}$ close to the mass of an unstable state with appropriate
quantum numbers, then
they can form that state and remain in it for a time
roughly equal to its lifetime, before decaying. Consequently, the
transition amplitude shows a bump at the appropriate
energy, which normally
corresponds to a complex simple pole in the
$S$-matrix amplitude. This pole is located on the second Riemann
sheet of the complex (Mandelstam) $s$-plane, and its position can be
conveniently written as $s_{\rm R}=(M_{\rm R}-i\Gamma_{\rm R}/2)^2$.
In this way, $\tau=\hbar/\Gamma_{\rm R}$ measures the lifetime of
the unstable particle, and the form of the resonance pole is given by the
Breit-Wigner formula
\begin{equation}
S\approx 
1-i\> \frac{2M_{\rm R}\Gamma_{\rm R}}%
{(s-s_R)}~~,\qquad
s_R\equiv(M_{\rm R}-i\Gamma_{\rm R}/2)^2
\>.
\lab{BW}
\end{equation}
Notice that the bump in the scattering probability $|1-S|^2$ occurs
around $s={\rm Re\/}(s_{\rm R})$,
justifying the usual identification of
$M_\rho\equiv \sqrt{{\rm Re\/}(s_{\rm R})}$
with the physical mass of the unstable particle
(see for example
the discussions in~\cite{DATA}, especially \cite{DATAa}).
Another definition sometimes used for this mass
is $M_{\rm R}={\rm
Re\/}(\sqrt{s_{\rm R}})$. If the lifetime is large, which
translates into the condition\footnote{For real
physical unstable particles, the ratio $\Gamma_{\rm R}/M_{\rm R}$ runs
from
$\sim 10^{-2}$--$10^{-1}$ for hadron resonances and the $Z^0$ and
$W^\pm$ bosons, to much smaller values for other electroweakly
decaying particles, like $\sim 10^{-7}$ or $\sim 10^{-15}$ for the
pions $\pi^0$ and $\pi^\pm$, respectively~\cite{DATA}.}
$\Gamma_{\rm R}\ll M_{\rm R}$ and
corresponds to the situation when the pole is close to the
real (physical) axis of the complex $s$-plane, then $M_{\rm R}\simeq
M_\rho$. Otherwise, when
$\Gamma_{\rm R}$ is larger, the lifetime is short and the unstable
particle does not have a definite physical mass, as a consequence
of the uncertainty principle. This has made the proper
definition of the masses of unstable particles a subject of
debate, which becomes of phenomenological relevance when the experimental
data is accurate enough, as is exemplified by the cases of the $Z^0$
boson~\cite{DATAb} or the baryon resonances~\cite{DATAc} (see
also~\cite{UNSTABLE}). In particular, there is no
general consensus as to how to choose between
$M_\rho$ and
$M_{\rm R}$ to characterise the mass of the unstable state, even though
they differ significantly when
$\Gamma_{\rm R}/M_{\rm R}$ is large.

In the HSG models this question can be studied in a context
where the S-matrix and related observables are known exactly.
Even though it was the correspondence between
unstable particles and resonance poles in the semiclassical limit
that provided the starting-point for the S-matrix
elements conjectured in~\cite{SMAT}, the fact that they are thought to
be exact even away from this limit allows non-trivial predictions to be
made and tested. For the scattering between particles
$(i,1)$ and $(j,1)$, with $i$ and $j$ neighbouring nodes on the Dynkin
diagram of $g$ and with $(i,1)$ initially to the left of $(j,1)$,
the relevant amplitude
$S{_1^i}{_1^j}(\theta)$ has a resonance pole at the complex
rapidity value $\theta_{{\rm R}_{ij}}=\sigma_{ji}-i\pi/k$.
This corresponds to a pole
on the second sheet of the complex $s$-plane
at $s=s_{{\rm R}_{ij}}$, where
\begin{equation}
s_{{\rm R}_{ij}}=
M^2(m_i^2 +m_j^2 +2m_i m_j \cosh\theta_{{\rm R}_{ij}})\,.
\lab{SPole}
\end{equation}
This pole can only be associated with a physical unstable particle
if ${\rm Im\/}(s_{{\rm R}_{ij}})<0$~\cite{EDEN}, which requires
$\sigma_{ji}>0$. (If $\sigma_{ji}<0$, the pole is a `shadow
pole' whose existence is required by the Hermitian anlyticity condition
satisfied by the $S$-matrix amplitudes~\cite{SMAT}.)
In the semiclassical limit, $k$ is large and the parameters
$\Gamma_{{\rm R}_{ij}}$ and $M_{{\rm R}_{ij}}$
corresponding to $s_{{\rm R}_{ij}}$ satisfy the bound
$\Gamma_{{\rm R}_{ij}}/M_{{\rm R}_{ij}}< \pi/k$\,.
The consequent smallness of this ratio means that
the pole at $s_{{\rm R}_{ij}}$
can be immediately interpreted as a trace of a long-lived unstable
particle associated with the height-two root $\alv_i+\alv_j$.
In this regime the mass scale of the unstable particle is unambiguously
defined by
$M_{\rho_{ij}}\simeq M_{{\rm R}_{ij}} \simeq M m_{\alv_i+\alv_j}$, where
$m_{\alv_i+\alv_j}$ is given by~\rf{MassRoot}.

\begin{figure}[ht]
\begin{center}
\epsfig{file=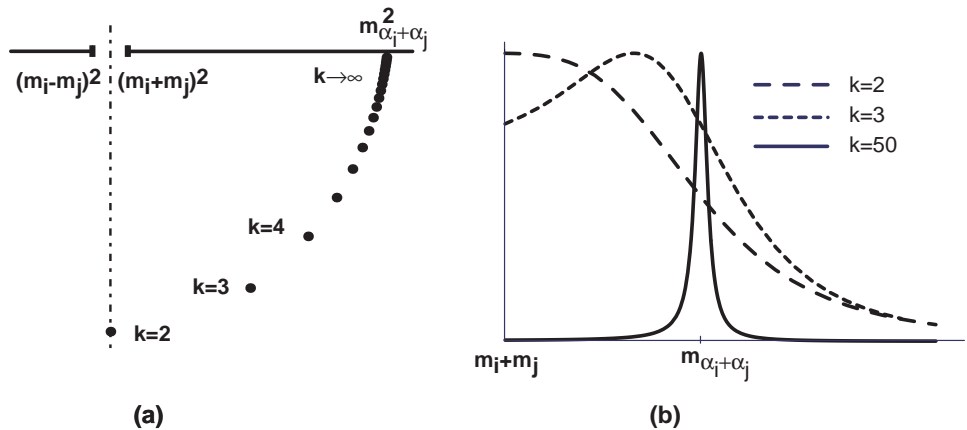,height=5.76truecm,width=12.73truecm}
\caption{\small
(a) The resonance pole of $S{_1^i}{_1^j}$ in the second
Riemann sheet of the complex $s$-plane for various values of $k$. (b)
The scattering probability $|\eta_{ij}{\rm
e\>}^{-i\pi/k}-S{_1^i}{_1^j}|^2$ as a function of the center-of-mass
energy
$\sqrt{s}/M$ for $\sigma_{ji}=5$, where $\eta_{ij}{\rm
e\>}^{-i\pi/k}=\lim_{\;\theta\rightarrow+\infty}S{_1^i}{_1^j}(\theta)$.}
\label{Pole}
\end{center}
\end{figure}

For small $k$, beyond the
semiclassical limit, the interpretation of this pole is not so clear.
It is so far from the physical real axis that the approximation
provided by the Breit-Wigner formula is less useful, as illustrated by
figure~\ref{Pole}. This is particularly clear for $k=2$, when the pole is
located at $s_{{\rm R}_{ij}}= M^2 (m_i^2+m_j^2 -im_i m_j {\rm
e\>}^{\sigma_{ji}})$. Then, for large enough values of the resonance
parameter, namely $\sigma_{ji}\gg
\ln\> (m_i^2+ m_j^2)/m_i m_j$, the ratio
$\Gamma_{{\rm R}_{ij}}/M_{{\rm R}_{ij}}\approx 2$ and,
not surprisingly, the two standard ways to characterise the mass
scale of the would-be unstable state lead to very different
values:
\begin{equation}
M_{\rho_{ij}}= M\sqrt{m_i^2+m_j^2} {\rm \quad and\quad} M_{{\rm
R}_{ij}}
\approx M\sqrt{\frac{m_i m_j}{2}}\> {\rm e\>}^{\sigma_{ji}/2} \>.
\lab{MassChoice}
\end{equation}
Letting $\sigma_{ji}$ tend to infinity, these two scales
can be made arbitrarily far apart.

Thus, even the unstable particles whose resonance poles can in
principle be seen in the elementary amplitudes $S{_1^i}{_1^j}(\theta)$
are hard to identify unambiguously once the semiclassical domain has
been left. The more general amplitudes $S{_a^i}{_b^j}(\theta)$ have a
plethora of resonance poles for generic
values of $k$ -- either ${\rm min\/}(a,b)$ or $k{-}{\rm max\/}(a,b)$\,,
depending on whether $a+b\leq k$ or $a+b\geq k$ -- and the
classification of the resonances in multiparticle scattering,
necessary to see the unstable particles of height greater than $2$,
becomes more and more involved as the number of particles increases.

Fortunately, there are other physical observables, such as
correlation functions and finite-size effects, where all types of
particles play similar roles, setting the scales of
crossover phenomena. This is because the effective
behaviour of the system at a given scale depends on
the number of particle states which are effectively light at
that scale, irrespective of their stability.
Examining the system at different
scales thus 
provides a well-defined method to detect the existence of
physical mass scales associated with
both stable and unstable particles.
There is just one limitation: the nature  of crossover
phenomena means that their study cannot provide the
values of mass scales with arbitrary
precision. In fact, this is not surprising:
while the masses of the {\em stable} particles
can alternatively be extracted, with in principle arbitrary
accuracy, from the far-infrared asymptotics
of correlation functions and finite-size data,
no such option can exist for the unstable particles,
given the uncertainty principle.

In the next sections we shall study the finite-size behaviour
of generic HSG models
using the thermodynamic Bethe ansatz, or TBA.
Mass scales can only be picked up approximately, and some
may be impossible to split apart, but
within these limitations we find complete agreement with
the idea that there is a scale in the quantum theory
associated with each positive
root of $g$,
given by eqs.~\rf{LambdaW} and~\rf{MassRoot}
as a function of the  $S$-matrix parameters
$\{m_i,\sigma_i\}$. This is consistent with the idea that all of the
semiclassical soliton particles survive in the quantum theory as stable
or unstable particles, for any value of
$k$. Moreover, taking~\rf{MassChoice} into
account, our results indicate that, at least for $k=2$, the
commonly-used formula
$M_\rho=\sqrt{{\rm Re\/}(s_{\rm R})}\,$ {\em fails}\/ to characterise
the mass scales of the unstable particles,
and favour the use of $M_{\rm R}$ instead.

\sect{Separating the classical mass scales}
\label{classicalscales}
\iindent
The classical theories depend on $2\rg{-}1$ independent
parameters, but have a set of solitonic
particle states for each of the $\rg\,\hg/2$ positive roots. For generic
values of the parameters, the masses assigned to these roots by the
formula \rf{MassRoot} will all be different, and if mass scales could
be discerned with arbitrary precision, one might hope to distinguish
$\rg\,\hg/2$ different values. However,
this is too much to expect in
the quantum theory: as just discussed, unstable particles
only show up in resonance poles or crossover phenomena, and these do not
yield sharply-defined mass scales.
The most that can be asked is to pick up
{\em well-separated} scales, and this raises the question of how many
of these can be manufactured by varying the $2\rg{-}1$
parameters at our disposal.
Even at the classical level this is a non-trivial problem,
the investigation of which forms the topic of the present section.
With this out of the way we shall return to the quantum theory in
section \ref{TBAsec}.

Before we begin, we should explain what we mean by `well-separated'.
The theory has $2\rg-1$ parameters; for two scales to be well-separated,
we mean that their ratio can be made arbitrarily
large by varying those parameters, while leaving the overall ordering of all
scales in the model unchanged. There may of course be many ways to
do this, but one general prescription will be given in eq.~\rf{rscl}
below. At some stages we shall also consider the logarithms
of well-separated quantities, whose {\em differences} can be made large;
to distinguish between the two concepts, we set up the following
notation:
\begin{eqnarray}
a\,\gg\,\, b &\Leftrightarrow& a-b >\kappa \\
A\ggg B &\Leftrightarrow& A/B >K
\end{eqnarray}
where the constants $\kappa$ and $K$ can be made arbitrarily large
by varying the available parameters, uniformly for all quantities
under discussion. In particular, $a\gg b \Leftrightarrow
e^a\ggg e^b$.

The classical discussion starts with the mass formula
\rf{MassRoot}, which we repeat here:
\begin{equation}
m^2_{\bfm\beta}=(\bfm\lambda_+\cdot
\bfm\beta)\,(\bfm\lambda_-\cdot\bfm\beta)~,
\lab{MassRoota}
\end{equation}
where
\begin{equation}
\bfm\lambda_{\pm}=\sum_{i=1}^{\rg}m_i\> e^{\pm\sigma_i}\>
\bfm\lambda_i~,\quad
\bfm\lambda_i\cdot \bfm\alpha_j=\delta_{ij}~.
\lab{LambdaWa}
\end{equation}
Our task is to characterise, for given values of the parameters $m_i$
and $\sigma_j$, how many separated scales appear in the set of numbers
$\{m^2_{\bfm\beta}\,,\,{\bfm\beta\in \Phi^+_g}\}\,$.

Expanding $\bfm\beta$ in the basis of simple roots $\{\bfm\alpha_i\}$,
\begin{equation}
\bfm\beta=\sum_{i=1}^{\rg}c_i(\bfm\beta)\,\bfm\alpha_i
\lab{Root}
\end{equation}
where the non-negative integers
$c_i(\bfm\beta)=\bfm\lambda_i\cdot\bfm\beta$ are all of order one (the
largest possible value, found for the highest root of the $e_8$
theories, is $6$). 
Substituting into 
\rf{MassRoota},
\begin{equation}
m^2_{\bfm\beta}=
\sum_{i,j=1}^{\rg}c_i(\bfm\beta)c_j(\bfm\beta) \>m_im_j\>
e^{\sigma_i-\sigma_j},
\lab{MassRootb}
\end{equation}
and thus all squared masses are linear
combinations 
of the $\rg(\rg{+}1)/2$ quantities
\begin{equation}
m_im_j\cosh(\sigma_i{-}\sigma_j)~,\quad i,j=1\dots \rg\,,
\lab{MassRootc}
\end{equation}
with coefficients $2c_i(\bfm\beta) c_j(\bfm\beta)$
that are fixed,
and so independent of the parameters $\{m_i,\sigma_j\}$, and are
the squares of numbers of order one. Therefore, we can be sure
that the model has no more than $\rg(\rg{+}1)/2$ separable scales,
given by the numbers
\begin{equation}
m_{ij}=\sqrt{m_im_j}\,e^{|\sigma_i-\sigma_j|/2}~,\quad i,j=1\dots
\rg\,.
\lab{Mdef}
\end{equation}
Only in the $a_n$ theories, for which $\hg=n{+}1=\rg{+}1$, is
$\rg(\rg{+}1)/2$ equal to the number of positive roots -- in all
other cases it is smaller. An immediate consequence is that
the maximal number of separated mass scales that a classical HSG model
can exhibit is generally less than the number of positive roots.
However, two more issues remain. First,
we should check that the
$m_{ij}$
really can be separated. Second,
since these scales only ever appear in
the linear combinations \rf{MassRootb}, for a given
configuration of the parameters, it could be that some numbers from
the set \rf{MassRootc} never occur as the largest term in these
sums, but rather are always swamped, or shielded, by other terms.
This would mean that the number of scales actually present was less
than a naive analysis of \rf{Mdef} would suggest.

The first issue is easily resolved by means of a specific example.
Take the particular choice $m_j=\ex{j a}$ and $\sigma_j= j b$, for
two real numbers $a,b\gg0$. Then
\begin{equation}
m_im_j\cosh(\sigma_i{-}\sigma_j)\approx  \fract{1}{2}\,\ex{(i+j)\>a}\>
\ex{|i-j|\>b}\>.
\lab{DifScales}
\end{equation}
For generic $a$ and $b$ with $a/b$ irrational this
provides $\rg(\rg+1)/2$ different scales,
which can be made of
arbitrarily different magnitude by
choosing $a$ and $b$ large enough.

The second question is more subtle, and requires some more
detailed properties of root systems. Consider a particular number
from the set \rf{MassRootc}, say
$m_km_l\cosh(\sigma_k{-}\sigma_l)$\,.
If $k=l$ then this can always be realised as the squared mass
of a classical particle simply by taking $\bfm\beta=\bfm\alpha_k$ in
\rf{MassRootb}, and so we can take $k\neq l$.
For 
$m_km_l\cosh(\sigma_k{-}\sigma_l)$
to appear in the sum
\rf{MassRootb} for a specific root
$\bfm\beta$, it must be true that $c_k(\bfm\beta)\neq 0$ and
$c_l(\bfm\beta)\neq 0$\,. Now for any root $\bfm\beta$, the set of
simple roots $\bfm\alpha_i$ such that $c_i(\bfm\beta)\neq 0$
is connected on the Dynkin diagram of $g$.\foot{Suppose there are
$s$ connected components to the set of
$\bfm\alpha_i$ with $c_i(\bfm\beta)$ nonzero. Let $\bfm\beta_t$ be the
sum of the $c_i(\bfm\beta)\bfm\alpha_i$ with $\bfm\alpha_i$ restricted
to the $t^{\rm th}$ component, so that
$\bfm\beta=\sum_{t=1}^s\bfm\beta_t$\,. Since each $\bfm\beta_t$ lies
on the root lattice, $|\bfm\beta_t|^2\ge 2$\,; and since the $\bfm\beta_t$
are sums of roots on mutually disconnected portions of the Dynkin
diagram, $\bfm\beta_t\cdot\bfm\beta_{t'}=0$ for $t\neq t'$. Thus
$|\bfm\beta|^2=\sum_{t=1}^s|\bfm\beta_t|^2\geq 2s$\,. But $\bfm\beta$
is a root, so $|\bfm\beta|^2=2$ and hence $s=1$, as
claimed.\label{RootsChains}}
Hence, there is a chain of nodes
$\{{i_1}\ldots i_n\}$\,, on the Dynkin diagram of $g$ with
$\bfm\alpha_{i_1}=\bfm\alpha_k$\,,
$\bfm\alpha_{i_n}=\bfm\alpha_l$\,,
$c_{i_p}(\bev)\not=0$ for $p=1\dots n$\,, and
with $\{{i_p}\,,\,{i_{p+1}}\}$
neighbouring nodes on the Dynkin diagram for $p=1\dots n{-}1$.
This means that whenever
$m_km_l\cosh(\sigma_k{-}\sigma_l)$
appears in one of the
sums \rf{MassRootb}, it is inevitably accompanied by the terms
\begin{equation}
m_{i_p}m_{i_q}\cosh(\sigma_{i_p}{-}\sigma_{i_q})\,,\quad p,q=1\dots
n\,,
\lab{MassRoote}
\end{equation}
and $m_km_l\cosh(\sigma_k{-}\sigma_l)$
must be larger than all of these
numbers if it is not to be swamped.
Taking square roots,
the condition for the scale $m_{kl}$ not to be hidden by the
other scales that always appear with it is that
\begin{equation}
m_{kl}\ggg m_{i_pi_q}\quad \forall~p,q\in\{1\dots n\}
\quad{\rm with}\quad
\{p,q\}\not=\{1,n\}\,,
\lab{cond}
\end{equation}
where the roots $\{\alv_{i_1}\dots\alv_{i_n}\}$ form the
unique chain of simple roots on the Dynkin diagram of $g$ joining
$\alv_k$ to $\alv_l$.
(The chain is unique since non-affine Dynkin diagrams are trees.)

Conversely, suppose that \rf{cond} is satisfied for $(k,l)$. Then, with
$\{\alv_{i_1}\dots\alv_{i_n}\}$ again the chain of roots joining
$\alv_k$ to $\alv_l$,
$\bev=\alv_{i_1}+\cdots +\alv_{i_n}$ is a root of $g$, and
$m_km_l\cosh(\sigma_k{-}\sigma_l)$
is realised as the squared mass scale for the
classical particles associated with $\bev$ -- by \rf{cond}, it
dominates all other
terms in the expression \rf{MassRootb} for $m^2_{\bev}$.
Thus, \rf{cond} gives a necessary and sufficient condition for
$m_{kl}$ to be realised as the mass scale of a set of classical particle
states in the model.

There is a simple graphical method to check all of the
conditions~\rf{cond} at once. Start by drawing a series of horizontal
lines, or `telegraph wires', one for each root in the chain
$\{\alv_{i_1}\dots\alv_{i_n}\}$. Give each wire a coordinate $x$
running from $-\infty$ to $+\infty$,
and for each
$p=1\dots n$, paint those parts of the $p^{\rm th}$ wire with
$|x|>\ln(m_{kl})-\ln(m_{i_p})$ a different colour, red say.

Without loss of generality, assume that
$\sigma_k\ge\sigma_l$\,; otherwise relabel the chain so that $i_1=l$ and
$i_n=k$.
Now draw a zig-zag line between the wires, starting at the point
$x_1=-\ln(m_{kl})+\ln(m_{i_1})$ on the first wire ($p=1$, $i_1=k$) and
then
moving horizontally by an amount $\sigma_{i_p}-\sigma_{i_{p+1}}$ going
from the $p^{\rm th}$ to the $(p{+}1)^{\rm th}$ wire.
This way, the $p^{\rm th}$ segment of the line joins the point
$x_p=x_1-\sigma_{i_p}+\sigma_{i_1}$ on the $p^{\rm th}$ wire to the
point $x_{p{+}1}=x_1-\sigma_{i_{p{+}1}}+\sigma_{i_1}$ on the
$(p{+}1)^{\rm th}$ wire.
By the time the last wire has been reached, the total horizontal shift is
$\sigma_{i_1}-\sigma_{i_n}=\sigma_k-\sigma_l$, and the zig-zag terminates
at $x_n=x_1-\sigma_l+\sigma_k=+\ln(m_{kl})-\ln(m_{i_n})$, on the
$n^{\rm th}$ wire.

Then condition \rf{cond} is equivalent to the following
demand:
\begin{equation}
\fbox{\parbox{4.5in}{Apart from at its beginning and end,
the zig-zag remains far from all
red-painted regions of wire.}}
\lab{telcond}
\end{equation}
This can be proved as follows.

Moving from wire $p$ to wire $q$, the total horizontal shift of the zig-zag
is $\sigma=\sigma_{i_p}-\sigma_{i_q}$\,. For this to fit
between the red-painted regions of wires $p$ and $q$, the absolute value
of $\sigma$ must be less than the horizontal separation of these
two regions, which is $2\ln(m_{kl})-\ln(m_{i_p})-\ln(m_{i_q})$. Hence
\begin{equation}
|\sigma_{i_p}-\sigma_{i_q}|
\ll 2\ln(m_{kl})-\ln(m_{i_p})-\ln(m_{i_q})
\lab{ca}
\end{equation}
and so, recalling \rf{Mdef},
\begin{equation}
2\ln(m_{i_pi_q})\ll 2\ln(m_{kl})
\lab{cb}
\end{equation}
For the zig-zag to remain far from all red-painted regions, this
condition must be met for all $1\le p,q\le n$ with $\{p,q\}\neq \{1,n\}$.
(Note we include the cases $p=q$, for which \rf{ca} reduces to the
requirement that there be a non-zero
gap between the red-painted regions of the
$p^{\rm th}$ wire.)
Dividing \rf{cb} by $2$ and exponentiating, \rf{cond} is recovered.

\newcommand{\sts}{\footnotesize}
\setlength{\unitlength}{1mm}
\thicklines
\newsavebox{\An}
\sbox{\An}{\begin{picture}(52,5)(0,-3.5)
\put(10,6){\mathversion{bold}$a_n$}
\multiput(10,0)(10,0){5}{\circle*{1.75}}
\multiput(10,0)(10,0){2}{\line(1,0){10}}
\multiput(31,0)(1,0){9}{\circle*{.2}}
\put(40,0){\line(1,0){10}}
\put(10,-4){\makebox(0,0)[b]{{\sts 1}}}
\put(20,-4){\makebox(0,0)[b]{{\sts 2}}}
\put(30,-4){\makebox(0,0)[b]{{\sts 3}}}
\put(40,-4){\makebox(0,0)[b]{{\sts {\em n}--1}}}
\put(50,-4){\makebox(0,0)[b]{{\sts {\em n}}}}
\end{picture}}
\newsavebox{\Dn}
\sbox{\Dn}{\begin{picture}(70,15)(0,-5)
\put(10,8){\mathversion{bold}$d_n$}
\multiput(10,0)(10,0){5}{\circle*{1.75}}
\multiput(10,0)(10,0){2}{\line(1,0){10}}
\multiput(31,0)(1,0){9}{\circle*{.2}}
\put(40,0){\line(0,1){10}}
\put(40,0){\line(1,0){10}}
\put(10,-4){\makebox(0,0)[b]{{\sts 1}}}
\put(20,-4){\makebox(0,0)[b]{{\sts 2}}}
\put(30,-4){\makebox(0,0)[b]{{\sts 3}}}
\put(40,-4){\makebox(0,0)[b]{{\sts {\em n}--2}}}
\put(40,10){\circle*{1.5}}
\put(43,9){\makebox(0,0)[b]{{\sts {\em n}}}}
\put(50,-4){\makebox(0,0)[b]{{\sts {\em n}--1}}}
\end{picture}}
\newsavebox{\En}
\sbox{\En}{\begin{picture}(70,15)(0,-5)
\put(10,8){\mathversion{bold}$e_n$}
\multiput(10,0)(10,0){5}{\circle*{1.75}}
\multiput(10,0)(10,0){1}{\line(1,0){10}}
\multiput(21,0)(1,0){9}{\circle*{.2}}
\multiput(30,0)(10,0){2}{\line(1,0){10}}
\put(30,0){\line(0,1){10}}
\put(30,10){\circle*{1.5}}
\put(10,-4){\makebox(0,0)[b]{{\sts 1}}}
\put(20,-4){\makebox(0,0)[b]{{\sts 2}}}
\put(30,-4){\makebox(0,0)[b]{{\sts{\em n}--3}}}
\put(40,-4){\makebox(0,0)[b]{{\sts{\em n}--2}}}
\put(50,-4){\makebox(0,0)[b]{{\sts{\em n}--1}}}
\put(33,9){\makebox(0,0)[b]{{\sts{\em n}}}}
\end{picture}}

\begin{figure}[ht]
\begin{center}
\begin{picture}(130,40)(0,25)
\put(30,50){\usebox{\An}}
\put(0,25){\usebox{\Dn}}
\put(70,25){\usebox{\En}}
\end{picture}
\caption{\small
Dynkin diagrams of the simply-laced
Lie algebras. The numbers show our labelling convention for the nodes.}
\label{DynkinDiag}
\end{center}
\end{figure}
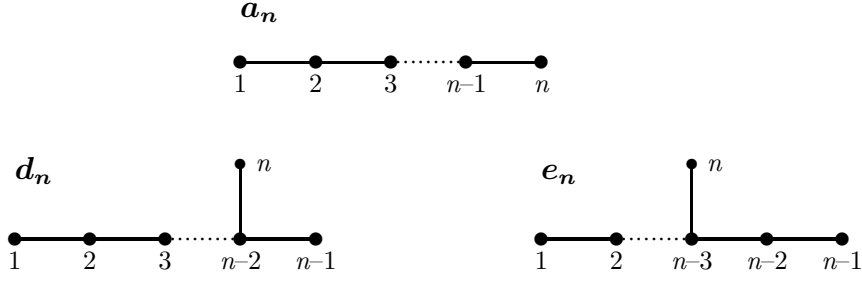


To see the condition in action,
consider the $d_4$, or $SO(8)$, case.
The Lie algebra $d_4$ has $4$ simple roots, $\alv_1\dots\alv_4$,
and $12$ positive roots; we label the
simple roots as in figure~\ref{DynkinDiag}.
For this example we shall take the masses of the four stable
particles equal, and set the parameters $\sigma_i$ as
follows:
\begin{eqnarray}
&&m_1=m_2=m_3=m_4=m~; \nonumber\\
&&\sigma_1=100\,,~~
\sigma_2=80\,,~~
\sigma_3=45\,,~~
\sigma_4=0\,.
\lab{d4example}
\end{eqnarray}
These correspond to the following values of the three original $S$-matrix
resonance parameters:
$\sigma_{12}=20$,
$\sigma_{23}=35$ and
$\sigma_{24}=80$\,.

Grouping the roots together according to their heights, the squared
masses implied by \rf{MassRoota} and~\rf{MassRootb} are
\begin{eqnarray}
\mbox{Height~1\,:}
&&\!\!\!m^2_{\alv_1}= m^2_{\alv_2}= m^2_{\alv_3}= m^2_{\alv_4}= m^2~;
\allowdisplaybreaks\nonumber\\[5pt]
\mbox{Height~2\,:}
&&\!\!\!m^2_{\alv_1+\alv_2}= m^2 (e^{20}+2+e^{-20})\,,
\nonumber\\[3pt]
&&\!\!\!m^2_{\alv_2+\alv_3}= m^2 (e^{35}+2+e^{-35})\,,
\nonumber\\[3pt]
&&\!\!\!m^2_{\alv_2+\alv_4}= m^2 (e^{80}+2+e^{-80})\,;
\allowdisplaybreaks\nonumber\\[5pt]
\mbox{Height~3\,:}
&&\!\!\!m^2_{\alv_1+\alv_2+\alv_3}= m^2
(e^{55}+e^{35}+e^{20}+3+e^{-20}+e^{-35}+e^{-55})\,,
\nonumber\\[3pt]
&&\!\!\!m^2_{\alv_1+\alv_2+\alv_4}= m^2
(e^{100}+e^{80}+e^{20}+3+e^{-20}+e^{-80}+e^{-100})\,,
\nonumber\\[3pt]
&&\!\!\!m^2_{\alv_3+\alv_2+\alv_4}= m^2
(e^{80}+e^{45}+e^{35}+3+e^{-35}+e^{-45}+e^{-80})\,;
\allowdisplaybreaks\nonumber\\[5pt]
\mbox{Height~4\,:}
&&\!\!\!m^2_{\alv_1+\alv_2+\alv_3+\alv_4}= m^2
(e^{100}+e^{80}+e^{55}+e^{45}+e^{35}+e^{20}+4\nonumber\\
&&\!\!\!\phantom{m^2_{\alv_1+\alv_2+\alv_3+\alv_4}= m^2)}
+e^{-20}+e^{-35}+e^{-45}+e^{-55}+e^{-80}+e^{-100})\,;\qquad\qquad~~
\allowdisplaybreaks\nonumber\\[5pt]
\mbox{Height~5\,:}
&&\!\!\!m^2_{\alv_1+2\alv_2+\alv_3+\alv_4}= m^2
(e^{100}+2e^{80}+e^{55}+e^{45}+2e^{35}+2e^{20}+7\nonumber\\
&&\!\!\!\phantom{m^2_{\alv_1+2\alv_2+\alv_3+\alv_4}= m^2)}
+2e^{-20}+2e^{-35}+e^{-45}+e^{-55}+2e^{-80}+e^{-100})\,.\qquad
\lab{sumsone}
\end{eqnarray}
Dropping subleading terms and taking square roots,
the theory therefore has six separated mass scales:
\begin{eqnarray}
m_{ii}\,= m~~~~~ 
&& 
(\,\sim\, m_{\alv_1}\,,~ m_{\alv_2}\,,~ m_{\alv_3}\,,~
m_{\alv_4}\, )
\nonumber\\
m_{12}=me^{10} ~\,
&& 
 (\,\sim\, m_{\alv_1+\alv_2}\,) \nonumber\\
m_{23}=me^{17.5} 
&& 
(\,\sim\, m_{\alv_2+\alv_3}\,)
\nonumber\\
m_{13}=me^{27.5}
&& 
(\,\sim\, m_{\alv_1+\alv_2+\alv_3}\,)
\nonumber\\
m_{24}=me^{40}~\,
&& 
(\,\sim\, m_{\alv_2+\alv_4}\,,~
m_{\alv_3+\alv_2+\alv_4}\,)
\nonumber\\
m_{14}=me^{50}~\,
&&
(\,\sim\, m_{\alv_1+\alv_2+\alv_4}\,,~
m_{\alv_1+\alv_2+\alv_3+\alv_4}\,,~
m_{\alv_1+2\alv_2+\alv_3+\alv_4}\,)
\lab{sumstwo}
\end{eqnarray}
Comparing with the masses following from \rf{Mdef}, one is
missing, namely $m_{34}=me^{22.5}$\,. The reason for its absence is
that, while $(m_{34})^2=m^2e^{45}$ does appear in certain of the sums in
\rf{sumsone}, 
it does not dominate any of them, and hence $m_{34}$ is
always hidden underneath other scales; in particular,
any $m_{\bev}^2$ which might contain $m_{34}^2$ also contains
$m_{24}^2$\,, and
$m_{24}\ggg m_{34}$. In other words, for the choice \rf{d4example}
of parameters, the
scale $m_{34}$ is {\em shielded} by
$m_{24}$. To see that this shielding also
follows from condition \rf{telcond}, figure \ref{chains}
below shows the `telegraph wire'
diagrams for the chains of simple
roots relevant for the mass scales $m_{14}$, $m_{13}$ and $m_{34}$.
(For those viewing the figures
in black-and-white, the red sections of wire have
been made thicker than the other parts.)

\begin{figure}[ht]
\[
\begin{array}{l}
\epsfig{file=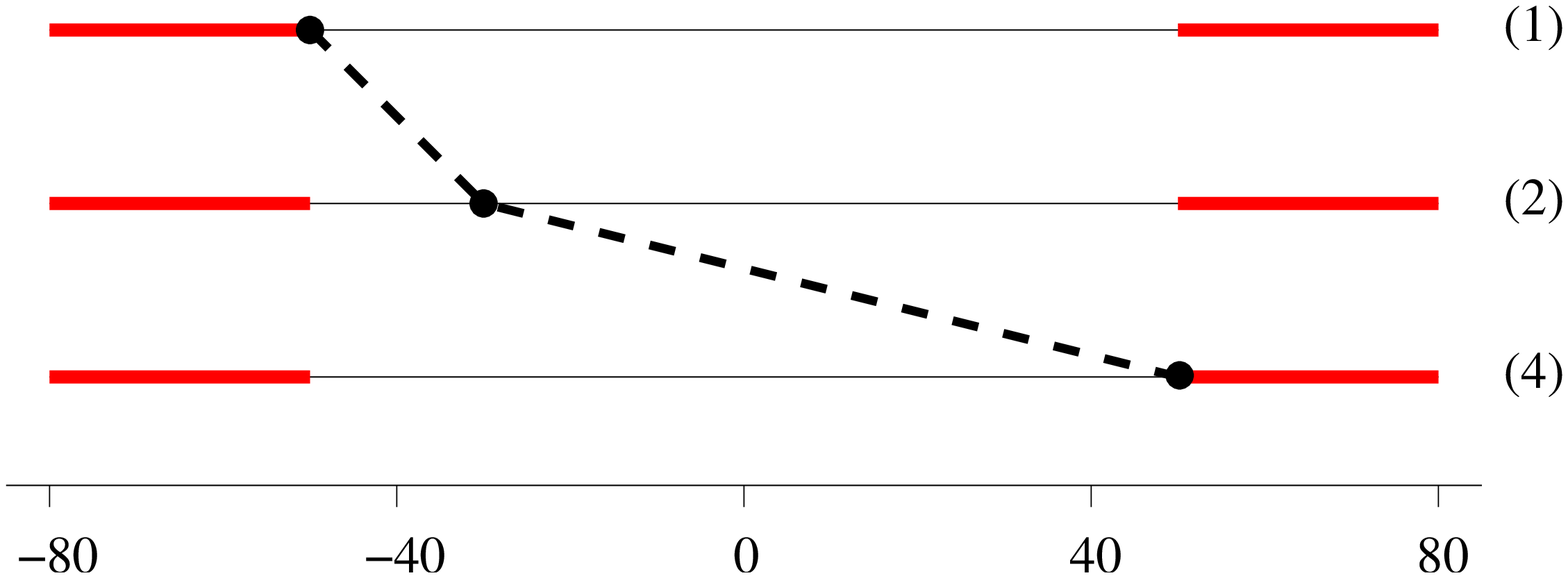,width=9.5truecm}\\[10pt]
\mbox{\small\ref{chains}a) The chain for $m_{14}\,.$}\\[20pt]
\epsfig{file=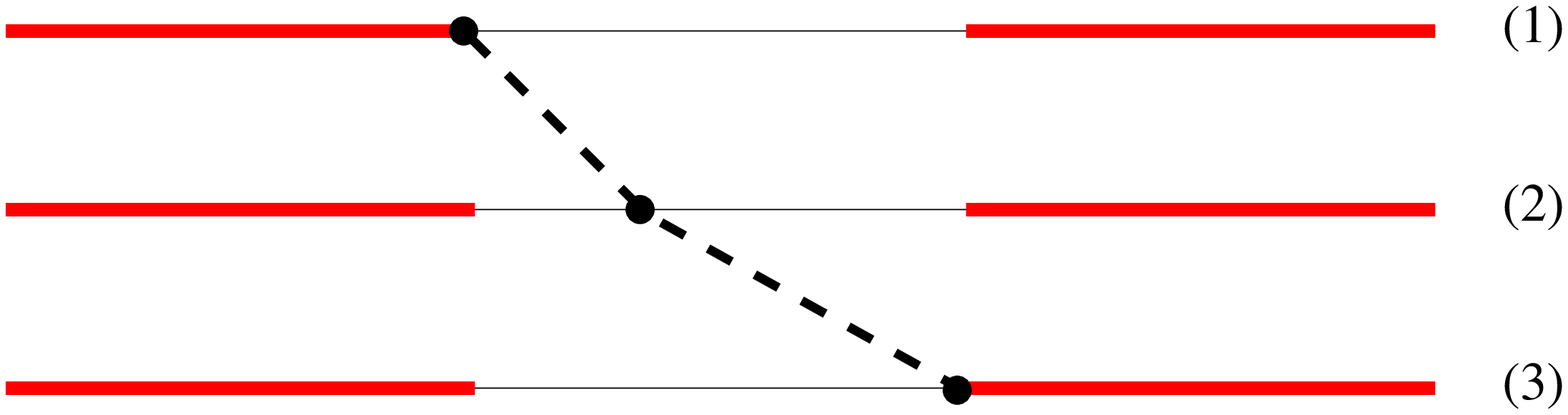,width=9.5truecm}\\[10pt]
\mbox{\small\ref{chains}b) The chain for $m_{13}\,.$}\\[20pt]
\epsfig{file=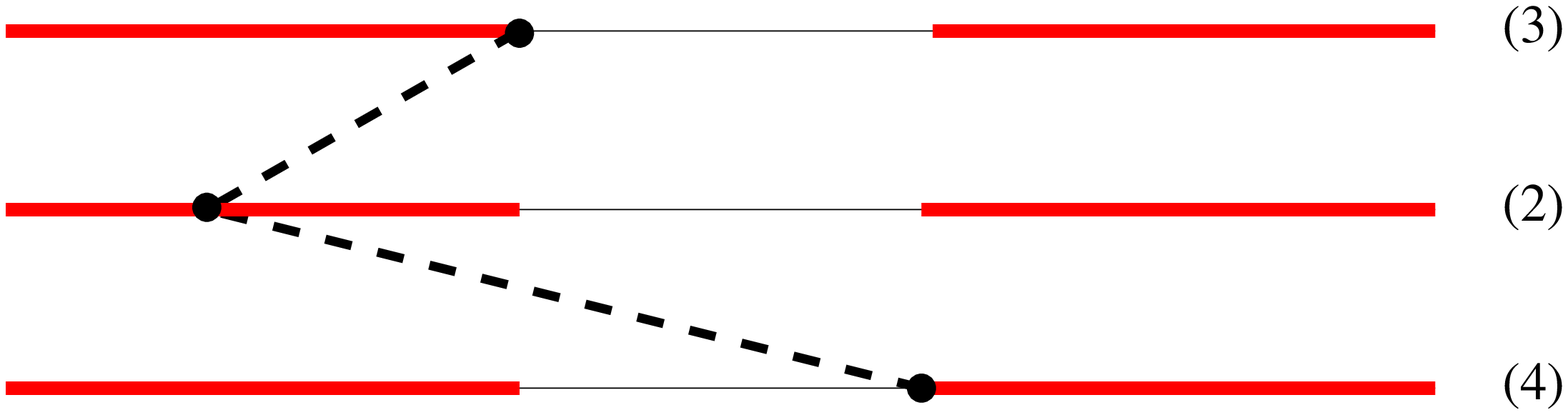,width=9.5truecm}\\[10pt]
\mbox{\small\ref{chains}c) The chain for $m_{34}\,.$}
\end{array}
\]
\label{chains}
\caption{\small Telegraph wire diagrams for the would-be
scales $m_{14}$, $m_{13}$ and $m_{34}$ in the $d_4$ example of
section \ref{classicalscales}.}
\end{figure}

Clearly, figures \ref{chains}a and \ref{chains}b meet
condition \rf{telcond}, while figure \ref{chains}c does not.
In the next section this rephrasing of the shielding criterion
will be used to show in complete
generality that the separated
mass scales in the quantum theory, as seen in
the finite-size crossover effects described by the TBA equations,
precisely match those of the classical theory.

\sect{The quantum theory}
\label{TBAsec}
\subsection{The TBA equations}
\iindent
The thermodynamic Bethe ansatz (TBA) is an exact method for the
calculation of 
the ground state energy of an integrable quantum field
theory on a circle of circumference $R$ or, equivalently, the free
energy of the same theory at finite temperature $T=R^{-1}$~\cite{TBAGEN}.
This allows the theory to be studied non-perturbatively at all length
scales, by varying the value of $R$.
The  key input to the method is the set of two-particle S-matrix
elements for the
scattering of stable particle states, and these were proposed for the
simply-laced HSG models in
\cite{SMAT}.
The masses of the one-particle states are given by
the formula \rf{Perron}, with $M$ an overall (quantum) mass scale.
The scattering is diagonal, and
to emphasise the similarities that the resulting TBA systems have with
those which had previously arisen
in the contexts of perturbed coset theories
\cite{ZAMOcoset,RAVA,DYNK} and staircase
models \cite{PAT,SPIRAL}, we shall rewrite the S-matrix elements of
\cite{SMAT}
in a slightly modified notation. Borrowing from \cite{PAT},
define two functions
\begin{equation}
S_{ab}^{\rm min}%
(\theta)=\prod_{x\in A_{ab}} \bigl\{x\bigr\}(\theta)~,\qquad
S^F_{ab}(\theta)= \prod_{x\in A_{ab}} \bigl(x\bigr)(\theta)\>,
\end{equation}
where
$A_{ab}$ is the set of integers $\{a+b+1-2l\}_{l=1}^{{\rm min}(a,b)}$\,,
and
the blocks
\begin{equation}
\bigl\{x\bigr\}= \bigl(x{-}1\bigr)\bigl(x{+}1\bigr)
\>,\qquad
\bigl(x\bigr) (\theta)=
\frac{\sinh\frac{1}{2}( \theta +i\frac{\pi x}{k})}{%
\sinh\frac{1}{2}(\theta -i\frac{\pi x}{k})}
\lab{Block}
\end{equation}
are as in~\cite{CORR}.
The two-particle scattering
amplitudes are then
\begin{eqnarray}
S{_a^i}{_b^i}(\theta) &=&
S_{ab}^{\rm min} (\theta)
\lab{MINS} \\
S{_a^i}{_b^j}(\theta) &=&
\left[(\eta_{ij})^{ab}\>
S^F_{ab}(\theta+\sigma_{ij})
\right]^{-I^{g}_{ij}}\quad {\rm for} \quad i\not=j\>,
\lab{NewS}
\end{eqnarray}
where $\theta$ is the rapidity, $I^{g}$ is the incidence
matrix of $g$, and $\eta_{ij}= \eta_{ji}^{-1}$ are arbitrary
(fixed) $k^{\rm th}$ roots of~$-1$.
Recall that the functions $S_{ab}^{\rm F}(\theta)$ sometimes
fail, by a sign, to satisfy
the bootstrap and
crossing equations holding for
$S_{ab}^{\rm min}(\theta)$~\cite{PAT}; in contrast, the scattering
amplitudes
$S{_a^i}{_b^j}(\theta)$ do satisfy them due to the constant factors
$\eta_{ij}$~\cite{SMAT}.
The numbers $\sigma_{ij}=-\sigma_{ji}= \sigma_i-\sigma_j$ are the
real-valued resonance parameters seen earlier; since they
need only be defined for $I^{g}_{ij}=1$, there
is an independent resonance
parameter for each of the $\rg-1$
links on the Dynkin diagram of $g$.
An integral representation for these scattering amplitudes was
given in~\cite{HSGTBA}.

Unlike the resonance parameters,
the relative mass scales do
not appear explicitly in the S-matrix. Instead,
they emerge when the TBA
equations are introduced.
These equations have the standard form for a diagonal scattering theory,
though care is needed in their derivation owing to the parity-breaking
of the model \cite{HSGTBA}.
There is a pseudoenergy $\varepsilon_a^{i}(\theta)$ for each of the
$(k{-}1)\times \rg$ stable particles;
the mass scales $m_i$ influence them via
$(k{-}1)\times \rg$ energy terms
\begin{equation}
\nu_{a}^i(\theta)=
 M_{a}^i R\> \cosh\theta =
 m_{i}\mu_a r\> \cosh\theta
\lab{EnergyTerms}
\end{equation}
where
$\mu_a=\sin(\pi a/k)/\sin(\pi/k)$ as before, and $r$ is a
dimensionless overall crossover scale:
\begin{equation}
r=MR\,.
\end{equation}
Defining $L_{a}^i (\theta)=
\ln (1+{\rm e}^{-\varepsilon_{a}^i (\theta)})$, the pseudoenergies
solve the
TBA equations
\begin{equation}
\varepsilon_{a}^i (\theta)= \nu_{a}^i(\theta) -
\sum_{b=1}^{k-1}\left(\phi_{ab}\ast L_{b}^i (\theta) +
\sum_{j=1}^{\rg}
\>I^g_{ij}\>
\psi_{ab}\ast L_{b}^j (\theta-\sigma_{ji})
\right)
\lab{TBAGen}
\end{equation}
where~`$\ast$' denotes
the rapidity convolution
\begin{equation}
f\ast g(\theta)= \int_{-\infty}^{+\infty}\> \frac{d\theta'}{2\pi}\>
f(\theta-\theta')\> g(\theta')\>,
\end{equation}
and the TBA kernel functions $\phi$ and $\psi$  are
\bea
&&\phi_{ab}(\theta)= -i\frac{{\rm d}}{{\rm d}\theta} \ln S{_a^i}{_b^i}
(\theta)
= -i\frac{{\rm d}}{{\rm d}\theta} \ln S_{ab}^{\rm min}(\theta)\>, \nn
\noalign{\vskip0.4truecm}
&&\psi_{ab}(\theta)= -i\frac{{\rm d}}{{\rm d}\theta}
\ln S{_a^i}{_b^j} (\theta+\sigma_{ji})
= +i\frac{{\rm d}}{{\rm d}\theta} \ln S_{ab}^{\rm
F}(\theta)\>,\quad {\rm for}\quad I^g_{ij}=1\>.
\ena
The dimensionless effective central charge $c(r)$
for the theory at scale $r$
is expressed in the standard way in terms of the energy terms and
the solutions to the TBA equations:
\begin{equation}
c(r)= \frac{3}{\pi^2}\> \sum_{i=1}^{\rg}\> \sum_{a=1}^{k-1}\>
\int_{-\infty}^{+\infty} d\theta\> \nu_{a}^i(\theta) \> L_a^i(\theta)\>,
\lab{effc}
\end{equation}
and the ground state energy $E(R)$
is related to this through
\begin{equation}
E(R)= -\frac{\pi}{6R}\,c(M\!R)\,.
\end{equation}
A bulk term linear in $R$ may also contribute to $E(R)$, but this
need not concern us here -- most of the physically-relevant
information from the point of view of RG flows is already contained
in $c(r)$.  Note that $c(r)$ depends not only on $r$, but also
on the $\rg$ mass scales $\{m_i\}$ and the $\rg-1$ resonance
parameters
$\{\sigma_{ij}\}$\,: $c(r)=c(r,\{m_i\},\{\sigma_{ij}\})$.
However, since
\begin{equation}
c(r,\{m_i\},\{\sigma_{ij}\})=
c(\alpha r,\{\alpha^{-1}m_i\},\{\sigma_{ij}\})
\end{equation}
the effective central charge
depends non-trivially on just $(2\rg-1)$ parameters, one of which can be
chosen to be the dimensionless overall crossover
scale.

The limiting value of $c(r)$ as
$r\rightarrow0$ with all other parameters fixed
is equal, in unitary cases such as
these, to the central charge of the
conformal field theory which is
the far UV limit of the theory.
For the HSG theories this was calculated
in~\cite{HSGTBA}, with the result
\begin{equation}
\lim_{r\rightarrow0} c(r)=\frac{k-1}{k+\hg}\> \hg \rg\>,
\lab{ConstantTBA}
\end{equation}
which is the central charge of the $G_k/U(1)^{\rg}$ coset CFT.
This holds for any fixed choice of the $\rg$ relative mass
scales $0<m_i<+\infty$ and the
$\rg-1$ resonance 
parameters $-\infty<\sigma_{ij}<+\infty$.
In the opposite, $r\to\infty$, limit, $c(r)$ tends to
zero, as expected for a massive theory.

\subsection{The staircase flow}
\iindent
The value of the central charge in the far UV
is not the only information hidden inside
the HSG TBA equations. For intermediate values of $r$, depending on the
values taken by the $\{m_i\}$ and the $\{\sigma_{ij}\}$,
numerical work has shown that the scaling function $c(r)$ can have
a characteristic `staircase' pattern, hinting at a renormalisation group
flow which passes close to a number of other fixed points.
In
contrast to Zamolodchikov's original staircase model~\cite{ZAMO} and
its generalizations in~\cite{PAT,MARTINS,SPIRAL}, the number of steps
is always finite. Furthermore, for the HSG models
the  staircase pattern can be understood physically, as
a consequence of the decoupling of those stable or
unstable particles that are effectively heavy at the relative
energy scale fixed by the temperature $r^{-1}$.
This was demonstrated for the $SU(3)_k/U(1)^2$ HSG models in
\cite{HSGTBA,FFRG}, but these cases are too simple to be affected by
the subtleties about separable
mass scales and shielding that were discussed in the last section.
Here we shall give a more general analysis, following
a line of argument used for other staircase models in \cite{PAT,SPIRAL}.
This will allow a full understanding
of the staircase pattern to be gained, subject only to
mild assumptions about the
form of the solutions to~\rf{TBAGen}.
These assumptions are no more severe than those made in
the analysis of the UV limit of more-usual TBA systems, but we have
nevertheless
verified our predictions numerically in a
number of particular cases. These checks, which also serve to illustrate
the patterns of flows,
will be reported in
section \ref{examplesection}.

We shall work at a fixed (finite) value of $k$. Since our
interest is in scales which can be made arbitrarily well-separated,
the constants $\mu_a=\sin(\pi a/k)/\sin(\pi/k)$ appearing in the
energy terms $\nu^i_a(\theta)=m_i\mu_ar\cosh(\theta)$ can be taken to
be of order one, and ignored for the rest of the discussion.
The pseudoenergies are then controlled by two sets of
numbers: the resonance parameters $\{\sigma_{ij}\}$, and
the values of the
stable mass scales relative to the (inverse)
system size, $m_ir$, which
are conveniently parametrised by defining
\begin{equation}
\theta_i(r)\equiv \ln(\fract{2}{m_ir})~.
\end{equation}

For $|\theta|\gg \theta_i(r)$, the energy terms $\nu^i_a(\theta)$
completely
dominate the TBA equations for the corresponding pseudoenergies, and
as a result 
\begin{equation}
\varepsilon_a^i(\theta) \ggg 1~~~\mbox{for}~~~
|\theta|\gg\theta_i(r)~,~~a=1\dots k{-}1\,.
\lab{nudec}
\end{equation}
This causes the functions $L_a^i(\theta)$ to suffer a double
exponential decay in this region, and to the level of approximation
to which we are working,
\begin{equation}
L_a^i(\theta) \approx0\
\quad {\rm for}\quad |\theta|\gg \theta_i(r)~,~~ a=1\dots
k{-}1\,.
\lab{Hypo}
\end{equation}
An important special case is
\begin{equation}
L_a^i(\theta) \approx0\
\quad \forall\, \theta\quad\mbox{if}~ \theta_i(r)\ll 0\,.
\lab{decup}
\end{equation}
In this event the pseudoenergy $\varepsilon^i_a(\theta)$
contributes neither directly to the effective central
charge, nor indirectly via any influence on
the values of the other pseudoenergies. This happens when $2/r\lll
m_i$\,; physically, it corresponds to the energy scale set by the
system size, $R^{-1}=M/r$, being so much less than the mass scale of the
(stable) particles
of type $i$, $Mm_i$, that these particles are effectively decoupled.
Such decouplings have the effect of splitting the original HSG
model into smaller HSG models.

Such cases apart, there will be a region
$|\theta|\ll\theta_i(r)$ within which the energy terms
$\nu^a_i(\theta)$ are exponentially small, allowing them to
be dropped from the TBA equations:
\begin{equation}
\nu_a^i(\theta) \approx 0~~~\mbox{for}~~~
|\theta|\ll\theta_i(r)~,~~a=1\dots k-1\,.
\end{equation}
In this region no immediate conclusion can be drawn about the values of
the pseudoenergies $\varepsilon^i_a(\theta)$, as they continue to
interact with other pseudoenergies via the convolution terms. The key
feature~\cite{PAT} 
of this interaction is that it is {\em localised}\/ in
rapidity-space:
for real values of $\theta$,
the kernels $\phi_{ab}(\theta)$ and $\psi_{ab}(\theta)$ are
peaked about
$\theta=0$, and fall exponentially to
zero outside a region of order one. In the absence of the
resonance parameters $\sigma_{ji}$\,, this implies that, apart from
the driving energy term, the value
of any pseudoenergy near a given value of $\theta$ is only influenced by
the values of the other pseudoenergies near that same value of $\theta$.
(This lies behind the presence of `kink' solutions in even
the simplest TBA systems \cite{TBAGEN}.)
For the HSG models, as for the earlier examples of staircase
models,
non-zero
values of the resonance parameters cause the interactions between
pseudoenergies to be shifted,
so that, for $I^g_{ij}\neq 0$, the TBA equations
\rf{TBAGen}
couple $\varepsilon^i_a(\theta_0)$
not to $\varepsilon^j_b(\theta)$
near $\theta=\theta_0$, but
near $\theta=\theta_0-\sigma_{ji}$\,.
In turn, each $\varepsilon^j_b(\theta_0-\sigma_{ji})$ interacts with
further
pseudoenergies $\varepsilon^k_c(\theta)$ near
$\theta=\theta_0-\sigma_{ji}-\sigma_{kj}=\theta_0-\sigma_{ki}$\,,
for all $k$ such that $I^g_{jk}\neq 0$.
Continuing, it is clear that in the absence of the energy terms
the TBA equations
couple all pairs of pseudoenergies, with
$\varepsilon_a^i(\theta)$ near $\theta=\theta_0$ interacting with
$\varepsilon_b^j(\theta)$ near $\theta=\theta_0-\sigma_{ji}$ via a
unique sequence of
pairwise interactions along the links of the Dynkin diagram.

The fact that the non-affine Dynkin diagrams are trees,
together
with the antisymmetry of $\sigma_{ij}$, means that the set of
rapidities $\theta_0-\sigma_{ji}$ with which the pseudoenergy
$\varepsilon_a^i(\theta)$ at $\theta\approx\theta_0$
interacts is finite.
This contrasts with the
original staircase models of
\cite{ZAMO,MARTINS,PAT,SPIRAL},
where for non-zero values of the resonance parameter and
in the absence of energy terms the
psedoenergies are coupled at infinitely-many
shifted values of~$\theta$.
This distinction is the reason why
the TBA equations for the original staircase models can
show an infinite number of steps, while
for the HSG models
the number of steps is always finite\footnote{It may be instructive
to make the relationship between the spiral staircase models constructed
in~\cite{SPIRAL} and the HSG models more explicit.
Each spiral staircase model is associated with a simply-laced Lie algebra
${\cal G}$ and a cyclic group
${\fl Z}_n$, which can be viewed as the Dynkin diagram of the affine
algebra
$a_{n-1}^{(1)}$. Then, for ${\cal G}=a_{k-1}$, the TBA equations
defining
the spiral staircase model can be seen as the TBA equations of the
$SU(n)_k/U(1)^{n-1}$ HSG model with a particular (limiting)
choice for the energy terms and
resonance parameters, and with $a_{n-1}$ replaced by
$a_{n-1}^{(1)}$. 
}.

We now return to the effect of the energy terms, which bring the
scale-dependence  into the TBA equations. Select a pair of nodes $i$
and $j$ on the Dynkin diagram, and suppose that $r$ is such that
\begin{equation}
\theta_i(r)\gg 0\quad {\rm and}\quad
\theta_j(r)
\gg 0\>,
\end{equation}
so that neither node is decoupled.
For $\theta\approx \pm\theta_i(r)$,
the energy term $\nu_a^i(\theta)$
entering the TBA equation for $\varepsilon_a^i$ is of
the same order as the convolution term;
at these values of $\theta$,
$\varepsilon_a^i(\theta)$ has a non-trivial behaviour, and at generic
values of $r$ has the form of a so-called `kink solution' \cite{TBAGEN}
of the TBA equations. Likewise,
$\varepsilon_a^j(\theta)$ generally has a kink behaviour for $\theta
\approx \mp\theta_j(r)$. However, these two would-be kinks
may influence each other via the chains of convolution terms
just discussed. For values of $r$ such that this occurs,
the solution of the TBA system will depend on $r$ in a non-trivial way,
causing the value of the effective central charge to
change and signalling a crossover in
the finite-size behaviour of the model.

Two conditions must be satisfied for the interaction to occur.
Taking the shift
$\sigma_{ij}$ moving from node $i$ to node $j$ into account,
the first is that
either
\begin{equation}
\theta_i(r)-\sigma_{ji}\approx -\theta_j(r)
\qquad {\rm or} \qquad
-\theta_i(r)-\sigma_{ji}\approx \theta_j(r)\>,
\lab{crosscond}
\end{equation}
depending on whether $\sigma_{ji}>0$ or $\sigma_{ji}<0$, respectively.
Both cases are summarised by
\begin{equation}
\theta_i(r)+
\theta_j(r)\approx |\sigma_{ij}|\>,
\end{equation}
where the presence of the modulus sign is consistent with the
requirement
that both $\theta_i(r)$ and $\theta_j(r)$ be
non-negative. Rearranging, the
condition is
\begin{equation}
\frac{2}{r}\approx\sqrt{m_im_j}\> {\rm e\>}^{|\sigma_{ij}|/2}\,.
\lab{Scales}
\end{equation}
In other words, $r\approx 2/m_{ij}$, and the
physical system size $R$ is of the
order of the length-scale set by $Mm_{ij}$. Taken over all values of
$i$ and $j$, this yields exactly the set
of crossover scales that
one would predict on the basis of a na\"\i ve analysis
of the set \rf{Mdef} of classical mass scales.

However, the simple picture of a crossover for every pair of
pseudoenegies, caused by the
interaction between the corresponding pairs of kink systems,
can break down once the effects of
other energy terms are taken into consideration.
According to~\rf{Hypo},
the energy terms force the functions $L^l_a(\theta)$ to zero for
$|\theta|\gg\theta_l(r)$, irrespective of the values of any
other pseudoenergies.
To take this into account,
an additional condition is required to ensure that the chain of
interactions connecting the kinks for $\varepsilon_a^i(\theta)$ and
$\varepsilon_a^j(\theta)$ near to
$\theta = \pm\theta_i(r)$ and $\theta =
\mp\theta_j(r)$ is actually effective.
Suppose that $\sigma_{ji}>0$,
and let $\{i_1\ldots i_n\}$
be the unique chain of adjacent nodes
on the Dynkin diagram joining nodes $i$ and $j$, so that $i_1=i$ and
$i_n=j$.
Then, for any $p=2\ldots n{-}1$, the value of
$\varepsilon_a^{i}(\theta)$ at $\theta \approx\theta_i(r)$ is
coupled to the value of $\varepsilon_b^{i_p}(\theta)$ at $\theta
\approx\theta_i(r)-\sigma_{i_p i}$ provided
that $|\theta_i(r)-\sigma_{i_p i}|\ll \theta_{i_p}(r)$.
Therefore, taking~\rf{Scales} into account, the required
interaction between kinks will only take place if
\begin{equation}
|\theta_i(r)-\sigma_{i_p i}|\ll \theta_{i_p}(r) \qquad \forall\>
p=2\ldots n{-}1 \qquad {\rm at}\quad r\approx \frac{2}{m_{ij}}\>.
\lab{Crossover}
\end{equation}
This is identical to the classical condition
summarised by eq.~\rf{telcond}, which ensures that the (classical)
scale $m_{ij}$
is not shielded, and really does appear as the dominant term in the
mass of some classical particle.  Therefore, we deduce
that there is a crossover in the finite-size behaviour of
the quantum 
model at $r\approx 2/m_{ij}$ for each {\em unshielded}\/
mass scale $m_{ij}$
within the set of numbers given by eq.~\rf{Mdef}, where $m_i$ and
$\sigma_i$ are now the (quantum) TBA
parameters.
This is one of our main results: for given values
of the parameters, the set of distinct
mass scales picked up by the finite-size
crossover behaviour as the system size varies from zero to infinity is
exactly the same as would have been predicted from an examination of
the full set of
classical particle masses, stable and unstable. This match includes the
shielding of classical scales
that was discussed 
in section \ref{classicalscales},
and holds for all values of $k$, and not just
semiclassically.
Otherwise stated, for the quantum theory, the set of scales at which
crossover phenomena occur is not $\{m_{ij},i,j=1\dots \rg\}$, but
rather $\{m_{\bev},\> \bev\in \Phi^+_g\}$, just as in the classical theory.

These non-perturbative results also provide a quantitative test
for the accuracy of the identifications of $M_\rho$ and $M_{\rm R}$, as
defined in section~\ref{PARTsec}, with the physical mass scales of the
unstable particles. Take two neighbouring nodes on the
Dynkin diagram of $g$, say $\{i,j\}$, and consider the
simple pole of the amplitude
$S{_1^i}{_1^j}(\theta)$ at
$\theta_{{\rm R}_{ij}}=\sigma_{ji}-i\pi/k$. As explained just
after~\rf{SPole}, when
$\sigma_{ji}>0$ this pole is expected to be
the trace of the unstable particle associated with the
root $\alv_i+\alv_j$, of height two. At
level $k=2$ in the limit
$\sigma_{ji}\gg \ln(m_i^2+m_j^2)/m_im_j$\,, the two standard
candidates to characterise the mass of this
particle, from \rf{MassChoice}, are:
\begin{equation}
M_{\rho_{ij}}= M\sqrt{m_i^2+m_j^2} {\rm \quad and\quad}
M_{{\rm
R}_{ij}}
\approx M\sqrt{\frac{m_i m_j}{2}}\> {\rm e\>}^{\sigma_{ji}/2}
=\frac{Mm_{ij}}{\sqrt{2}} \>.
\lab{MassChoiceB}
\end{equation}
For this range of parameters, the mass scale $m_{ij}$ is
unshielded, so $m_{\alv_i+\alv_j}\approx m_{ij}$. Moreover,
according to our results, the finite-size behaviour of the quantum model has
a crossover at $1/r\approx m_{ij}/2$, a consequence of
the decoupling of the unstable particles associated with the
root $\alv_i+\alv_j$. This singles out $Mm_{\alv_i+\alv_j}\approx Mm_{ij}$
as the physical mass scale of the
unstable particle and,
comparing with \rf{MassChoiceB} and taking into account the approximate
nature of the mass  scales provided by the study of finite-size
effects, shows that $M_{{\rm R}_{ij}}$ provides the correct value
for the  mass scale of this unstable particle. In contrast, the value
of $M_{\rho_{ij}}$ can be made arbitrarily far
from the value of $Mm_{\alv_i+\alv_j}$ by letting $\sigma_{ji}$ tend to
infinity. In this case, our results clearly favour the use of $M_{\rm
R}={\rm Re\/}(\sqrt{s_R})$ to characterise the mass scale of the unstable
particles against the more standard choice $M_\rho=\sqrt{{\rm Re\/}(s_R)}$.

For $k>2$, and for the same values of the parameters, the position
of the pole on the complex $s$-plane is
$s_{{\rm R}_{ij}} \approx (M m_{ij})^2 e^{-i{\pi\over k}}$
(see~\rf{SPole}). This leads to the following
values of the two candidate masses:
\begin{equation}
M_{\rho_{ij}}\approx Mm_{\alv_i+\alv_j}
\sqrt{\cos{\pi\over k}}\quad \text{and}\quad
M_{{\rm R}_{ij}}\approx Mm_{\alv_i+\alv_j} \cos{\pi\over
2k}\>,
\end{equation}
which coincide in the large~$k$ limit, as expected. However,
even at finite values of $k\ge 3$, and
in contrast to the $k=2$ case, it is not possible to
make an unambiguous choice between $M_{\rho}$ and $M_R$\,.

\subsection{The central charges on the plateaux}
\label{platsec}

\iindent
At a crossover, the value of the effective central charge $c(r)$
changes rapidly; between any two separated
crossovers it remains approximately
constant. The next task is to calculate this constant,
as it will help us to identify the fixed point being visited by the
staircase flow.  The calculation is largely standard; for the
aspects peculiar to staircase models,
see also \cite{PAT,SPIRAL}.

To place ourselves far from all crossovers, we suppose that
$|\ln(\frac{2}{r})-\ln m_{ij}|\gg 0$ for all $i,j$ such that the
scale $m_{ij}$ is unshielded. (Since we are anticipating that some
of the scales $m_{ij}$ may be well-separated, this does not exclude
$\ln(\frac{2}{r})$ lying between two different crossover scales.)
In addition, to simplify the initial analysis we suppose that the
parameters are such that there
are no `accidental' degeneracies between unshielded scales of
the form $m_{ii}\approx m_{ij}$.
(The precise reasons for these conditions, and what happens when
they are broken, will be discussed in section \ref{effsection} below.)

The effective central charge is given by \rf{effc},
which we rewrite as
\begin{equation}
c(r)=
\sum_{p=1}^{\rg}c\phup_p(r)
\lab{Pieces}
\end{equation}
where
\begin{equation}
c\phup_p(r)=
\frac{3}{\pi^2}
\sum_{a=1}^{k-1}
\int_{-\infty}^{+\infty} d\theta\> \nu_{a}^p(\theta) \> L_a^p(\theta)
\end{equation}
is the direct contribution of the $p^{\rm th}$ set of
pseudoenergies to $c(r)$.
Since
$\nu^p_a(\theta)\approx 0$ for $|\theta|\ll \theta\phup_p(r)$, and
$L_a^p(\theta)\approx 0$ for $|\theta|\gg \theta\phup_p(r)$, this integral
is dominated by the values taken by $L_a^p(\theta)$ at $\theta \approx
\pm\theta\phup_p(r)$ if $\theta\phup_p(r)\gg 0$, and is zero
if $\theta\phup_p(r)\ll 0$. (Note, $\theta\phup_p(r)\approx 0$ corresponds
to 
$|\ln(\frac{2}{r})-\ln m\phup_{pp}|$ being small.
Since $m\phup_{pp}$ is never shielded, this is already excluded by the
requirement that $r$ be
far from all crossover values.)

Suppose that $\theta\phup_p(r)\gg 0$.
Since $\nu_{a}^p(\theta) = m\phup_p\mu_ar\cosh\,\theta$, we have
\begin{equation}
c\phup_p(r)=c_p^-(r)+ c_p^+(r)\,,
\lab{PiecesB}
\end{equation}
where
\begin{equation}
c_p^{\pm}(r)=
\frac{3}{2\pi^2} 
\sum_{a=1}^{k-1}
\int_{-\infty}^{+\infty} d\theta\> m\phup_p\mu_are^{\pm\theta}
  L_a^p(\theta)
=
\frac{3}{\pi^2} 
\sum_{a=1}^{k-1}
\int_{-\infty}^{+\infty} d\theta\> \mu_ae^{-\theta\phup_p(r)\pm\theta}
  L_a^p(\theta)
\lab{LeftRight}
\end{equation}
are the `left' and `right' contributions of the $p^{\rm th}$
pseudoenergy to the
effective central charge: $c^-_p(r)$ is dominated by the values
of $L^p_a(\theta)$ near $\theta=-\theta\phup_p(r)$, and $c^+_p(r)$
by the values near $\theta=+\theta\phup_p(r)$.
The approximate values of $L^p_a(\theta)$ in these two regions are
determined
by a pair of `effective' kink TBA systems. These can
be found using again the facts that the kernel functions
couple pseudoenergies which are adjacent on the Dynkin diagram of
$g$ at values of $\theta$ shifted by the resonance
parameters, and that these chains of interacting pseudoenergies are
cut whenever other energy terms force the functions $L^j_a(\theta)$
to zero. To specify the effective systems precisely, let
$\widetilde{g}_p^{\pm}(r,\{m_i\},\{\sigma_{ij}\})$ be
the (possibly disconnected) Dynkin diagram obtained by deleting all
nodes $j$ on the Dynkin diagram of $g$ for which
$|\pm\theta\phup_p(r)-\sigma_{jp}|\gg \theta_j(r)$, and let
${g}_p^{\pm}(r,\{m_i\},\{\sigma_{ij}\})$ be the connected component of
$\widetilde{g}_p^{\pm}(r,\{m_i\},\{\sigma_{ij}\})$ containing the
node~$p$. 
Then, 
for all nodes $i\in g_p^{\pm}$ and to leading
approximation,
\begin{eqnarray}
L^i_a(\theta)|_{\theta\approx\pm\theta\phup_p(r)-\sigma_{ip}}&\approx&
 L^i_a(\theta\mp\theta\phup_p(r)+\sigma_{ip})^{\pm}~;\nonumber\\[2pt]
\varepsilon^i_a(\theta)|_{\theta\approx\pm\theta\phup_p(r)-\sigma_{ip}}
&\approx&
 \varepsilon^i_a(\theta\mp\theta\phup_p(r)+\sigma_{ip})^{\pm}~,
\lab{lapprox}
\end{eqnarray}
where 
$L^i_a(\theta)^{\pm}=\ln(1+e^{\varepsilon^i_a(\theta)^{\pm}})$
and
the effective TBA system solved by
the `kink pseudoenergies'
$\varepsilon^i_a(\theta)^{\pm}$
is found by substituting the definitions \rf{lapprox} into
\rf{TBAGen} and dropping all subleading terms:
\begin{equation}
\varepsilon_{a}^i(\theta)^\pm
= 
\delta_{ip}\,\mu_{a}e^{\pm\theta} -
\sum_{b=1}^{k-1}\left(\phi_{ab}\ast L_{b}^i (\theta)^\pm +
\sum_{j\in g^\pm_p}
\>I^{g^{\pm}_p}_{ij}\>
\psi_{ab}\ast L_{b}^j (\theta)^\pm
\right).
\lab{etba}
\end{equation}
Here $I^{g^{\pm}_p}_{ij}$ is the incidence matrix of the reduced Dynkin
diagram $g^{\pm}_p$\,. (Note, $g^{\pm}_p$
in fact depends on 
$r$, $\{m_i\}$ and $\{\sigma_{ij}\}$\,, but this  has been
left implicit to avoid overburdening
the notation.) In terms of the kink pseudoenergies, $c_p^{\pm}(r)$ is
simply
\begin{equation}
c_p^{\pm}(r)=
\frac{3}{\pi^2} 
\sum_{a=1}^{k-1}
\int_{-\infty}^{+\infty} d\theta\> \mu_ae^{\pm\theta}
  L_a^p(\theta)^{\pm}\,.
\lab{eceff}
\end{equation}
The rapidity shifts in \rf{lapprox} serve to eliminate all relative
rapidity shifts in the effective TBA system \rf{etba}, though
of course these shifts still influence the system indirectly, via
their role in the determination of the diagram $g^{\pm}_p$\,.
With all shifts removed, the effective TBA
is exactly the kink form of a `Dynkin'
TBA system, of the sort discussed in \cite{DYNK}. Notice that there is
no explicit $r$-dependence -- the value of
$r$ only enters via its effect on
${g}_l^{\pm}(r,\{m_i\},\{\sigma_{ij}\})$\,.
This does not change between crossovers, and so the
expected plateau structure is confirmed.

The integral in \rf{eceff} can be evaluated exactly as a sum of
dilogarithms. Such sums have been well-studied (see, for example,
\cite{TBArefs}), and their values
can be expressed in terms of Lie-algebraic data, as
follows. Let $g_p^{\pm}$ be defined as above, and let
$\widehat g_p^{\pm}$
be the (possibly disconnected) Dynkin diagram formed by
deleting the node $p$ from $g_p^{\pm}$. For any connected
Dynkin diagram $g\in a,d,e$
with rank $r$ and Coxeter number $h$, define $C_k(g)$ by
\begin{equation}
C_k(g)=\frac{k-1}{k+h}\,hr
\lab{cform}
\end{equation}
and if $g$ is disconnected, define $C_k(g)$ to be the sum of
\rf{cform} over all connected components of $g$. Then
\begin{equation}
c_p^{\pm}=\fract{1}{2}(C_k(g_p^{\pm})-C_k(\widehat g_p^{\pm}))\,.
\lab{ccform}
\end{equation}
Note, $C_k(g)$ is the central charge of the $G_k/U(1)^r$
coset; the factor of $\frac{1}{2}$ appears in \rf{ccform} because
the full effective central charge is the sum of two
contributions, one from the left and one from the
right kink system. As we shall
see in an example shortly, the parity-breaking of the HSG models means
that in general the individual terms
$c_p^+$ and $c_p^-$ are not directly related, a contrast to
the behaviour of more usual systems. However, the
effective central charge $c(r)$ is not sensitive to
parity-breaking, in the sense that
the  total `left' and `right' contributions are equal;
{\em i.e.\/}, 
\begin{equation}
\sum_{p=1}^{\rg} c_p^+(r) =\sum_{p=1}^{\rg} c_p^-(r)\>,
\lab{LRident}
\end{equation}
which is proved in appendix~\ref{Heterotic}.

The rules we have given here allow the unambiguous calculation
of $c(r)$ in generic situations, away from any crossovers.
The calculation of the plateau values is not exact
because, for finite values of the $\theta_i(r)$ and
$\sigma_{ij}$, the pseudoenergies approximated by the different effective
kink TBA  systems actually interact with each other, as they correspond to
the behaviours in different regions of a single set of functions. If a
limit could be taken such that the separation between these different
regions
became infinite, then the plateau value for the effective central
charge would be exact. This cannot be achieved simply by taking $r$
to zero -- this would just reproduce the far UV
central charge of the model in every case. Instead, to capture the
intermediate plateaux, a more subtle
`multiple scaling limit' should be taken. The simplest choice is to
settle on a finite set of parameters
$\{\theta_i(r),\sigma_{ij}\}$
away from any crossover, and then rescale as
\begin{equation}
\{\theta_i(r),\sigma_{ij}\}\to
\{\rho\theta_i(r),\rho\sigma_{ij}\}
\end{equation}
for some positive real number $\rho$. In the limit $\rho\to\infty$,
the plateau values of the effective central charge, calculated using
the above rules, become exact. In terms of the original infrared
parameters this limit is essentially
\begin{equation}
\{r,m_i,\sigma_{ij}\}\to
\{2\left(\frac{r}{2}\right)^{\rho},m_i^{\rho},\rho\sigma_{ij}\}~~;~~\rho\to
\infty.
\lab{rscl}
\end{equation}
It is interesting that a single TBA system can hide such a variety
of exact limits. Similar remarks in the simpler
context of the traditional staircase
models were made in \cite{PAT,SPIRAL}.

\subsection{Effective TBA systems for the crossovers}
\label{effsection}
\iindent
The appearance of just a single `driving term'
$\delta_{ip}\,\mu_{a}e^{\pm\theta}$ in each kink TBA
system \rf{etba} is a consequence of
the conditions that $r$ should be far from any crossovers,
and that there should be no accidental degeneracies
between unshielded scales of the form $m_{ii}\approx m_{ij}$.
This can be shown as follows.
Consider the value of $\varepsilon^p_a(\theta)$ near
$\theta=\pm\theta\phup_p(r)$, where the balance between the
energy and convolution terms in its TBA equation causes it to have a
non-trivial form.  Via a chain of links on the Dynkin diagram of $g$,
this form might be influenced by
$\varepsilon^q_a(\theta)$ near
$\theta=\pm\theta\phup_p(r)-\sigma_{qp}$ (here $q$ labels another node on
the Dynkin diagram  of $g$).
Now if $|{\pm}\theta\phup_p(r)-\sigma_{qp}|\gg\theta_q(r)$ then
$\varepsilon^q_a(\theta)$ is completely dominated by the energy term
at these values of $\theta$, and the chain is cut; and if
$|{\pm}\theta\phup_p(r)-\sigma_{qp}|\ll\theta_q(r)$, then the energy
term is effectively zero at these same values of $\theta$, and so no
energy term for $\varepsilon^q_a(\theta)^{\pm}$ should be included in
the effective TBA system. The extra energy term needs only be included
explicitly if
$|{\pm}\theta\phup_p(r)-\sigma_{qp}|\approx\theta_q(r)$, with
the chain linking $p$ to $q$
uncut.
There are two  cases:
(a)
$\theta\phup_p(r)\mp\sigma_{qp}\approx\theta_q(r)$\,,
or (b)
$-\theta\phup_p(r)\pm\sigma_{qp}\approx\theta_q(r)$.
Case (b)
has already appeared in eq.~\rf{crosscond} above, and is ruled out
if $r$ is far from any crossover scale. Case (a)
translates as $m_q/m_p e^{\mp\sigma_{qp}}\approx 1$. Since
$(m_{pq}/m_{qq})^2=(m_p/m_q)e^{|\sigma_{qp}|}$ and
$(m_{pp}/m_{pq})^2=(m_p/m_q)e^{-|\sigma_{qp}|}$, this is ruled out by
the condition on coincidental mass scales.

When the conditions are not met, the decoupling of the full TBA into
separate effective systems, one for each still-active
energy term, is not complete. The effective TBA systems for any set
of energy terms which have not been disentangled must be combined into
one, which itself has more than one driving term.
Suppose for illustration that just one extra driving term, coming from
$\varepsilon^q_a(\theta)$, needs to be included in the effective TBA
system governing the form of $\varepsilon^p_a(\theta)$ near
$\theta=\theta_p(r)$.
(The modifications to the discussion
when an extra term is instead needed for the system
governing $\varepsilon^p_a(\theta)$ near
$\theta=-\theta_p(r)$, or when larger numbers of
extra energy terms are involved, should then be clear.)
Case (a) 
corresponds to $\theta_p(r)-\sigma_{qp}\approx\theta_q(r)$, and
it follows from the definitions made just before eq.~\rf{lapprox}
that the reduced diagrams
$g^{+}_p(r,\{m_i\},\{\sigma_{ij}\})$
and
$g^{+}_q(r,\{m_i\},\{\sigma_{ij}\})$
coincide.
The effective TBA systems governing
$\varepsilon^p_a(\theta)$ near $\theta=\theta_p(r)$
and
$\varepsilon^q_a(\theta)$ near $\theta=\theta_q(r)$ should therefore
be merged.  Defining kink pseudoenergies as in \rf{lapprox}, there are two
energy terms which cannot be discarded and the
effective system is
\begin{equation}
\varepsilon_{a}^i(\theta)^+
= 
\delta_{ip}\,\mu_{a}e^{\theta} +
\delta_{iq}\,
\mu_{a}
\frac{m_q}{m_p}e^{-\sigma_{qp}}
e^{\theta} -
\sum_{b=1}^{k-1}\left(\phi_{ab}\ast L_{b}^i (\theta)^+ +
\sum_{j\in g^\pm_l}
\>I^{g^{\pm}_l}_{ij}\>
\psi_{ab}\ast L_{b}^j (\theta)^+
\right)
\lab{eetba}
\end{equation}
and the separate contributions of
$c_p^+(r)$ and 
$c_q^+(r)$ to the total effective central charge should be replaced
by
\begin{equation}
c_{p+q}^+(r)=
\frac{3}{\pi^2} \sum_{a=1}^{k-1} \int_{-\infty}^{+\infty} d\theta\>
\left( \mu_ae^{\theta} L_a^p(\theta)^+ + \mu_{a}
\frac{m_q}{m_p}e^{-\sigma_{qp}} e^{\theta}
  L_a^q(\theta)^+ \right).
\lab{eeceff}
\end{equation}
Note that the `coincidence condition'
$\theta_p(r)-\sigma_{qp}\approx\theta_q(r)$ implies that
$(m_q/m_p)e^{-\sigma_{qp}}$ is of order $1$.
This term cannot be eliminated from the equations
without reintroducing a
rapidity shift for the second convolution term in \rf{eetba}, but the
apparent asymmetry between $p$ and $q$ can be removed by making
an overall shift $\theta \to \theta+\ln m_p+\sigma_{qp}/2$.
The value of $c^+_{p+q}$ is calculated as for the generic
plateau case discussed in the last section, the only difference being
that the diagram $\widehat g_p^{+}$ is now found by deleting both
nodes $p$ and $q$ from $g_p^+$, instead of just node $p$.
{}From this it follows that the value of $c_{p+q}^{\pm}$ is in
fact independent of $(m_q/m_p)e^{-\sigma_{qp}}$,
and it is easy to check from the rules
for calculation given above that its value is consistent with the
value of $c_p^\pm+c_q^\pm$
found when $(m_q/m_p)e^{-\sigma_{qp}}$ becomes large (or small)
and the $p$ and $q$ kink systems
decouple. Physically this is as it should be --
the approximate equality of the scales $m_{pp}$ and
$m_{pq}$ will only be seen in the finite-size behaviour of the system
at the corresponding crossovers, and
the scale $r$ at which this calculation has been performed is
away from all crossovers.

For case (b), the story is different and gives the archetypal
approximation for the finite-size behaviour during a
crossover. We have $\theta_p(r)-\sigma_{qp}\approx -\theta_q(r)$,
with $\sigma_{qp}>0$ since we have already supposed that both
$\theta_p(r)$ and $\theta_q(r)$ be positive.
The reduced diagrams $g^+_p$ and $g^-_q$ coincide, and
the effective TBA systems governing
$\varepsilon^p_a(\theta)$ near $\theta=\theta_p(r)$ and
$\varepsilon^q_a(\theta)$ near $\theta=-\theta_q(r)$ should be merged.
As the first system involves $e^{\theta}$ in its energy term and the
second $e^{-\theta}$, it is no longer possible to eliminate all
$r$-dependence by an overall shift in $\theta$. If
kink pseudoenergies are again defined as in \rf{lapprox}, the
effective system is
\begin{equation}
\varepsilon_{a}^i(\theta)^+
= 
\delta_{ip}\,\mu_{a}e^{\theta} +
\delta_{iq}\,\mu_{a}\frac{m_pm_q}{4}e^{\sigma_{qp}}r^2\,e^{-\theta} -
\sum_{b=1}^{k-1}\left(\phi_{ab}\ast L_{b}^i (\theta)^+ +
\sum_{j\in g^\pm_p}
\>I^{g^{\pm}_p}_{ij}\>
\psi_{ab}\ast L_{b}^j (\theta)^+
\right)
\lab{eeetbaa}
\end{equation}
Using $m_{pq}^2=m_pm_qe^{\sigma_{qp}}$ (recall that $\sigma_{qp}$ is
positive), shifting $\theta\to\theta+\ln(m_{pq}/2)$ and redefining the
pseudoenergies appropriately,
this can be put in the more symmetrical
form
\begin{equation}
\varepsilon_{a}^i(\theta)^+
= 
\fract{1}{2}\delta_{ip}\,\mu_{a}m_{pq}r\,e^{\theta} +
\fract{1}{2}\delta_{iq}\,\mu_{a}m_{pq}r\,e^{-\theta} -
\sum_{b=1}^{k-1}\left(\phi_{ab}\ast L_{b}^i (\theta)^+ +
\sum_{j\in g^\pm_p}
\>I^{g^{\pm}_p}_{ij}\>
\psi_{ab}\ast L_{b}^j (\theta)^+
\right).
\lab{eeetba}
\end{equation}
The contribution to the effective central charge, which replaces
$c^+_p(r)+c^-_q(r)$, is then\foot{Beware that the
`+' label we associated with this effective TBA
system is potentially misleading, as it actually
arises from the merging of a `+' kink system with a `$-$' kink
system, but for the sake of simplicity we do not introduce a further
notation.}
\begin{equation}
c_{p+q}^+(r)=
\frac{3}{\pi^2} \sum_{a=1}^{k-1} \int_{-\infty}^{+\infty} d\theta\>
\left( 
\fract{1}{2}\mu_{a}m_{pq}r\,e^{\theta}
L_a^p(\theta)^+ + 
\fract{1}{2}\mu_{a}m_{pq}r\,e^{-\theta}
L_a^q(\theta)^+ \right).
\lab{eeeceff}
\end{equation}
These equations make it particularly clear that the crossover scale is
$m_{pq}$. Considered on their own, these HSG crossover TBA
systems
generalise massless TBAs discussed in \cite{DYNK}, in that there is no
requirement for the nodes $p$ and $q$ to be symmetrically-placed on the
Dynkin diagram of $g$. This greater freedom
is related to the fact that the HSG models can
break parity, an option not treated in \cite{DYNK}.
Note also that we only discussed the simplest cases here; by suitably
tuning the parameters it can be arranged for more driving terms to
be present in the effective TBA systems, giving, for example, new
multiparameter families of massless flows which may or may not break
parity.

\subsection{Examples}
\label{examplesection}
\begin{figure}[ht]
\begin{center}
\epsfig{file=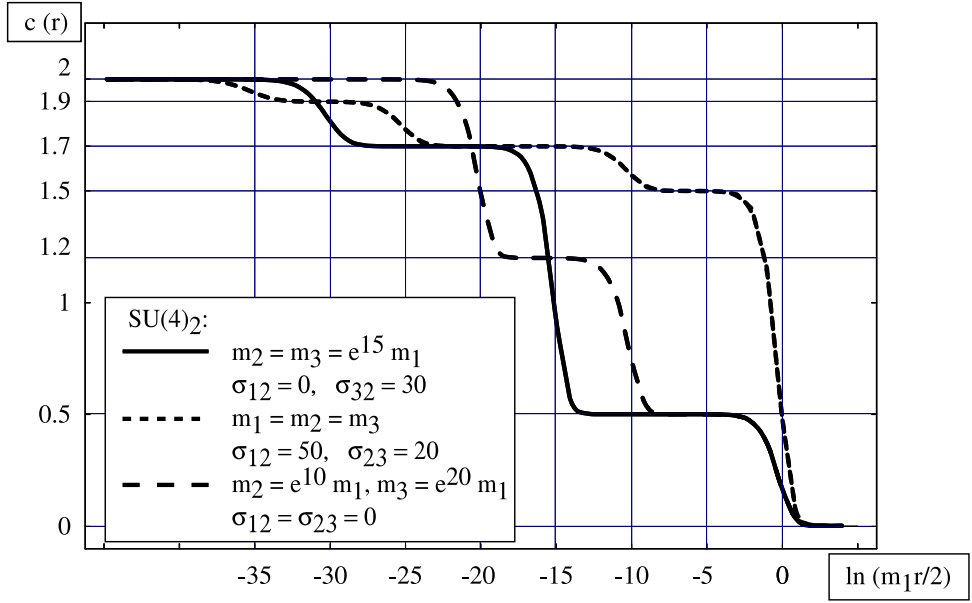,width=0.8\linewidth}
\caption{\small
The TBA scaling functions for the $SU(4)_2/U(1)^3$ HSG model.}
\label{fig1}
\end{center}
\end{figure}

\iindent
We now outline some specific
examples, starting with the $SU(4)_2/U(1)^3$ HSG model, for which a number
of
numerically-obtained plots are shown in figure \ref{fig1}.

Simplest to describe is the flow
with both resonance parameters zero:
$\sigma_{12}=\sigma_{23}=0$, drawn as a dashed line
on the plot. The mass scales are
such that $m_1\lll m_2\lll m_3$.\footnote{We
continue to use the conventions of fig.~\ref{DynkinDiag} for numbering the
nodes of the Dynkin diagram of $g$.} As
$r$ varies from $0$ and $+\infty$, the effective central charge
exhibits three plateaux  corresponding to the regions
$2r^{-1}\ggg m_3$ (the deep UV limit), $m_2\lll 2r^{-1}\lll m_3$, and
$m_1\lll 2r^{-1}\lll  m_2$, before it reaches the massive region for
$2r^{-1}\lll  m_1$ where
$c(r)$ vanishes. Within each region, $c(r)$ matches the central
charges of
the following coset CFTs:
\begin{equation}
\mbox{(UV)}\>\>\>\>\frac{SU(4)_k}{U(1)^{3}} \>
\buildrel m_3\over{\hbox to 30pt{\rightarrowfill}}\>
\frac{SU(3)_k}{U(1)^{2}} \>
\buildrel m_2\over{\hbox to 30pt{\rightarrowfill}}\>
\frac{SU(2)_k}{U(1)}\>
\buildrel m_1\over{\hbox to 30pt{\rightarrowfill}}\>
{\rm Massive}\>\>\>\> \mbox{(IR)}\>.
\lab{Mass1}
\end{equation}
The central charges can be recovered using the rules given above as
follows.  We return to the `telegraph wire' diagrams of the last section,
but this time allow them to depend on $r$ (the previous `chain'
diagrams occur 
as subdiagrams when $r$ is placed at the relevant
crossover scale). Draw a wire for every node of the Dynkin diagram of
$g$ (in this case, that for $a_3$), give each wire a coordinate
$\theta$, and paint red those parts of the $l^{\rm th}$ wire with
$|\theta|>\theta_l(r)$\,, for $l=1\dots \rg$\,.
The diagram $g_l^{\pm}$ is found by first drawing a node for the
`driving' point $\theta=\pm\theta_l(r)$ on the $l^{\rm th}$ wire.
(If $\theta_l(r)$ is negative, then $\varepsilon^l_a(\theta)$ is
already decoupled, and no node need be drawn.)
Then move from wire to wire of the diagram according to the
connectivity of the Dynkin diagram, shifting in $\theta$ by
$\sigma_{ij}$ when moving from wire $i$ to wire $j$. So long as the
points reached lie on unpainted sections of wire, then they should be
included in $g^{\pm}_l$. It is easily checked that this graphical
technique matches the rule for the construction of $g^{\pm}_l$ given
just before \rf{lapprox}.

\begin{figure}[ht]
\[
\begin{array}{l}
\epsfig{file=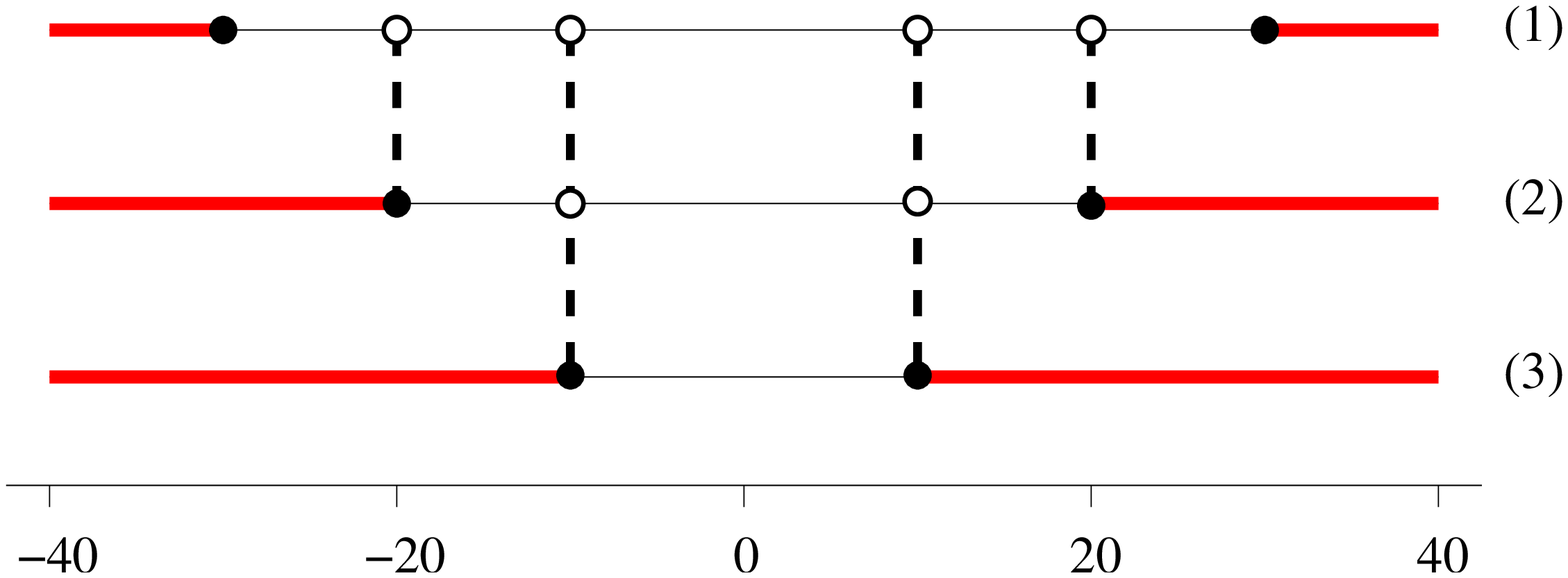,height=0.197\linewidth}\\[10pt]
\mbox{\small\ref{calc1}a)~$\ln(m_1r/2)=-30$.}\\[15pt]
\epsfig{file=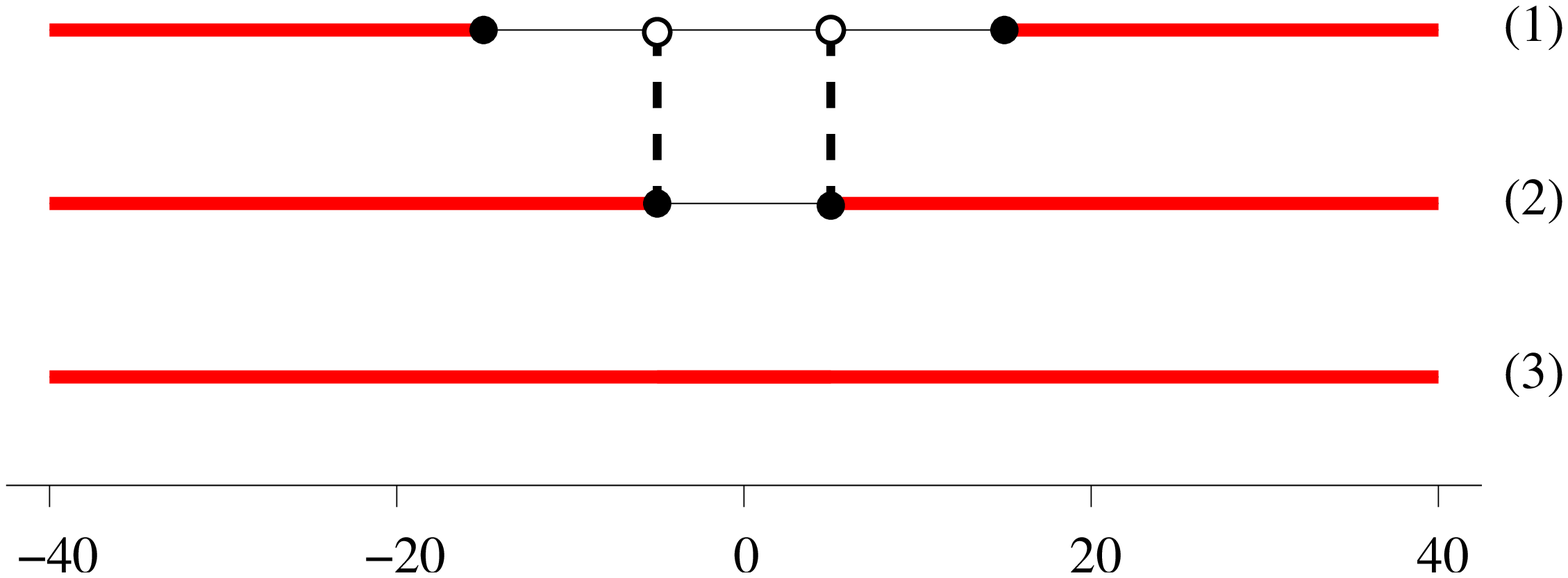,height=0.197\linewidth}\\[10pt]
\mbox{\small\ref{calc1}b)~$\ln(m_1r/2)=-15$.}\\[15pt]
\epsfig{file=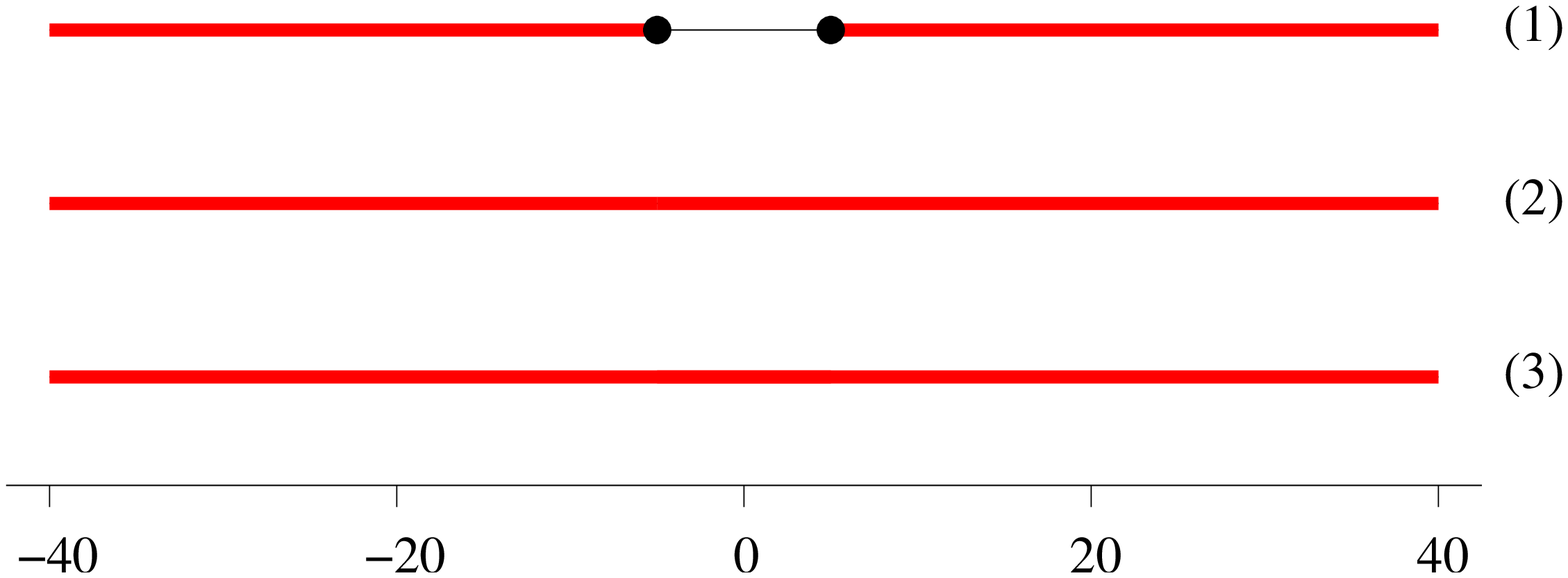,height=0.197\linewidth}\\[10pt]
\mbox{\small\ref{calc1}c)~$\ln(m_1r/2)=-5$.}
\end{array}
\]
\caption{\small Effective TBA systems for the dashed-line flow
in figure \ref{fig1}.}
\label{calc1}
\end{figure}

Figure \ref{calc1}a shows the resulting collection of kink TBA
systems for the dashed-line flow at $\ln(m_1r/2)=-30$. The systems are
symmetrical between
left and right, so $c_l^+=c_l^-$ for each $l$ and
$c_l\phup=2c_l^{\pm}=C_k(g_l^{\pm})-C_k(\widehat g_l^{\pm})$.
Hence the total central charge is
\begin{equation}
c=
C_k(a_1)+
(C_k(a_2){-}C_k(a_1))+
(C_k(a_3){-}C_k(a_2))=C_k(a_3)
\end{equation}
as expected for the far UV limit.

If instead $\ln(m_1r/2)=-15$, the relevant diagram is drawn in figure
\ref{calc1}b. The pseudoenergies $\varepsilon^3_a$
have decoupled, and the central charge is
\begin{equation}
c=
C_k(a_1)+
(C_k(a_2){-}C_k(a_1))
=C_k(a_2)\,.
\end{equation}

Finally, at $\ln(m_1r/2)=-5$ all pseudoenergies but $\varepsilon^1_a$
have decoupled; the diagram is shown in figure \ref{calc1}c and the
central charge is
\begin{equation}
c=
C_k(a_1).
\end{equation}

In contrast, had the mass scales been chosen such that $m_1\simeq
m_3\lll m_2$, then $c(r)$ would have exhibited
only two plateaux with effective central charges
matching a flow
\begin{equation}
\mbox{(UV)}\>\>\>\>\frac{SU(4)_k}{U(1)^{3}} \>
\buildrel m_2\over{\hbox to 30pt{\rightarrowfill}}\>
\frac{SU(2)_k}{U(1)}\times {SU(2)_k\over U(1)}\>
\buildrel m_1\simeq m_3\over{\hbox to 45pt{\rightarrowfill}}\>
{\rm Massive}\>\>\>\> \mbox{(IR)}\>.
\lab{Mass2}
\end{equation}
We leave it to the reader to verify this using the diagrammatic
approach.

Parenthetically, for any HSG model, we remark that if
all the resonance parameters vanish then
the mass scales can be adjusted so as
to permit the existence of a regime in the RG flow from UV
to IR where
\begin{equation}
m_{i_1},\ldots,m_{i_l}\lll 2r^{-1} \lll m_{i_{l+1}},\ldots,m_{i_{\rg}}\>.
\lab{MassOrder}
\end{equation}
In this portion of the flow all the particles of
types `$i_{l+1}$', $\ldots$, `$i_{\rg}$' have already decoupled, with
$L_a^{i_{l+1}}(\theta), \ldots, L_a^{i_{\rg}}(\theta)\approx0$. This
leaves us
with the TBA equations corresponding to the HSG model
associated with the coset
$G^{[i_{l+1}\ldots i_{\rg}]}_k/ U(1)^{\rg}$, where
$G^{[i_{l+1}\ldots i_{\rg}]}$ denotes the subgroup of
$G$ associated with the (possibly disconnected) Dynkin diagram obtained by
removing the
$i_{l+1},\ldots, i_{\rg}\text{--th}$ nodes from the Dynkin diagram of $g$,
times a
$U(1)^{\rg-l}$ factor associated with those nodes.
Physically, these staircase patterns reflect the
decoupling of the stable particles when they are heavy compared to
the relative energy scale fixed by the temperature $r^{-1}$, and the
possible splitting of the initial HSG model into a number of
decoupled components.

\begin{figure}[ht]
\[
\begin{array}{l}
\epsfig{file=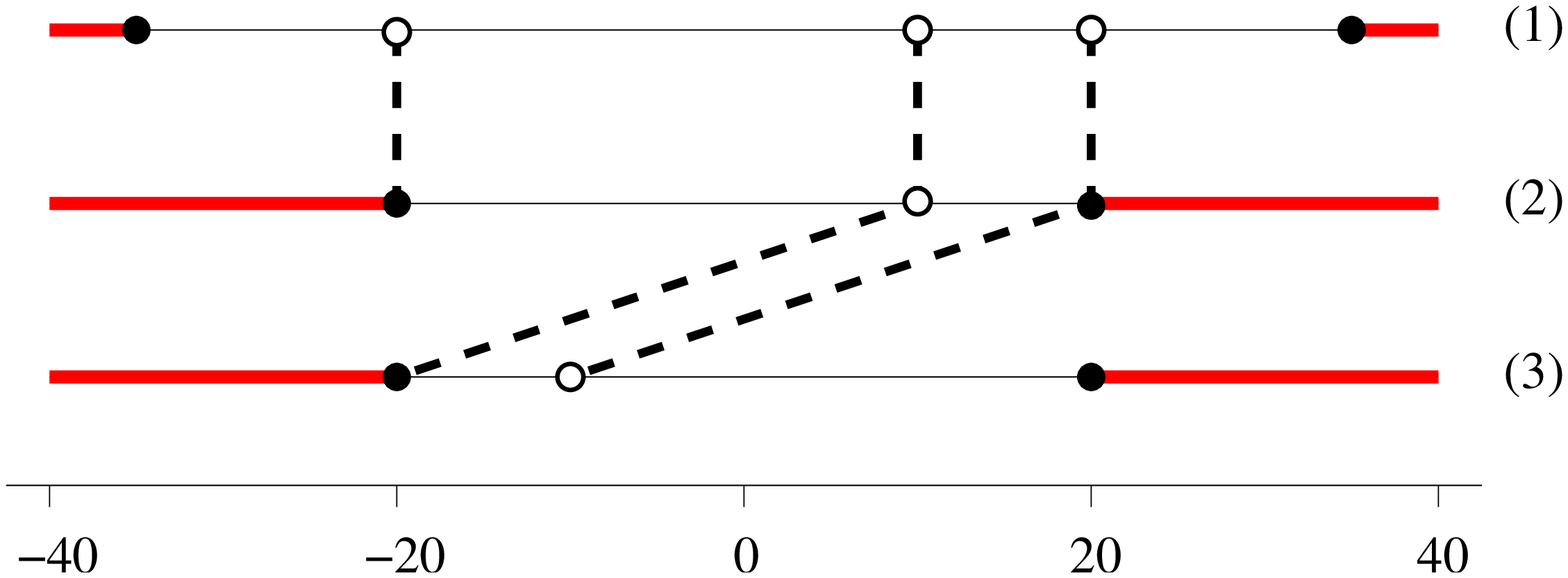,height=0.197\linewidth}\\[10pt]
\mbox{\small\ref{calc2}a)~$\ln(m_1r/2)=-35$.}\\[15pt]
\epsfig{file=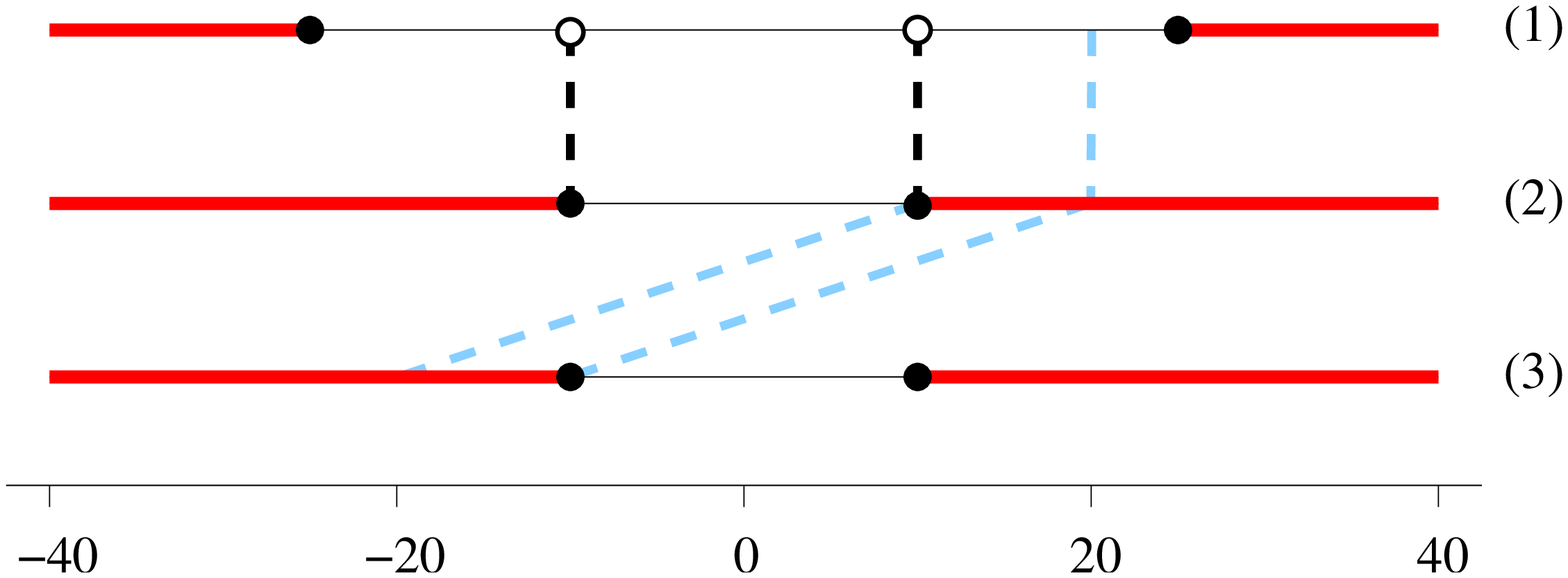,height=0.197\linewidth}\\[10pt]
\mbox{\small\ref{calc2}b)~$\ln(m_1r/2)=-25$.}
\end{array}
\]
\caption{\small Effective TBA systems for the solid-line flow
in figure \ref{fig1}.}
\label{calc2}
\end{figure}

Returning to the 
$SU(4)_2/U(1)^3$ examples of figure \ref{fig1}, we now
analyse the flow shown as a solid line on figure \ref{fig1}, which
breaks parity. In the far UV, which for this set of parameters can
be found for $\ln(m_1r/2)=-35$, the set of kink TBA systems is shown
in figure \ref{calc2}a.
Notice that, this time, the evident left-right symmetry has been lost.
The calculation of $c$ goes otherwise as before, with the same
(expected) result: $c=C_k(a_3)$.

Increasing $r$ through the first crossover, there is no decoupling of
stable pseudoenergies, but the effective kink systems determining the
values of
$c_3^-$ and $c_2^+$ change, as can be seen on figure \ref{calc2}b.
(We have also shown, lightly-shaded, the parts of the kink systems
which have been lost in the crossover.)
Now $c=C_k(a_1)+(C_k(a_2){-}C_k(a_1))+C_k(a_1)=C_k(a_2)+C_k(a_1)$.
Again this is easily understood physically -- the crossover
corresponds to the unstable particle corresponding to the root
$\alv_2+\alv_3$ becoming relatively heavy and decoupling; this
splits the $SU(4)_k/U(1)^3$ HSG model into two decoupled parts,
HSG models for $SU(3)_k/U(1)^2$ and $SU(2)_k/U(1)$, and makes it natural
to conjecture that
the full flow is
\begin{equation}
\mbox{(UV)}\>\>\>\>\frac{SU(4)_k}{U(1)^{3}} \>
\buildrel m_{23}\over{\hbox to 30pt{\rightarrowfill}}\>
\frac{SU(3)_k}{U(1)^{2}}\times
\frac{SU(2)_k}{U(1)}
 \>
\buildrel m_2=m_3\over{\hbox to 45pt{\rightarrowfill}}\>
\frac{SU(2)_k}{U(1)}\>
\buildrel m_1\over{\hbox to 30pt{\rightarrowfill}}\>
{\rm Massive}\>\>\>\> \mbox{(IR)}\>.
\lab{Mass3}
\end{equation}
Notice that the effective TBA system \rf{eeetba}
governing the crossover at
$r^{-1}\approx m_{23}$
is of massless type, but, since the driving terms are
asymmetrically placed on the $a_3$ Dynkin diagram, it lies outwith the
class of massless Dynkin TBA systems discussed in \cite{DYNK}.

\begin{figure}
\[
\begin{array}{l}
\epsfig{file=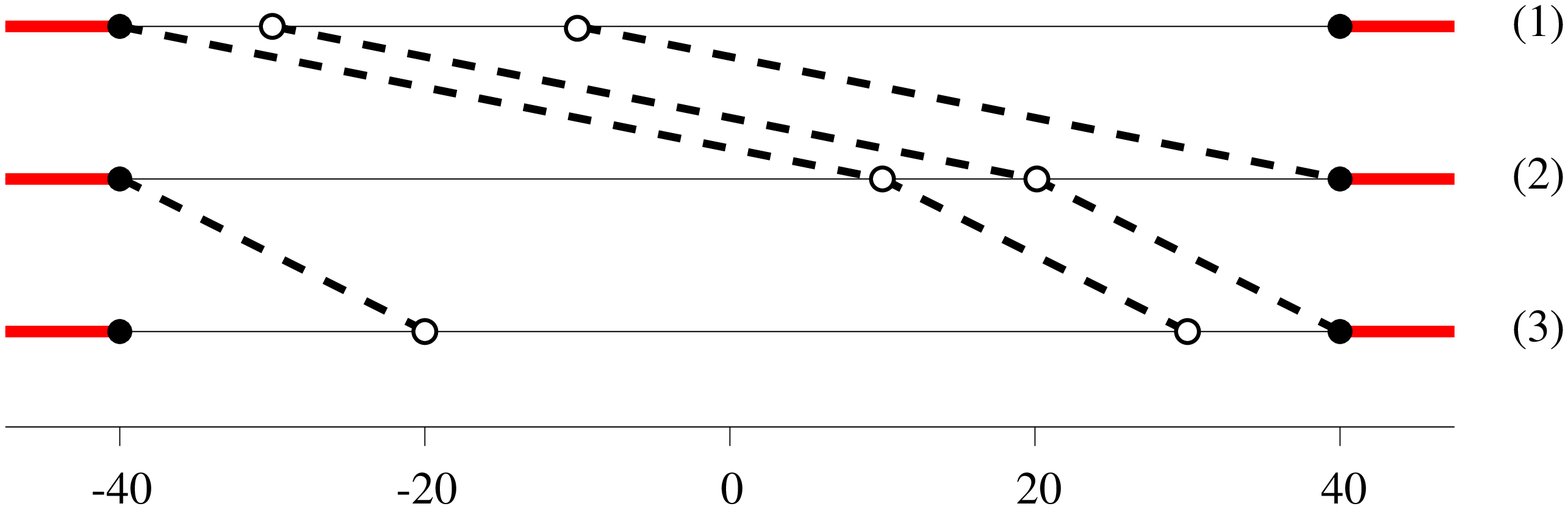,height=0.197\linewidth}\\[10pt]
\mbox{\small\ref{calc3}a)~$\ln(m_1r/2)=-40$.}\\[15pt]
\epsfig{file=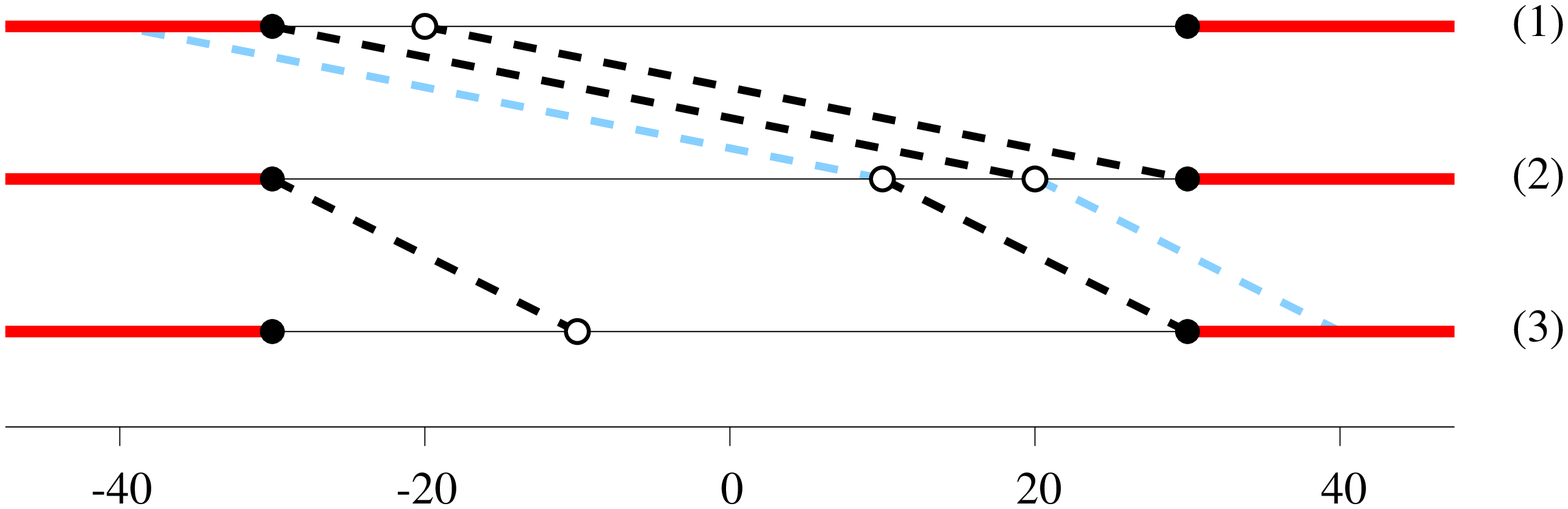,height=0.197\linewidth}\\[10pt]
\mbox{\small\ref{calc3}b)~$\ln(m_1r/2)=-30$.}\\[15pt]
\epsfig{file=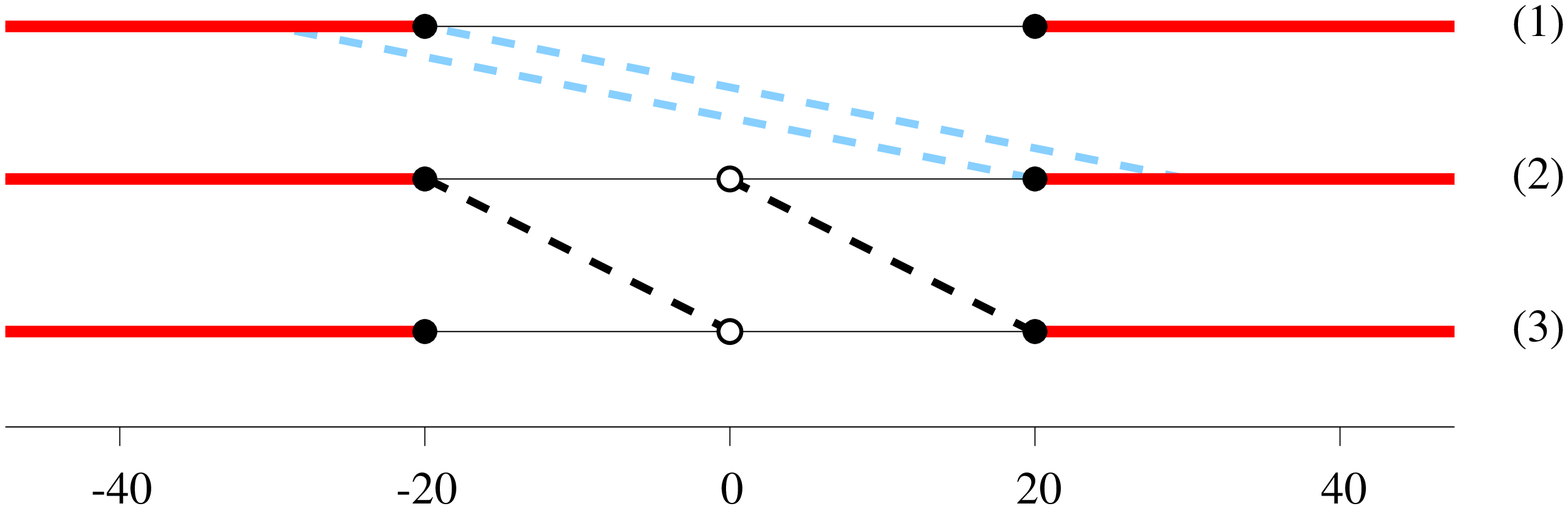,height=0.197\linewidth}\\[10pt]
\mbox{\small\ref{calc3}c)~$\ln(m_1r/2)=-20$.}\\[15pt]
\epsfig{file=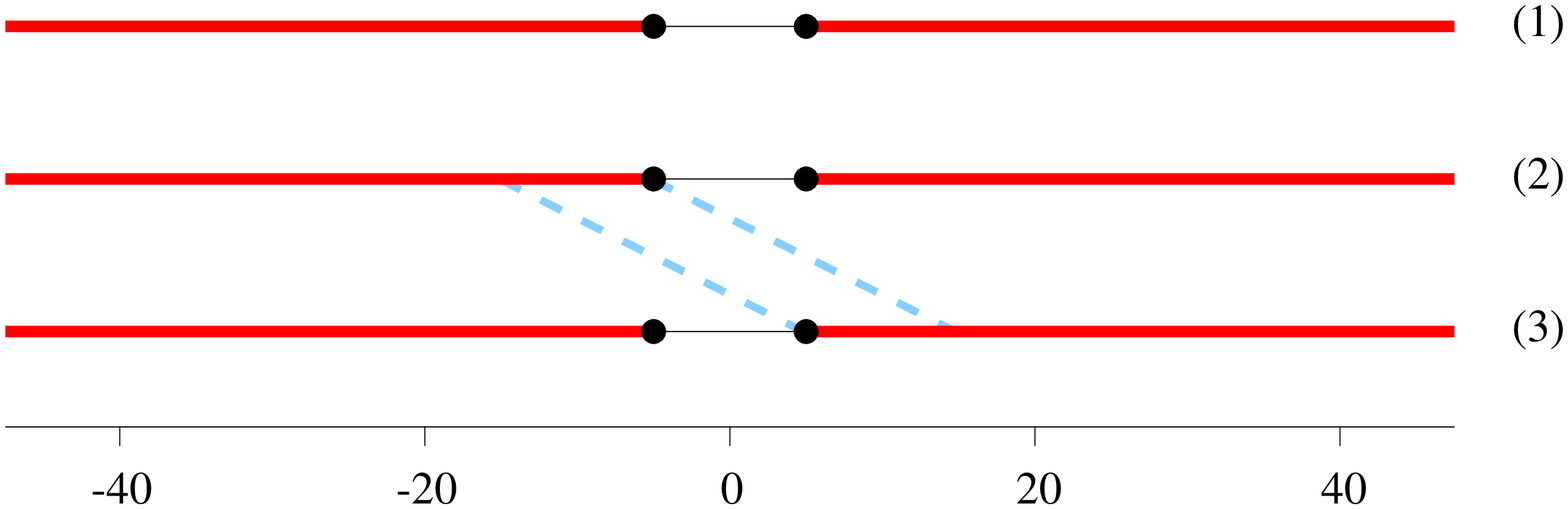,height=0.197\linewidth}\\[10pt]
\mbox{\small\ref{calc3}d)~$\ln(m_1r/2)=-5$.}
\end{array}
\]
\caption{\small Effective TBA systems for the dotted-line flow
in figure \ref{fig1}.}
\label{calc3}
\end{figure}

Finally, 
figure \ref{calc3} shows the sets of effective kink TBA
systems for the four plateaux of the flow shown as a
dotted line in figure \ref{fig1}, which are separated by
crossovers at $\ln(m_1r/2)\approx -35$, $-25$, $-10$ and $0$.

A coset identification for the fixed points visited by the flow is
\bea
\mbox{(UV)}\>\>\>\>\frac{SU(4)_k}{U(1)^{3}}
&&\hskip-0.6cm\buildrel m_{13}\over{\hbox to
30pt{\rightarrowfill}}\> \
\frac{SU(3)_k}{SU(2)_k\times
U(1)}\times \frac{SU(3)_k}{U(1)^2}
\buildrel m_{12}\over{\hbox to
30pt{\rightarrowfill}}\> \frac{SU(2)_k}{U(1)}\times\frac{SU(3)_k}{U(1)^2}
\allowdisplaybreaks\nn
\noalign{\vskip0.3truecm}
&&\hskip-1.5truecm\buildrel m_{23}\over{\hbox to
30pt{\rightarrowfill}}\> \left[\frac{SU(2)_k}{U(1)}\right]^{\times3}
\buildrel m_{ii}\over{\hbox to
30pt{\rightarrowfill}}\>
{\rm Massive}\>\>\>\> \mbox{(IR)}\>.
\lab{SU4flow}
\ena
This is the simplest example which
inevitably involves a coset CFT not of parafermionic type: here,
${SU(3)_k\over SU(2)_k\times U(1)}$. By level-rank duality~\cite{DUALITY},
this is $\frac{SU(k)_2\times SU(k)_1}{SU(k)_3}$\,, the coset which
might have been more
naturally suggested by a comparison of the effective TBA
governing the crossover with the results of~\cite{ZAMOcoset,RAVA}.
Note also that the step at highest
energy (that furthest into the UV) involves the mass scale $m_{13}$
associated with the root $\alv_1+\alv_2+\alv_3$,
which is of height 3, and so corresponds to an unstable
particle which is not seen directly in the two-particle S-matrix
elements. 
Even so, the
TBA picks it up with no problems, as
expected given our general analysis.

In general, a knowledge of the effective central charge is not enough
to identify a coset unambiguously. In particular,
parity-breaking in the HSG models brings with it a number of
interesting phenomena, to which we shall
return in section~\ref{LAGsec} and appendix~\ref{Heterotic}. However,
the examples discussed
above are sufficiently simple that these issues do not arise.

\begin{figure}[ht]
\begin{center}
\epsfig{file=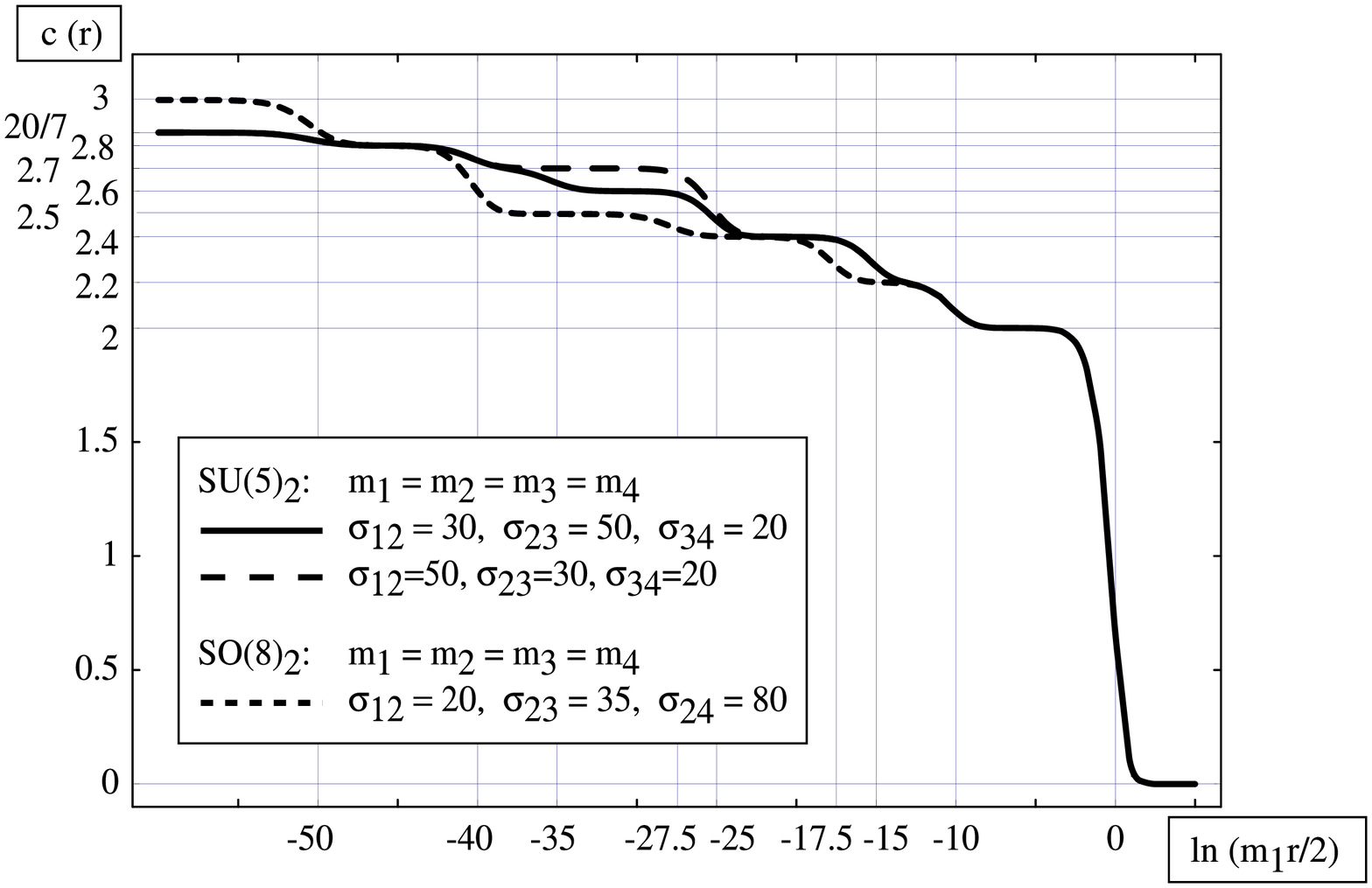,width=0.9\linewidth}
\caption{\small The TBA scaling function for the $SU(5)_2/U(1)^4$  and
$SO(8)_2/U(1)^4$ HSG models.}
\label{fig2}
\end{center}
\end{figure}

Some flows for the two rank
$4$ simply-laced Lie algebras,
$a_4\equiv SU(5)$ and $d_4\equiv SO(8)$, are shown in figure \ref{fig2}.
Again,
the locations and heights of the steps all agree with the predictions of the
rules formulated above. These flows were already presented
in~\cite{Ltalk,Dorey:2002sc}, where conjectures for the fixed points visited
by
each of them can be found. More about the
$d_4$ case can be found around eq.~\rf{SO8flow} below,
in the context of the
Lagrangian treatment of the flows.

Notice that the two $a_4$ flows
have a different number of steps:
$7$ for the solid line, and only~$6$ for the dashed line.
This is a simple consequence
of the fact that $m_{12}=m_{24}$ for the choice of parameters corresponding
to the dashed-line flow, while the mass scales $m_{ij}$,
$i\not=j$, are all different for the solid line.
A more subtle feature, not seen on the
previous collection of plots, concerns the
$d_4$ flow, for which the quantum parameters were chosen to match
the set of classical parameters analysed in section \ref{classicalscales},
eq.~\rf{d4example}. All of the mass scales $m_{ij}$,
$i\not=j$, are different, and yet there are only $6$ steps. A
na\"\i ve analysis would have predicted a crossover at
$\ln(m_1r/2)=\ln(m_1/m_{34})=-22.5$, but
this step is absent. This absence is exactly as predicted
by our analysis of shielding.

By separating the stable mass scales we can add $3$ more steps to each
flow, making a total of $10$ for the $a_4$ example corresponding to the
solid line but only $9$ for $d_4$.
Now, in spite of the phenomenon of shielding,
one might have expected that the {\em maximal}\/ number of steps
for a given algebra would equal
the maximal possible number of distinct scales $m_{ij}$ in every
case, which for
rank $4$ is $\rg(\rg{+}1)/2=10$. Might
a different choice of parameters for $d_4$
avoid the shielding, and produce a flow with the $10$ steps we
have already seen for $a_4$? In fact the answer to
this question is no, as we now show.

\sect{The maximal number of steps}
\label{steps}
\iindent
In this section, we find,
for a generic simply-laced Lie algebra,
the maximal number of well-separated
scales that can be found
in the set of numbers $\{m_{\bev}\,,\>
\bev\in \Phi^+_g\}$, as the $2\rg-1$ parameters are varied.
This is equal to the maximal number of
well-separated steps that can be found in the flow of the effective
central charge of the corresponding quantum field theories.

By the results of section~\ref{classicalscales}, the number of
well-separated scales
in $\{m_{\bev}\,,\> \bev\in \Phi^+_g\}$
for any given set of values of the parameters
$\{m_i\}$ and $\{\sigma_j\}$
is equal to the number of
{\em unshielded}\/
well-separated scales in $\{m_{ij}\,,\> i,j=1\ldots \rg\}$.
Our strategy will be to start by maximising the number
of scales from this latter set which are
unshielded, and only then to worry about separating the
unshielded scales so as to avoid any degeneracies.

We first take $g=a_n$, whose Dynkin diagram is simply a chain
$\{1\ldots n\}$ of $n$ nodes,  where $\{j,j+1\}$ are neighbours
for $j=1\ldots n-1$ (see fig.~\ref{DynkinDiag}).  This case is rather
trivial: consider 
the particular choice
$m_j=\ex{ja}$ and
$\sigma_j=jb$, for two real numbers $a,b\gg0$ such that $a/b$ is
irrational. If we take $a\ll b$, it is
straightforward to check that the
constraints~\rf{cond} are satisfied for all $k$, $l$, and so all of
the $m_{kl}$ are unshielded. Furthermore, as in~\rf{DifScales}, these
scales can be made arbitrarily far apart by taking $a$ and $b$ large.
Therefore, for $g=a_n$,
the maximal number of well-separated scales is equal to
$n(n+1)/2$, which is also the number of positive roots.

The $d$ and $e$ cases are more tricky, and we shall exploit the fact that
their Dynkin diagrams always contain $a$--type subdiagrams. As a
preliminary, we claim that a necessary condition for all of the scales
$m_{kl}$ associated with an $a_n$ diagram (or subdiagram)
to be unshielded is that the resonance
parameters $\{\sigma_i\}$ and the stable-particle mass scales
$\{m_i\}$ should satisfy
\begin{eqnarray}
&\bullet &
\bigl|\ln\left(m_{i}/ m_{i+1}\right)| \ll
\bigl|\sigma_{i}-\sigma_{i+1}\bigr|
\quad \forall\>
i=1\ldots n-1\>
\lab{SizeAn}\\
\noalign{and}
&\bullet &
{\rm either}\quad \sigma_{1}\gg \sigma_{2}\gg \cdots \gg\sigma_{n}
\quad {\rm or}\quad
\sigma_{1}\ll \sigma_{2}\ll \cdots \ll\sigma_{n}\>
\lab{OrderAn}
\end{eqnarray}
In order to prove this, notice that, from~\rf{cond}, the
relevant conditions to ensure that all the scales
$m_{kl}$ be unshielded are
\begin{equation}
m_{kl}\ggg m_{pq} \quad \forall\> k,l,p,q ~~ \text{such
that}~~ 1\leq k\le p\le q\le l\leq n ~~{\rm with}\quad
\{p,q\}\not=\{k,l\}\>.
\lab{CondAn}
\end{equation}
These ensure that each
$m_{kl}$ appears as the dominating term in $m_{\bev}$\,
for the positive root $\bev=\sum_{i=k}^l \alv_i$\,.
Taking logs 
and rearranging,
the inequality in \rf{CondAn} is equivalent to
\begin{equation}
|\sigma_k{-}\sigma_l|
-|\sigma_p{-}\sigma_q|
\gg
\ln(m_p/m_k)
+\ln(m_q/m_l)
\lab{recond}
\end{equation}
Then, the need for \rf{SizeAn} follows from \rf{recond} on setting
$k=p=q=i$, $l=i{+}1$, and then $k=i$, $p=q=l=i{+}1$. The first choice
gives
\begin{equation}
|\sigma_i-\sigma_{i+1}|\gg\ln(m_i/m_{i{+}1})\,\,
\end{equation}
and the second
\begin{equation}
|\sigma_i-\sigma_{i+1}|\gg\ln(m_{i{+}1}/m_i)\,.
\end{equation}
Combining the two, \rf{SizeAn} is recovered. To see the need
for \rf{OrderAn}, suppose
that the condition does {\it not}\/ hold. Then there must be a
sequence $\sigma_i$, $\sigma_{i+1}$, $\sigma_{i+2}$ such that either
\begin{equation}
\sigma_i\gg\sigma_{i+1}\ll\sigma_{i+2}
\lab{casa}
\end{equation}
or
\begin{equation}
\sigma_i\ll\sigma_{i+1}\gg\sigma_{i+2}
\lab{casb}
\end{equation}
(Note, 
$\sigma_i\sim\sigma_{i+1}$ and
$\sigma_{i+1}\sim\sigma_{i+2}$ are excluded by \rf{SizeAn}, which has
already been established.) Consider \rf{casa}, and suppose
that $\sigma_i\ge\sigma_{i+2}$.
Then
\begin{equation}
|\sigma_i-\sigma_{i+2}|=
\sigma_i-\sigma_{i+2}=
\sigma_i-\sigma_{i+1}+
\sigma_{i+1}-\sigma_{i+2}
=|\sigma_i-\sigma_{i+1}|-
|\sigma_{i+1}-\sigma_{i+2}|
\end{equation}
and so
\begin{equation}
|\sigma_i-\sigma_{i+2}|-|\sigma_i-\sigma_{i+1}|=
-|\sigma_{i+1}-\sigma_{i+2}|\ll -|\ln(m_{i+1}/m_{i+2})|
\lab{lastc}
\end{equation}
using \rf{SizeAn} for the final inequality. On the other hand,
\rf{recond} for $k=p=i$, $q=i{+}1$, $l=i{+}2$ is
\begin{equation}
|\sigma_i-\sigma_{i+2}|-|\sigma_i-\sigma_{i+1}|\gg
\ln(m_{i+1}/m_{i+2})
\end{equation}
The contradiction with \rf{lastc} rules out \rf{casa} with
$\sigma_i\ge\sigma_{i+2}$\,; the other options are dealt with
similarly, establishing the necessity of \rf{OrderAn}.

Parenthetically we remark that not
only are \rf{SizeAn} and \rf{OrderAn} necessary conditions for the
$n(n+1)/2$ scales $\{m_{kl}\}$ to be unshielded; they are also sufficient.
Suppose that the first option of \rf{OrderAn} holds. (The
argument is trivially rewritten for the second option.) Then the LHS
of \rf{recond} is equal to $\sigma_k-\sigma_p+\sigma_q-\sigma_l$\,,
while, using~\rf{SizeAn},
\begin{eqnarray}
\ln(m_p/m_k)&\leq& |\ln(m_k/m_p)| \nn
&\leq & |\ln(m_k/m_{k+1})|+ |\ln(m_{k+1}/m_{k+2})|+\dots
+|\ln(m_{p-1}/m_p)|
\nn
&\ll& \sigma_k-\sigma_p\,.
\end{eqnarray}
Likewise, $\ln(m_q/m_l)\ll\sigma_q-\sigma_l$\,, and so
$\ln(m_p/m_k)
+\ln(m_q/m_l)\ll\sigma_k-\sigma_p+\sigma_q-\sigma_l$\,,
which is the required result.

Once the necessity of
\rf{SizeAn} and~\rf{OrderAn} has been established, it is easy to see that at
least one scale
$m_{kl}$ must be shielded for the $d$ and $e$ cases.
Condition \rf{OrderAn} implies that the sequence $\{\sigma_{i_k}\}$
for any chain of nodes $\{i_k\}$ on the Dynkin diagram must be
monotonic, since otherwise one of the scales $m_{i_ki_l}$ associated
with that chain will be shielded. Now the $d$ and
$e$ Dynkin diagrams are forked, and it is clearly impossible to enforce
simultaneous monotonicity for the three maximal chains of nodes
including the fork node. (For $g=d_n$, these are the chains $\{1\ldots
n{-}1\}$, $\{1\ldots n{-}2,n\}$, and $\{n{-}1,n{-}2,n\}$ in the
labelling of fig.~\ref{DynkinDiag}; for $g=e_n$, $\{1\ldots
n{-}1\}$, $\{1\ldots n{-}3,n\}$, and $\{n{-}1,n{-}2,n{-}3,n\}$.) Hence, the
number of unshielded scales certainly cannot exceed $n(n{+}1)/2-1$.
To show that this number can be attained, and that the resulting
$n(n{+}1)/2-1$
unshielded scales can be separated, we resort again to explicit
examples. For $g=d_n$, one can take
\begin{eqnarray}
&&\sigma_i = i b \quad \forall i=1\ldots n{-}1\>, \qquad
\sigma_n= \sigma_{n-2}\>,\nn
&&m_i = \ex{ja} \quad \forall j=1\ldots n{-}1\>, \qquad m_n=
\ex{(n+1)a}\>,\nn
\lab{DnParams}
\end{eqnarray}
and for $g=e_n$,
\begin{eqnarray}
&&\sigma_i = i b \quad \forall i=1\ldots n{-}1\>, \qquad
\sigma_n=\sigma_{n-3}\>,\nn
&&m_j = \ex{ja} \quad \forall j=1\ldots n{-}1\>, \qquad m_n=
\ex{na}\>,
\lab{EnParams}
\end{eqnarray}
where in all cases
$a$ and $b$ are real numbers, $\gg0$,
such that $a/b$ is irrational, and $b\gg
4a$.
It is easy to check that the constraints~\rf{cond} are satisfied for all
$k$,
$l$, with the only exceptions being $\{k,l\}=\{n{-}2,n\}$ for $g=d_n$,
or $\{k,l\}=\{n{-}3,n\}$ for $g=e_n$. This means that the
scale $m_{n{-}2,n}$ (for $g=d_n$) or
$m_{n{-}3,n}$ (for $g=e_n$) is shielded.
Moreover, all the unshielded scales $m_{kl}$
can be given arbitrarily
well-separated
magnitudes by choosing~$a$ and~$b$ large enough. Therefore, for
$g=d_n$ and $g=e_n$, the maximal number of well-separated mass scales is
$n(n+1)/2-1$.

For $g=d_n$, there is another generic way to arrange
the parameters such that only one scale is shielded:
suppose that
\begin{eqnarray}
&{\rm either}&\quad \sigma_{1}\ll \sigma_{2}\ll \cdots
\ll\sigma_{n-1}\quad {\rm and}\quad \sigma_{n-2} \ll\sigma_n\nn
&{\rm or}&\quad
\sigma_{1}\gg \sigma_{2}\gg \cdots
\gg\sigma_{n-1}\quad {\rm and}\quad \sigma_{n-2} \gg\sigma_n\>,
\lab{OrderDn}
\end{eqnarray}
and in addition
\begin{equation}
\bigl|\ln\left(m_{i}/ m_{j}\right)| \ll
\bigl|\sigma_{i}-\sigma_{j}\bigr|
\lab{SizeDn}
\end{equation}
for each pair of neighbouring nodes $\{i,j\}$. Then, all the scales
$m_{ij}$ with \hbox{$i,j=1\ldots n-1$}, and $m_{in}$ with $i=1\ldots n-2,n$
are unshielded.
In contrast, the
number $m_{n-1\, n}$ is always swamped by other scales in the
sums~\rf{MassRootb}. All these numbers can be separated,
as shown by the particular choice of parameters used
in~\rf{DifScales} with
$a\ll b$. 

A similar trick does not succeed for $g=e_n$ because the shortest
maximal chain including the fork node has length $4$ rather than $3$.
In this case, the only choices of
$\{m_i,\sigma_i\}$ leading to the maximal number of unshielded scales are
those
that satisfy the following constraints. First, the parameters associated
with the chain of nodes $\{1\ldots n-1\}$ have to
satisfy the conditions~\rf{SizeAn} and~\rf{OrderAn}, to ensure that all the
scales $m_{kl}$ with $k,l=1\ldots n-1$ are unshielded. Second,
$\sigma_n$ and $m_n$ have to be chosen such that
\begin{eqnarray}
&&m_n\ggg m_{n-3}\>, \quad |\sigma_{n-3}-\sigma_n|\lesssim
|\ln(m_{n-3}/m_n)|\>, \nn[2pt]
&&|\sigma_{n-1}-\sigma_n|- |\sigma_{n-2}-\sigma_n|\gg
\ln(m_{n-2}/m_{n-1})\>,\nn[2pt]
&&|\sigma_i-\sigma_n|\gg \ln(m_n/m_i)\quad i=n-4,n-2\>,\nn[2pt]
&& |\sigma_i-\sigma_n|- |\sigma_i-\sigma_{n-3}| \gg \ln(m_{n-3}/m_n)\quad
\forall
i=1\ldots n-1\>, \; i\not={n-3}\>, \nn[2pt]
&& |\sigma_i-\sigma_n|- |\sigma_{i+1}-\sigma_n| \gg \ln(m_{i+1}/m_i)\quad
\forall
i=1\ldots n-5\>.
\lab{En}
\end{eqnarray}
The proof of these conditions is based on the characterisation of the
choices of parameters such that all but one of the scales associated with
a given chain are unshielded, which can be easily derived from~\rf{SizeAn}
and~\rf{OrderAn}. Since it is rather involved, it will be omitted. It is
straightforward to check that the values~\rf{EnParams}
satisfy~\rf{En}.

Our results for the maximal number of separable mass scales are
summarised in
table~\ref{MaxNumb}. (Note that this table corrects an error in the
result for $g=e_n$ reported in~\cite{Dorey:2002sc}.)

\begin{table}[ht]
\begin{center}
\vskip0.2truecm
\begin{tabular}{|c|l|}
\hline 
\quad $g$~\qquad & \quad Maximal number of
separable scales~\qquad~\qquad
\\
\hline \hline
\quad $a_n$~\qquad & \quad $n(n+1)/2\>, \quad n\geq 1$ \\
\hline
\quad $d_n$~\qquad & \quad $n(n+1)/2-1\>, \quad n\geq 4$ \\
\hline
\quad $e_n$~\qquad & \quad $n(n+1)/2-1\>, \quad n=6,7,8$ \\
\hline
\end{tabular}
\end{center}
\caption{\small The maximal number of separable scales for the HSG models
associated with the different simply-laced Lie groups (see also the comments
in the last two paragraphs of section~\ref{steps}).}
\label{MaxNumb}
\end{table}

Sometimes, it is of interest to consider choices
of the parameters with all the stable-particle mass scales equal,
so that $m_i=m_j\;\; \forall i,j$. For such cases it is clear that the
maximal number of well-separated scales  cannot be larger that the number
quoted in table~\ref{MaxNumb}, minus $(n{-}1)$.
For $g=a_n$ and~$d_n$, this number can be attained,
as can be shown via an explicit
example. In both cases, take
\begin{equation}
\sigma_i =  2^i b \quad \forall i=1\ldots n\>,
~\lab{RedMax}
\end{equation}
where $b$ is a real number $\gg0$. For $g=a_n$, this choice
satisfies~\rf{SizeAn} and~\rf{OrderAn}, and for $g=d_n$\,,
\rf{OrderDn} and~\rf{SizeDn}. All of the resulting unshielded scales
$m_{kl}$ can be given arbitrarily well-separated values by choosing $b$
large enough. 

However, for $g=e_n$, eq.~\rf{En} shows that $m_n\ggg m_{n-3}$ is a
necessary condition to achieve the maximal number of unshielded scales.
Therefore, by taking all the stable-particle mass scales to be
equal we lose at least one extra unshielded scale, and the resulting
maximal
number of well-separated scales cannot be larger than the number quoted in
table~\ref{MaxNumb} minus $n$. For an explicit example where this
number is attained, one can again take~\rf{RedMax}, which makes all
of the
scales $m_{kl}$ with $k\not=l$ unshielded with the only exception of
$m_{n-2\>n}$ and $m_{n-1\>n}$, and well-separated, by choosing $b$
large enough.

\sect{Crossovers from the Lagrangian approach}
\label{LAGsec}

\iindent
The original formulation of the HSG theories was
 in terms of a gauged WZW action
modified by a potential~\cite{HSG,QHSG}.
This explicit Lagrangian definition provides a more physical interpretation
of the crossovers observed in the study of finite-size effects using the
TBA,
allowing them to be seen as
consequences of changes in the number of the field configurations
that remain effectively massless at the given finite-size (RG)
scale. In addition, this interpretation turns out to be
very useful in elucidating the precise nature of the effective
field theories between well-separated crossovers. While such a semiclassical
analysis is rather na\"\i ve, it turns out to be surprisingly powerful.

The HSG theories corresponding to
perturbations of the coset $G_k/U(1)^{\rg}$ have
actions
\begin{equation} 
S_{\rm HSG}[{\gamma},A_\pm]= k\Bigl(S_{\rm gWZW}[{\gamma},A_\pm]
\>-\int d^2x\> V({\gamma})\Bigr)\>.
\lab{Action}
\end{equation}
Here, ${\gamma}={\gamma}(t,x)$ is a
bosonic field that takes values in some faithful representation of the
compact Lie group $G$, and $A_\pm$ are non-dynamical gauge fields taking
values in the Cartan subalgebra of $g$ associated with $H\simeq U(1)^{\rg}$,
a maximal torus of $G$. Then, $S_{\rm gWZW}$ is the
gauged  WZW action corresponding to the coset
$G/H$~\cite{GWZWGK,GWZW}.
The potential is
\begin{equation} 
V({\gamma}) =\frac{m_0^2}{4\pi} \> \langle
\Lambda_+ , {\gamma}^\dagger \Lambda_- {\gamma}\rangle\>,
\lab{Potential}
\end{equation}
where $m_0^2$ is a bare overall mass scale, $\langle\;,\>\rangle$ is the
Killing form of $g$, and
$\Lambda_\pm=i\bfm{\lambda}_\pm\cdot\bfm{h}$ are two arbitrary elements
in the Cartan subalgebra of $g$ associated with the maximal torus
$H$, specified by two $\rg$-dimensional vectors $\bfm{\lambda}_+$ and
$\bfm{\lambda}_-$.  In this context, $S_{\rm HSG}$ is
a Lagrangian action defined on 1+1 Minkowski space. (In contrast,
eq.~\rf{PCFTaction} defines the model as a perturbed conformal field
theory in two-dimensional Euclidean space with role of the potential
$V(\gamma)$ being taken by the perturbing operator
$\mu\> \phi_{\bfm\lambda,\overline{\bfm\lambda}}$.)
The field $\gamma$ in \rf{Action} has ${\rm dim}(G)=
(\hg{+}1)\rg$ degrees of freedom. However, $S_{\rm
HSG}$ is invariant under a group of abelian gauge transformations
generated by $H$. This built-in gauge symmetry
removes $\rg$ degrees of freedom, and $S_{\rm HSG}$ is actually defined
on the coset manifold $G/H$ of dimension $\rg\hg$.
The positive integer~$k$ is known as the
`level'~\cite{WZW}. Classically, it plays the role of an
inverse coupling constant, and both the
weak-coupling (perturbative) and semiclassical regimes are recovered
when $k$ is large. $kS_{\rm gWZW}$ provides an
action for the $G_k/U(1)^{\rg}$ coset conformal field theory,  and the
potential is a composite field that can be identified with a (gauge
invariant) matrix element of the WZW field
$\gamma$ taken in the adjoint representation~\cite{KZ}. This is the
spinless relevant primary field that defines the perturbation, and the
resulting theory is massive
for any choice of
$\bfm{\lambda}_+$ and $\bfm{\lambda}_-$ such that $\bfm{\lambda}_\pm\cdot
\alv\not=0$ for all roots $\alv$ of $g$; for these cases quantum
integrability was checked in~\cite{QHSG}.

Without loss of generality,
$\lav_+$ and
$\lav_-$ can be restricted to live in the same Weyl
chamber~\cite{HSG},
which ensures that
$V(\gamma)$ has a minimum at $\gamma=\one\;$\foot{The group of gauge
transformations can be taken to be of axial form, {\em i.e.\/},
$\gamma\rightarrow
\alpha \gamma \alpha$ with
$\alpha=\alpha(t,x)$ taking values in $H$, which makes the minimum of the
potential at
$\gamma=\one$ unique, modulo gauge transformations.}.
In the following, we shall assume that
the simple roots have been chosen so that
this chamber is 
the fundamental
Weyl chamber.
Then, $\lav_+$ and $\lav_-$
specify the classical masses and resonance parameters of the solitons
associated with the positive roots $\bev$ via the
formulae~\cite{SMAT,HSGSOL}
\begin{equation}
M_{\bev}^2= m_0^2 (\bev\cdot\lav_+)(\bev\cdot\lav_-) \quad\text{and} \quad
\sigma_{\bev}={1\over2} \ln{\bev\cdot\lav_+\over \bev\cdot\lav_-}\>,
\lab{ClassMass}
\end{equation}
and, up to the overall bare mass scale, $M_{\bev}^2$ coincides
with~\rf{MassRoot}. These equations can be solved for $\lav_+$ and $\lav_-$
as functions of
$m_i=M_{\alv_i}/m_0$ and $\sigma_i=\sigma_{\alv_i}$
to recover \rf{LambdaW}. When substituted
in~\rf{Potential}, this provides the following decomposition of
the potential:
\begin{equation}
V({\gamma})= {m_0^2\over 4\pi}\> \sum_{i,j=1}^{\rg}\>
{\mu}_{ij}^2\> \Gamma_{ij}({\gamma})\>,
\lab{Split}
\end{equation}
where
\begin{equation}
{\mu}_{ij}^2= {m}_i\>
{m}_j\> \ex{{\sigma}_i-{\sigma}_j}
\qquad {\rm and}\qquad \Gamma_{ij}({\gamma}) =
-\Big\langle
(\lav_i\cdot\bfm{h})\>,\; {\gamma}^\dagger \; (\lav_j\cdot\bfm{h})\;
{\gamma} \Big\rangle\>.
\end{equation}
This shows that the potential depends on the
$\rg\hg$ gauge-invariant degrees of freedom of the fundamental field
$\gamma$ through the $\rg^2$ composite fields $\Gamma_{ij}({\gamma})$.
All the dependence on the parameters
$\{{m}_i,{\sigma}_i\}$ is concentrated in the
real positive coupling constants ${\mu}_{ij}$.
Notice that $m_{ij}$, as defined
in~\rf{Mdef}, is equal to $\max(\mu_{ij},\mu_{ji})$.
Recall also that the parity symmetry of the
unperturbed gauged WZW theory is implemented by the simultaneous
transformations $x\ra -x$, $\gamma\ra \gamma^\dagger$.
Since $\Gamma_{ij}({\gamma^\dagger})=\Gamma_{ji}({\gamma})$,
the condition for this symmetry to be respected by the perturbation
is ${\mu}_{ij}= {\mu}_{ji}$.
The composite fields $\Gamma_{ij}(\gamma)$ are independent of the
parameters and, since the group $G$ is compact, their
size is bounded and can be taken to be of order one.
Eq.~\rf{Split} provides an explicit realisation of the general
parametrisation of the perturbing operator given in~\rf{PertOp}, as
\begin{equation}
V({\gamma})=  {m_0^2\over 4\pi}\> \sum_{i,j=1}^{\rg}\>
{\mu}_{ij}^2\> \Gamma_{ij}({\gamma}) \>\equiv\> \mu \sum_{p,q=1}^{\rg}
\lambda_p
\overline{\lambda}_q\> \phi{}^{\rm
adj,\>adj}_{\;p\>,\;q} = \mu\>\phi_{\bfm\lambda,\overline{\bfm\lambda}}\>.
\lab{PlotSplit}
\end{equation}
Noticing that
$\mu_{ij}^2=(\bfm{\lambda}_+\cdot\bfm{\alpha}_i)
(\bfm{\lambda}_-\cdot\bfm{\alpha}_j)$, this shows that the two vectors
$\bfm{\lambda}$ and $\overline{\bfm\lambda}$ correspond to $\bfm{\lambda}_+$
and $\bfm{\lambda}_-$ in the semiclassical limit.

The decomposition~\rf{Split} provides a
simple physical explanation for the pattern of crossover effects observed in
the TBA analysis. All of the coupling constants
$(m_0^2/4\pi)\>{\mu}_{ij}^2$ have the dimension of a squared mass scale.
Therefore, along the RG trajectory corresponding to the
action~\rf{Action}, they are relatively
small in the UV region, and large in the IR.
In massive theories with just one coupling constant, of the
type most often discussed in the perturbed conformal field theory
literature, there is only one independent mass scale and the
passage from the UV to the IR involves just one transition.
In contrast, for the HSG theories, the various mass scales
$(m_0^2/4\pi)\>{\mu}_{ij}^2$ can be given very different values by
varying the parameters $\{{m}_i,{\sigma}_i\}$. As the RG scale passes
any of these scales, a new transition may potentially occur; from the
Lagrangian point of view, this underlies the observed
staircase trajectories.

However, 
when the scale associated with a given
coupling is reached, it could be that all fields involved in the
corresponding interaction have already decoupled. To avoid this issue
in our initial
discussion, suppose that the values of the parameters are such that one
coefficient, say $\mu_{ij}$, is much larger than all of the others.
One way to achieve this is to choose the
parameters
\begin{equation}
m_k=\ex{a \delta_{k,i}}\>, \quad \sigma_k=0, \qquad \forall\> k=1\ldots
\rg
\lab{gamone}
\end{equation}
for $i=j$, or
\begin{equation}
m_k=1\>, \quad \sigma_k= a\bigl(\delta_{k,i} - \delta_{k,j}\bigr), \qquad
\forall\>  k=1\ldots \rg
\lab{gamtwo}
\end{equation}
for $i\not=j$\,,
with $a\gg0$ in both cases.
Then, starting from the UV where all couplings are
effectively zero and all fields effectively massless, at the squared mass
scale
$(m_0^2/4\pi)\>{\mu}_{ij}^2$\,, only $\mu_{ij}$ has become effectively
nonzero. This coupling governs the separation between the UV and the IR
regions for all configurations of the field $\gamma$ such that
$\Gamma_{ij}(\gamma)$ is non-trivial; {\em i.e.\/}, for those
$\gamma$ such that
$\Gamma_{ij}(\gamma)\not\approx\Gamma_{ij}(\one)$, where $\gamma=\one$ is
the absolute minimum of the potential.
Defining a dimensionless RG scale
$\overline\Lambda ={\sqrt{4\pi}\over m_0}\> \Lambda $,
the effect of
$\Gamma_{ij}(\gamma)$ is negligible for
$\overline\Lambda\ggg{\mu}_{ij}$,
 and
only those  field
configurations such that
$\Gamma_{ij}(\gamma)\approx\Gamma_{ij}(\one)$ remain
in the effective theory for $\overline\Lambda\lll{\mu}_{ij}$.
Since $\mu_{ij}$ was assumed to be
larger than all of the other coefficients, it is
in particular larger than $\mu_{ji}$, and so $\mu_{ij}=m_{ij}$ and
this crossover happens at one of the classical mass scales.

To solve the condition $\Gamma_{ij}(\gamma)\approx\Gamma_{ij}(\one)$,
we first recall that every element $\gamma$ of a compact connected Lie
group
lies  in a one-parameter subgroup, so that $\gamma=\exp X$ for some
$X\in g$.
Introducing a
Cartan-Weyl basis for the complexification of $g$,
consisting of a Cartan subalgebra
$\{h_1\ldots h_{\rg}\}$ and step generators $E_{\alv}$\,,
$X$ is of the general form
\begin{equation} X= i\bfm{t}\cdot \bfm{h} + \sum_{\alv\in
\Phi^+} (\phi_{\alv} E_{\alv} -
\phi_{\alv}^\ast E_{-\alv})=-X^\dagger\>,
\lab{SolvingB}
\end{equation}
where $\bfm{t}$ is 
real, and the $\phi_{\alv}$ are complex. Then
\begin{eqnarray}
&&\hskip-2truecm
\Gamma_{ij}(\exp X)- \Gamma_{ij}({\one})
=
 - \sum_{n\geq2} \frac{1}{n!}\Big\langle
({\rm ad\/}X)^n(\lav_i\cdot\bfm{h})\>,\;
\lav_j\cdot\bfm{h}\Big\rangle\> \nn[3pt]
&=& 
 \sum_{\alv\in\Phi^+} (\alv\cdot \lav_i)(\alv\cdot
\lav_j)\left\langle
E_{\alv},E_{-\alv}\right\rangle|\phi_{\alv}|^2
- \sum_{n\geq3} \frac{1}{n!}\Big\langle
({\rm ad\/}X)^n(\lav_i\cdot\bfm{h})\>,\;
\lav_j\cdot\bfm{h}\Big\rangle\>.
\lab{SolvingAm}
\end{eqnarray}
At the level of the quadratic (leading) term, the condition
$\Gamma_{ij}(\gamma)=\Gamma_{ij}(\one)$ is therefore
solved by restricting the
sum in~\rf{SolvingB} to the roots $\alv$ of $g$ that satisfy $(\alv\cdot
\lav_i)(\alv\cdot \lav_j)=0$,
and this 
gives the possible flat (effectively massless) directions
away from the identity. Notice that the condition to quadratic order
is symmetrical
in $i$ and $j$: the flat directions for $\Gamma_{ij}$ and
$\Gamma_{ji}$ coincide at the identity.

However, to match the non-perturbative
TBA results, the full vacuum manifold for the effective theory should
be mapped out.
In order
to do this, we first recall the notation
introduced just after~\rf{MassOrder}:
$g^{[i_1,\ldots,i_n]}$ denotes
the Dynkin diagram obtained by deleting the nodes
$i_1\ldots i_n$ from the Dynkin diagram of $g$, and
$G^{[i_1,\ldots,i_n]}$ the compact subgroup of $G$ associated with
$g^{[i_1,\ldots,i_n]}$ times a
$U(1)^{n}$ factor associated with those nodes. We shall
also use the superscript $^{[i_1,\ldots,i_n]}$ to indicate that a field
configuration takes values in this group. With this notation in place,
we show that $\Gamma_{ij}(\gamma)$ is invariant
under the action of $G^{[j]}_L\times G^{[i]}_R$\,, by which we mean that
\begin{equation}
\Gamma_{ij}(\phi^{[j]}\> \gamma \> \psi^{[i]}) = \Gamma_{ij}(\gamma),
\qquad \forall \phi^{[j]}\in G^{[j]}\quad \text{and}\quad \psi^{[i]}\in
G^{[i]}.
\lab{SolvingA}
\end{equation}
This follows from the fact that, if $\psi=\exp X$ with $X$
of the form \rf{SolvingB}, then
\begin{equation}
\psi (\lav_i\cdot\bfm{h}) \psi^\dagger =\lav_i\cdot\bfm{h}
-\sum_{\alv\in\Phi^+} (\alv\cdot \lav_i) \sum_{n\geq1} \frac{1}{n!}
({\rm ad\/}X)^{n-1}(\phi_{\alv} E_{\alv} +
\phi_{\alv}^\ast E_{-\alv})\>.
\end{equation}
If $\psi=\psi^{[i]}\in G^{[i]}$, the sum~\rf{SolvingB}
for the corresponding $X$ is
restricted to the roots
$\alv$ of $g$ such that $\alv\cdot
\lav_i=0$, and so $\psi^{[i]} (\lav_i\cdot\bfm{h})
{\psi^{[i]}}^\dagger =\lav_i\cdot\bfm{h}$. Similarly, if
$\phi^{[j]}\in G^{[j]}$ then
${\phi^{[j]}}^\dagger (\lav_j\cdot\bfm{h}) \phi^{[j]}
 =\lav_j\cdot\bfm{h}$. Combining these two facts is enough
to prove~\rf{SolvingA}, and to show, in
particular, that $\Gamma_{ij}(\phi^{[j]}\> \one \>
\psi^{[i]}) = \Gamma_{ij}(\one)$.
Adding the already-performed identification of the flat directions
at the identity, the final conclusion is that the configurations
connected to the identity which satisfy
$\Gamma_{ij}(\gamma)=\Gamma_{ij}(\one)$
live in the submanifold of $G$
\begin{equation}
G^{[j]}\cdot G^{[i]}= \{\phi^{[j]}\>\psi^{[i]}\,|\,
\phi^{[j]} \in G^{[j]}, \psi^{[i]} \in G^{[i]} \}.
\end{equation}
These fields participate in
the effective theory in the
regime ${\mu}_{ij}\ggg\overline\Lambda\ggg{\mu}_{pq}$ in situations where
${\mu}_{ij}\ggg{\mu}_{pq}$ for all $(p,q)\not=(i,j)$.
Near to the identity they
correspond to the effectively massless fluctuations, and are
symmetrical in $i$ and $j$.
However, it is important to appreciate that the full manifolds of
effective field configurations left unfrozen by $\Gamma_{ij}$ and
$\Gamma_{ji}$ can be different, reflecting the breaking of parity, even
though
this asymmetry never shows up in the quadratic approximation
to~\rf{SolvingAm}.

This characterisation of the effective theory provides useful
information about the fixed point visited by the RG flow after the
first crossover has occurred.
This fixed point is associated with the gauged WZW action
in~\rf{Action}, 
$S_{\rm gWZW}$, restricted to the manifold $G^{[j]}\cdot G^{[i]}$,
whose conformal field theory interpretation
will be investigated in the following paragraphs.

We start with the case when $i=j$. Then,  the manifold of
effective field configurations is $G^{[i]}\cdot
G^{[i]}= G^{[i]}$, which is a subgroup of $G$ that splits into a product of
simple groups associated with the (possibly disconnected) Dynkin diagram
$g^{[i]}$, times a $U(1)$ factor. Recall that the HSG model was
originally defined not on $G$, but on the coset manifold $G/H$, where
$H\simeq U(1)^{\rg}$ is a maximal torus of $G$. Therefore,
the fixed point is specified by the restriction of
$S_{\rm gWZW}$ to the manifold $G^{[i]}/H$, which is in
agreement  with the TBA results presented just after~\rf{MassOrder}.
Since, by construction, the $U(1)$ factor in
$G^{[i]}$ is contained in $H$, the
restriction of the gauged WZW action to $G^{[i]}/H$ splits into a number of
decoupled parafermionic theories, one for each simple factor in
$G^{[i]}$, and all of them at level $k$.

The case with $i\not=j$ is more involved. We start by
characterising the manifold of effective field configurations $G^{[j]}\cdot
G^{[i]}$ as a coset
manifold. First, we note that there is an obvious map from $G^{[j]}\times
G^{[i]}$ into $G^{[j]}\cdot
G^{[i]}$ given by
$(\phi^{[j]},\psi^{[i]})\rightarrow \phi^{[j]}\psi^{[i]}$. However this is
not one-to-one,  since $\phi^{[j]}\psi^{[i]}$ is invariant under
$\phi^{[j]}\rightarrow \phi^{[j]}\rho^{-1}$ and
$\psi^{[i]}\rightarrow \rho \psi^{[i]}$, for any $\rho\in
G^{[i,j]}=G^{[j]}\cap G^{[i]}$, and so this action should be factored out.
This provides the following coset realisation of the manifold of
effective field configurations:
\begin{equation}
G^{[j]}\cdot G^{[i]} \simeq {G^{[j]}\times
G^{[i]}\over G^{[i,j]}}\>,
\end{equation}
where the action of $G^{[i,j]}$ on $G^{[j]}\times G^{[i]}$ is
$(\phi^{[j]},\psi^{[i]})\rightarrow (\phi^{[j]}\rho^{-1},\rho \psi^{[i]})$,
for any $\rho\in G^{[i.j]}$. Parenthetically, we point out that this coset
is a particular example of those considered by Guadagnini {\em et al.\/}
in~\cite{GMM} (see also~\cite{VOLKER}), which is one of the first papers
dealing with $N/D$ coset models where different left and right actions of
$D$ on $N$ are gauged: the so-called `asymmetric cosets'.
Models of this type, and their generalisations,
are examples of `heterotic' conformal field
theories and have been of some
interest in string theory; in addition to
the papers just mentioned,
refs.~\cite{BARS,GANNON,Walton:2002db,Sarkissian:2002nq,Johnson:2004zq,%
Israel:2004vv} give a sample of work in this direction.

However, in our case, not all the field configurations $\gamma \in
G^{[j]}\cdot G^{[i]}$ are physical. The actions
$S_{\rm gWZW}$ and $S_{\rm HSG}$ in~\rf{Action}
are invariant under a group of abelian gauge
transformations generated by
a maximal torus $H\simeq U(1)^{\rg}$ of $G$, and they
are therefore defined on $G/H$.
The precise form of the gauge group is $\gamma\rightarrow \alpha\> \gamma\>
\widehat\tau(\alpha^{-1})$, $A_\pm\ra \alpha(A_\pm +
\partial_\pm)\alpha^{-1}$, which is parametrised by
the lift $\widehat\tau$ of a suitable
orthogonal $O(\rg)$ transformation $\tau$ acting on the Cartan subalgebra
into $H$~\cite{HSG,QHSG}.\footnote{In particular, taking
$\tau=+I$ or~$-I$ leads to gauge transformations of vector or axial type,
respectively.} Taking all of
this into account, the full manifold of {\em physical} effective
field configurations left unfrozen by $\Gamma_{ij}$ can be identified with
the coset
\begin{equation}
{G^{[j]}\times G^{[i]}\over G^{[i,j]}\times H}
\>\simeq\>
{G^{[j]}\times
G^{[i]}\over G^{[i,j]}\times U(1)^{\rg}}\>,
\lab{NewCos}
\end{equation}
where  the action of $G^{[i,j]}\times H$ on $G^{[j]}\times G^{[i]}$ is
\begin{eqnarray}
&&\hskip-1cm\bigl(\phi^{[j]}, \psi^{[i]}\bigr)\rightarrow \bigl(\alpha\>
\phi^{[j]}\>
\rho^{-1}, \rho\> \psi^{[i]}\> \widehat\tau(\alpha^{-1})\bigr)\nn[3pt]
\hskip-1cm&&= \bigl(\alpha,\rho\bigr)\> \bigl(\phi^{[j]}\>
,\psi^{[i]}\bigr)\> \bigl(\rho^{-1}, \widehat\tau(\alpha^{-1})\bigr)
\>, \quad
\forall \> \rho \in G^{[i,j]}\;\; \text{and}\;\; \alpha \in H\>.
\lab{CosDef}
\end{eqnarray}
Next, consider the gauged WZW action corresponding to
the coset~\rf{NewCos} and the (anomaly free) group of gauge
transformations
$h \rightarrow \epsilon_L(u)\> h\> \epsilon_R(u^{-1})$, where
$h=(\phi,\psi)\in G^{[j]}\times G^{[i]}$, $u=(\rho,\alpha)\in
G^{[i,j]}\times H$, and $\epsilon_{L/R}:G^{[i,j]}\times
H\rightarrow G^{[j]}\times G^{[i]}$ are two group homomorphisms
defined by
\begin{equation}
\epsilon_L(\rho,\alpha) = (\alpha, \rho)\quad \text{and}\quad
\epsilon_L(\rho,\alpha) = \bigl(\rho,\widehat{\tau}(\alpha)\bigr)\>,
\lab{AsymGauge}
\end{equation}
which descend to embeddings of the corresponding Lie algebras.
The action is given by~\cite{VOLKER,BARS,GWZWGK}
(see also~\cite{QHSG} for the normalisation)
\begin{eqnarray}
&& kS_{\rm gWZW}^{[i,j]}[h,{\cal A}_\pm]= kS_{\rm
WZW}^{[i,j]}[h] +{k\over\pi} \int d^2x\>
\Bigl(-\bil{\epsilon_L({\cal A}_+)}{\partial_-h h^{-1}} \nn[3pt]
&& 
\hskip 1truecm + \bil{\epsilon_R({\cal A}_-)}{h^{-1}\partial_+
h} +
\bil{h^{-1}\epsilon_L({\cal A}_+)h}{\epsilon_R({\cal A}_-)} -
\bil{\epsilon_L({\cal A}_+)}{\epsilon_L({\cal A}_-)} \Bigr)\>,
\lab{gWZWNew}
\end{eqnarray}
where $kS_{\rm WZW}^{[i,j]}[h]$ is the WZW action at level $k$ for
the field $h$ in $G^{[j]}\times G^{[i]}$,
and ${\cal A}_\pm=(a_\pm, A_\pm)$ are non-dynamical gauge fields taking
values in the Lie algebra of $G^{[i,j]}\times H$.
More precisely, $a_\pm \in
g^{[i,j]}\oplus u(1)^2$ and $A_\pm \in u(1)^{\rg}$,
where all the explicit $u(1)$ factors are embedded in
the Cartan subalgebra into $H$, and we have denoted by $g^{[i,j]}$ both the
Dynkin diagram defined just after~\rf{MassOrder} and the corresponding
(compact semisimple) Lie algebra. We have also used
the same notation for the invariant bilinear form of
$g$ and its trivial extension to
$\bigl(g^{[i,j]}\oplus u(1)^2\bigr) \oplus u(1)^{\rg}$:
\begin{equation}
\bil{(u,v)}{(r,s)} \equiv \bil{u}{r} + \bil{v}{s}\>.
\end{equation}
Writing $h=(\phi,\psi)$ and expanding~\rf{gWZWNew}, we obtain
\begin{eqnarray}
&&\hskip -1.5truecm  kS_{\rm gWZW}^{[i,j]}[h,{\cal A}_\pm]=
kS_{\rm WZW}[\phi] + kS_{\rm WZW}[\psi]\nn[3pt]
&&\hskip 1.7truecm
+{k\over\pi}
\int d^2x\>
\Bigl(-\bil{a_+}{\partial_-\psi \psi^{-1}}
+ \bil{a_-}{\phi^{-1}\partial_+ \phi}
+\bil{a_+}{\psi \tau(A_-) \psi^{-1}}\nn[3pt]
&&\hskip 2.7truecm 
+ \bil{a_-}{\phi^{-1} A_+ \phi} - \bil{a_+}{a_-}\nn[3pt]
&&\hskip 2.9truecm 
-\bil{A_+}{\partial_-\phi \phi^{-1}}
+ \bil{\tau(A_-)}{\psi^{-1}\partial_+ \psi} - \bil{A_+}{A_-}\Bigr)\>,
\end{eqnarray}
where $kS_{\rm WZW}$ is the WZW action at level $k$ corresponding to the
group $G$. The dependence of this action on the non-dynamical gauge fields
$a_\pm$ is very simple, allowing them to be integrated out by solving their
equations of motion. The result is
\begin{eqnarray}
&&\hskip -1.5truecm k\widetilde{S}_{\rm gWZW}^{[i,j]}[h,A_\pm]=
kS_{\rm WZW}[\phi] + kS_{\rm WZW}[\psi]
-{k\over\pi} \int d^2x\>
\bil{\phi^{-1}\partial_+ \phi}{\partial_-\psi \psi^{-1}}
\nn[3pt]
&&\hskip 1.7truecm
+{k\over\pi}
\int d^2x\>
\Bigl(-\bil{A_+}{\partial_-(\phi\psi) (\phi\psi)^{-1}}
+ \bil{\tau(A_-)}{(\phi\psi)^{-1}\partial_+ (\phi\psi)} \nn[3pt]
&&\hskip 3.5truecm 
+ \bil{(\phi\psi)^{-1} A_+(\phi\psi)}{ \tau(A_-) } - \bil{A_+}{A_-}\Bigr)\>.
\end{eqnarray}
Finally, using the Polyakov-Wiegmann
formula
\begin{equation}
kS_{\rm WZW}[\phi\psi] = kS_{\rm WZW}[\phi] + kS_{\rm WZW}[\psi] -
{k\over\pi} \int d^2x\> \bil{\phi^{-1}\partial_+ \phi}{\partial_- \psi
\psi^{-1}}\>,
\lab{PW}
\end{equation}
it is straightforward to check that
$k\widetilde{S}_{\rm gWZW}^{[i,j]}[h,A_\pm]$ coincides with the gauged WZW
action in~\rf{Action}, $S_{\rm gWZW}[\gamma,A_\pm]$, for $\gamma=\phi\>
\psi$\,.

Therefore, for $i\not=j$, the fixed point visited by the RG flow after the
first crossover has occurred corresponds to the
gauged WZW action of the `asymmetric coset'~\rf{NewCos},
where different left and right actions of $G^{[i,j]}\times
H$ on 
$G^{[j]}\times G^{[i]}$, specified by the two group
homomorphisms~\rf{AsymGauge}, are gauged. This means that the holomorphic
and anti-holomorphic sectors of the resulting effective
field theory correspond to different cosets sharing the same
central charge~\cite{VOLKER,BARS,GWZWGK}. Taking the form of
$\epsilon_{L/R}$ into account, the left and right cosets can be written as
\begin{equation}
{G^{[j]}_k\over U(1)^{\rg}} \times {G^{[i]}_k\over G^{[i,j]}_k} \quad\text{
and}\quad
{G^{[j]}_k\over G^{[i,j]}_k} \times {G^{[i]}_k\over U(1)^{\rg}}\>,
\lab{LagCosets}
\end{equation}
which are uniquely defined by the inclusion of $g^{[i,j]}$ into $g^{[j]}$
and $g^{[i]}$, respectively.
In retrospect,
this result is not surprising:
the HSG models generically break parity,
and this
can make the relationship between the
holomorphic and anti-holomorphic sectors in any intermediate-scale
conformal field theories non-trivial.
Our discussion has made this explicit by showing
that, for $i\not=j$, and $\mu_{ij}\ggg\mu_{pq}$ $\forall (p,q)\neq
(i,j)$, the fixed point visited
by the RG flow after the first crossover has occurred corresponds to a
`heterotic' conformal field theory where parity is broken, unless
the two cosets in~\rf{LagCosets} are isomorphic.

\newsavebox{\Cut}
\sbox{\Cut}{\begin{picture}(64,20)(0,-3.5)
\put(0,0){\mathversion{bold}$g=$}
\put(10,-3){\framebox(10,6){A}}
\put(20,0){\line(1,0){7}}
\put(27,0){\circle*{1.75}}
\put(27,-5){\makebox(0,0)[b]{{\sts$\bfm{i}$}}}
\put(27,0){\line(1,0){7}}
\put(34,-3){\framebox(10,6){B}}
\put(44,0){\line(1,0){7}}
\put(51,0){\circle*{1.75}}
\put(51,-5){\makebox(0,0)[b]{{\sts$\bfm{j}$}}}
\put(51,0){\line(1,0){7}}
\put(58,-3){\framebox(10,6){C}}
\put(27,0){\line(0,1){7}}
\put(24,7){\framebox(6,10){D}}
\put(51,0){\line(0,1){7}}
\put(48,7){\framebox(6,10){E}}
\end{picture}}
\newsavebox{\Pieces}
\sbox{\Pieces}{\begin{picture}(137,20)(0,-3.5)
\put(-35,0){\mathversion{bold}$g^{[j]_i}=$}
\put(-22,-3){\framebox(10,6){A}}
\put(-12,0){\line(1,0){7}}
\put(-5,0){\circle*{1.75}}
\put(-5,-5){\makebox(0,0)[b]{{\sts$\bfm{i}$}}}
\put(-5,0){\line(1,0){7}}
\put(2,-3){\framebox(10,6){B}}
\put(-5,0){\line(0,1){7}}
\put(-8,7){\framebox(6,10){D}}
\put(22,0){\mathversion{bold}$g^{[i]_j}=$}
\put(35,-3){\framebox(10,6){B}}
\put(45,0){\line(1,0){7}}
\put(52,0){\circle*{1.75}}
\put(52,-5){\makebox(0,0)[b]{{\sts$\bfm{j}$}}}
\put(52,0){\line(1,0){7}}
\put(59,-3){\framebox(10,6){C}}
\put(52,0){\line(0,1){7}}
\put(49,7){\framebox(6,10){E}}
\put(79,0){\mathversion{bold}$g^{[i\cap j]}=$}
\put(96,-3){\framebox(10,6){B}}
\end{picture}}
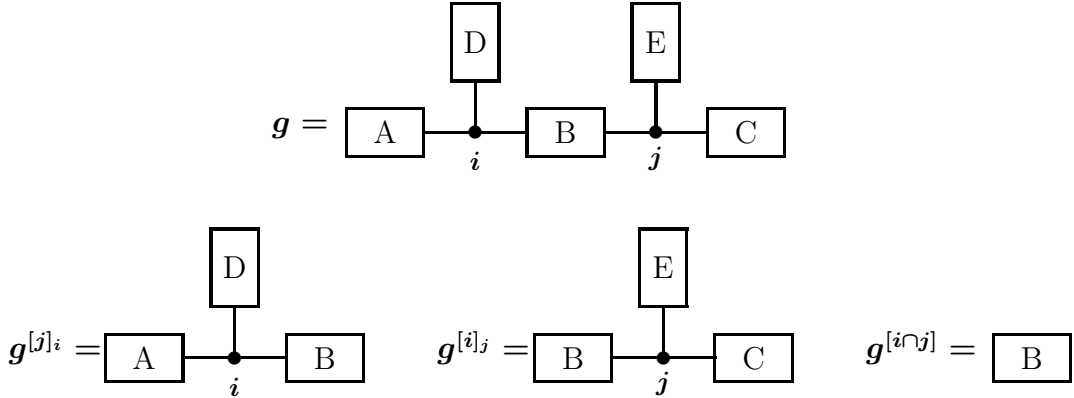
\begin{figure}[ht]
\begin{center}
\begin{picture}(130,60)(0,25)
\put(30,60){\usebox{\Cut}}
\put(30,30){\usebox{\Pieces}}
\end{picture}
\caption{\small
The generic form of the Dynkin diagrams of $g$, $g^{[j]_i}$, $g^{[i]_j}$,
and
$g^{[i\cap j]}$ defined just before eq.~\rf{ProdGroup}. The boxes
$A,B,C,D,E$ stand for Dynkin subdiagrams
connected to
either node~`$i$' or~`$j$'. Some of these boxes can be empty. In particular,
$B$
is empty if~$i$ and~$j$ are neighbouring nodes.}
\label{CutDynk}
\end{center}
\end{figure}

Although the notation used in~\rf{LagCosets} looks simple,
it can be misleading because the subgroups $G^{[j]}$,
$G^{[i]}$, and $G^{[i,j]}$ may
split into products of simple groups and $U(1)$
factors, yielding common factors in the numerators and denominators
that can be cancelled out. This prompts us to write the cosets in a more
detailed way, to clarify their structure. We start by introducing the
following notation: let
$g^{[i]_j}$ be the connected component of the Dynkin diagram $g^{[i]}$
containing the node
$j$, $r^{[i]_j}$ be the rank of $g^{[i]_j}$, and $g^{[i\cap j]}=
g^{[i]_j}\cap
g^{[j]_i}$. Correspondingly, let $G^{[i]_j}$, $G^{[j]_i}$,  and $G^{[i\cap
j]}$
denote the simple compact Lie groups whose Dynkin diagrams are $g^{[i]_j}$,
$g^{[j]_i}$,  and
$g^{[i\cap j]}$, respectively. Taking into account the form of the Dynkin
diagram of
$g$, which is sketched in figure~\ref{CutDynk} in a non-standard but
convenient
way for our purposes, it is easy to prove that any field configuration of
the
form
$\gamma=\phi^{[j]}\psi^{[i]}$ can always be written as the product of a
element in $G^{[j]_i}\times U(1)^{\rg-r^{[j]_i}}$ times an element
in
$G^{[i]_j}\times U(1)^{\rg-r^{[i]_j}}$, where $U(1)^{\rg-r^{[j]_i}}$ and
$U(1)^{\rg-r^{[i]_j}}$ are subsets of $H$; {\em i.e.\/},
\begin{equation}
G^{[j]}\cdot G^{[i]} =\{\phi^{[j]}\psi^{[i]}\}
=\{\widetilde{\phi}^{[j]_i}\>\widetilde{\psi}^{[i]_j}\}
=\bigl(G^{[j]_i}\times U(1)^{\rg-r^{[j]_i}}\bigr)\cdot
\bigl(G^{[i]_j}\times U(1)^{\rg-r^{[i]_j}}\bigr)
\>.
\lab{ProdGroup}
\end{equation}
As a direct consequence of this,
\begin{equation}
\Gamma_{ij}(\gamma)= \Gamma_{ij}(\one) \Rightarrow
\Gamma_{kl}(\gamma)= \Gamma_{kl}(\one)\quad
\forall\> k\notin g^{[i]_j}\quad \text{and}\quad l\notin g^{[j]_i}\>.
\lab{LagShield}
\end{equation}
This ensures that, once the crossover at the scale $\mu_{ij}$ has taken
place, no further crossovers will be observed at $\mu_{kl}$ for any $k\notin
g^{[i]_j}$ and $l\notin g^{[j]_i}$, and provides the Lagrangian
version of the shielding mechanism.

Notice that
$\widetilde{\phi}^{[j]_i}\>\widetilde{\psi}^{[i]_j}$ is invariant under
$\widetilde{\phi}^{[j]_i}\ra
\widetilde{\phi}^{[j]_i} \rho$ and $\widetilde{\psi}^{[i]_j}\ra
\rho^{-1}\widetilde{\psi}^{[i]_j}$ for any $\rho\in G^{[i\cap j]}\times
U(1)^{\rg-r^{[i\cap j]}}$, where $r^{[i\cap j]} = r^{[i]_j} +
r^{[j]_i}-\rg$ is the rank of $g^{[i\cap j]}$. Then, repeating the
arguments used in the previous paragraphs, the cosets~\rf{LagCosets} that
specify the holomorphic and anti-holomorphic sectors of the
effective field theory can be written as
\begin{equation}
{(G^{[j]_i})_k\over U(1)^{r^{[j]_i}}}\times
{(G^{[i]_j})_k\over (G^{[i\cap j]})_k \times
U(1)^{\rg-r^{[j]_i}}}
\quad \text{and}\quad
{(G^{[j]_i})_k\over (G^{[i\cap j]})_k \times
U(1)^{\rg-r^{[i]_j}}}\times
{(G^{[i]_j})_k\over
U(1)^{r^{[i]_j}}}
\>,
\lab{LagCosetsB}
\end{equation}
and these cosets are uniquely defined by the inclusions
$g^{[i\cap j]}\subset g^{[i]_j}$ and
$g^{[i\cap j]}\subset g^{[j]_i}$.
When $i$ and $j$ are
neighbours on the Dynkin diagram of $g$\,, the box $B$ in
figure~\ref{CutDynk} is empty, $G^{[i\cap j]}$ is trivial, and the
conjectured
effective field theory consists of two decoupled parafermionic theories
specified by the cosets
${(G^{[i]_j})_k/ U(1)^{r^{[i]_j}}}$ and ${(G^{[j]_i})_k/
U(1)^{r^{[j]_i}}}$. This particular decoupling could have been anticipated
from the structure of the two-particle scattering amplitudes that define
the $G_k/U(1)^{\rg}$ HSG model, given by~\rf{NewS}. In the limit
$|\sigma_{ij}|\ra\infty$, they split into the amplitudes corresponding to
the
direct sum of the ${(G^{[i]_j})_k/ U(1)^{r^{[i]_j}}}$ and ${(G^{[j]_i})_k/
U(1)^{r^{[j]_i}}}$ HSG models, provided we restrict ourselves to
rapidity values $|\theta|\ll |\sigma_{ij}|$.
Eq.~\rf{LagCosetsB} also shows that whenever $\{i,j\}$ are not
neighbours on the Dynkin diagram of $g$, the effective field theory
involves a coset CFT that is not of parafermionic type.

The predictions obtained using the explicit
formulation of the HSG models in terms of gauged WZW models can be checked
against the TBA results.
The TBA equations can also be used to find candidates for the effective
field theory, as explained in appendix~\ref{Heterotic}.
As a manifestation of parity breaking, the
TBA equations~\rf{TBAGen} are not always symmetrical under
$\theta\rightarrow -\theta$ and, consequently, they generically lead to
different `left' and `right' candidates that share the same central charge.
In the cases that we have analysed, this matches the results
obtained using the gauged WZW formulation, where different cosets specify
the holomorphic and anti-holomorphic sectors of the conformal field theory
corresponding to the fixed point visited.

More quantitative tests are provided by the central charges of the fixed
points visited by the RG flow, which can
be calculated as in section~\ref{TBAsec}.
Consider a set of TBA parameters chosen as in
\rf{gamone} or \rf{gamtwo}, so
that one scale, $m_{ij}$, is larger than all the others.
Then $2r^{-1}\ggg m_{ij}$ corresponds to
the deep
UV limit, where $c(r)\approx C_k(g)$, and increasing the value of $r$
through the first crossover, we get to the regime where $m_{pq}\lll
2r^{-1}\lll m_{ij}$, for all $(p,q)\not=(i,j)$. The
effective value of $c(r)$ is specified by
$\widetilde{g}_l^{\pm}(r,\{m_i\},\{\sigma_{ij}\})$, which are
the subsets of the Dynkin diagram of $g$ defined just before
eq.~\rf{lapprox}.
Taking into account that for this range of values of
$r$ the pseudoenergies
$\varepsilon_a^i$ are effectively independent of the pseudoenergies
$\varepsilon_b^j$, the Dynkin diagram $\widetilde{g}_l^{\pm}$ satisfies the
following relations:
\begin{eqnarray}
i)&& \hskip-0.25truecm {\rm if}\quad i\in \widetilde{g}_l^{\pm}\>,\quad {\rm
then}\quad j\not\in
\widetilde{g}_l^{\pm}
\>,\quad \widetilde{g}_l^{\pm} = \widetilde{{g^{[j]}}}_l^\pm\>, \quad{\rm
and}\quad \widetilde{{g^{[i]}}}_l^\pm = \widetilde{{g^{[i,j]}}}_l^\pm\>, \nn
ii)&& \hskip-0.25truecm {\rm if}\quad j\in \widetilde{g}_l^{\pm}\>,\quad
{\rm then}\quad i\not\in \widetilde{g}_l^{\pm} \>,\quad
\widetilde{g}_l^{\pm} = \widetilde{{g^{[i]}}}_l^\pm\>, \quad{\rm and}\quad
\widetilde{{g^{[j]}}}_l^\pm = \widetilde{{g^{[i,j]}}}_l^\pm\>, \nn
iii)&& \hskip-0.25truecm {\rm if}\quad i,j\not\in
\widetilde{g}_l^{\pm}\>,\quad {\rm then}\quad \widetilde{g}_l^{\pm} =
\widetilde{{g^{[i]}}}_l^\pm = \widetilde{{g^{[j]}}}_l^\pm =
\widetilde{{g^{[i,j]}}}_l^\pm\>,
\end{eqnarray}
for all $l=1\ldots\rg$. Therefore, using~\rf{Pieces},
\rf{PiecesB}, and~\rf{ccform}, the TBA equations lead to the behaviour
\begin{equation}
c(r)\approx 
\begin{cases}
C_k\bigl(g\bigr)\;, & \text{for $\; 2r^{-1}\ggg
{m}_{ij}$},\\
\noalign{\vskip 0.4truecm}
C_k\bigl(g^{[i]}\bigr) + C_k\bigl(g^{[j]}\bigr)-
C_k\bigl(g^{[i,j]}\bigr)\;, &
\text{for
$\; {m}_{pq}\lll 2r^{-1}\lll {m}_{ij}\>, \quad
\forall\; \{p,q\}\not=\{i,j\}$},
\end{cases}
\lab{CrossoverN}
\end{equation}
where $C_k\bigl(g\bigr)$ has been defined in~\rf{cform} for $g\in a,d,e$
with rank~$\rg$; it denotes the central charge of the $G_k/U(1)^{\rg}$
coset conformal  field theory. For
$g$ disconnected, it is the sum of \rf{cform} over all
connected components of $g$. Notice that $C_k\bigl(g^{[i]}\bigr) +
C_k\bigl(g^{[j]}\bigr)- C_k\bigl(g^{[i,j]}\bigr)$ is the central charge of
the conformal field theory specified by the cosets~\rf{LagCosets}.
Therefore, if we identify the
Lagrangian parameters $\{{m}_i,{\sigma}_i\}$ with the TBA parameters such
that $m_{ij}=\mu_{ij}$, and the dimensionless RG scale $\overline{\Lambda}$
with $2r^{-1}$, the behaviour predicted using the gauged WZW formulation
matches the plateau value of $c$ calculated from the TBA.

Eq.~\rf{CrossoverN} resembles a decoupling rule suggested in~\cite{CDF}
to describe the behaviour of the models in the $\sigma_{ij}\ra\infty$ limit.
However, the embeddings necessary to define the cosets were not
discussed in that paper, and in
particular the possible
left-right asymmetry of the relevant theories
was missed.

So far in this section we have restricted ourselves to
cases when one coefficient, say $\mu_{ij}$, is larger than all
of the others, and we have discussed the fixed point
visited by the RG flow just after the crossover associated with $\mu_{ij}$
has occurred.
More generally, in this formalism, crossovers
are associated with the decoupling
of field configurations that become effectively heavy at a given RG
scale. The relevant mass scales are just the classical
masses~\rf{MassRoot},
since, using~\rf{SolvingAm}, they determine the (quadratic)
leading term of the potential
\begin{equation}
V(\exp X)=  {m_0^2\over 4\pi}\> \sum_{i,j=1}^{\rg}\>
{\mu}_{ij}^2\> \Gamma_{ij}({\exp X}) ={m_0^2\over
4\pi}\>\sum_{\alv\in\Phi^+} m_{\alv}^2\left\langle
E_{\alv},E_{-\alv}\right\rangle|\phi_{\alv}|^2
+\cdots\>.
\end{equation}
Consequently, the resulting pattern of crossover scales is the same as
deduced in section~\ref{classicalscales}, and we will not discuss it
again here. Instead, we shall briefly discuss the identification of the
fixed points visited by the RG flow corresponding to the plateaux observed
in the effective central charge $c(r)$. According to the
results in sections~\ref{classicalscales} and~\ref{TBAsec}, a plateau
will
occur whenever, for particular values of the parameters, the
dimensionless mass scales $\{m_{\alv},\; \alv\in\Phi^+\}$ split into two
well-separated sets
${\cal E}_{\rm light}$ and ${\cal E}_{\rm heavy}$, with
$m_{\alv} \lll m_{\bev}$ for any $m_{\alv} \in
{\cal E}_{\rm light}$ and $m_{\bev}\in  {\cal E}_{\rm heavy}$.
The plateau occurs for RG scales between these two sets,
\begin{equation}
m_{\alv} \lll \overline\Lambda\lll  m_{\bev}\qquad
\forall\;
m_{\alv}\in  {\cal E}_{\rm light}\quad \text{and}\quad
m_{\bev}\in  {\cal E}_{\rm heavy}\>,
\lab{EffScales}
\end{equation}
and the corresponding fixed point is expected to be specified by the
field configurations
which are left unfrozen by the
components of the potential associated with ${\cal E}_{\rm heavy}$. This
is, by the field configurations that satisfy the conditions
\begin{equation}
\Gamma_{ij}(\gamma)=\Gamma_{ij}(\one)\qquad \forall\> (i,j)
\quad \text{such that}\quad
\mu_{ij}\sim m_{\alv}\quad \text{with}\quad m_{\alv}\in {\cal E}_{\rm
heavy}\>.
\lab{MoreThanOne}
\end{equation}
The general solution of these conditions and their detailed
comparison with the TBA results deserve further study, which is beyond
the scope of this paper.

In the following, we will just illustrate their
use in a concrete example. Consider the same
$d_4\equiv SO(8)$ case studied in section~3, corresponding to the
parameters in~\rf{d4example}. The scales $\mu_{ij}=m_{ij}$ are ordered as
follows
\begin{equation}
{\mu}_{14} \ggg {\mu}_{24} \ggg {\mu}_{13}  \ggg
{\mu}_{34} \ggg
{\mu}_{23} \ggg {\mu}_{12} \ggg {\mu}_{11}=
{\mu}_{22}= {\mu}_{33}= {\mu}_{44}\;,
\end{equation}
and the relationship between the classical mass scales $m_{\alv}$ and
the coefficients $\mu_{ij}$ can be worked out from~\rf{sumsone}
and~\rf{sumstwo}. 
Consider
first the regime
$\overline\Lambda\ggg {m}_{14}={\mu}_{14}$.
This corresponds to the deep UV limit,
where all the field configurations are effectively massless, the potential
can be completely neglected, and the effective field theory is described by
the unperturbed
$SO(8)_k/U(1)^4$ coset conformal field theory.

Notice that the coefficient $\mu_{14}$ is larger
than all of the others, which means that the effective theory after the
first crossover has occurred is determined by left and right
cosets of the form~\rf{LagCosetsB}. To be concrete, for
the range of energies
$ {m}_{24}={\mu}_{24}\lll \overline\Lambda\lll  {m}_{14}$
only the field configurations that satisfy
$\Gamma_{14}(\gamma)=\Gamma_{14}(\one)$ remain effectively
massless, which are of the form
$\gamma=\phi^{[4]}\>
\psi^{[1]}$. In this case, the Dynkin diagrams $g^{[1]_4}$ and $g^{[4]_1}$
defined just before~\rf{ProdGroup} are the
$a_3$ Dynkin subdiagrams associated to the nodes $\{2,3,4\}$ and
$\{1,2,3\}$,
respectively, while $g^{[1\cap4]}$ is the $a_2$ subdiagram corresponding to
$\{2,3\}$. Then, the two cosets~\rf{LagCosetsB} coincide with
\begin{equation}
{SU(4)_k\over SU(3)_k\times U(1)}\times {SU(4)_k\over U(1)^3}\>.
\end{equation}
Taking~\rf{sumstwo} into account, in this regime all the field
configurations associated with the roots $\alv_1+\alv_2+\alv_4$,
$\alv_1+\alv_2+\alv_3+\alv_4$, and $\alv_1+2\alv_2+\alv_3+\alv_4$ are
decoupled.

Next, we consider the energy scales
$ {m}_{13}={\mu}_{13}\lll
\overline\Lambda\lll {m}_{24}$, where all the field
configurations associated with the roots $\alv_2+\alv_4$ and
$\alv_3+\alv_2+\alv_4$ become decoupled too.
According to~\rf{MoreThanOne}, the
corresponding fixed point is determined by the simultaneous solutions to
$\Gamma_{14}(\gamma)=\Gamma_{14}(\one)$ and
$\Gamma_{24}(\gamma)=\Gamma_{24}(\one)$.
Remarkably, taking~\rf{LagShield} into account, the condition
$\Gamma_{24}(\gamma)=\Gamma_{24}(\one)$ already implies
$\Gamma_{14}(\gamma)=\Gamma_{14}(\one)$ and
$\Gamma_{34}(\gamma)=\Gamma_{34}(\one)$, which has two direct consequences.
First, the effective theory in this regime is specified by left and right
cosets of the form~\rf{LagCosetsB} with $i=2$ and $j=4$, which in our case
read
\begin{equation}
{SU(4)_k\over U(1)^3}\times {SU(2)_k\over U(1)}\>,
\end{equation}
and correspond to two decoupled parafermionic theories.

The second consequence is that
there is no crossover at the scale $m_{34}=\mu_{34}$, exactly as predicted
by our analysis of shielding in section~\ref{classicalscales}.
Therefore, the next regime is $ {m}_{23}={\mu}_{23}\lll
\overline\Lambda\lll {m}_{13}$, where all the field
configurations associated with the root $\alv_1+\alv_2+\alv_3$ become
decoupled. In this case, the fixed point is determined by the solutions
to $\Gamma_{24}(\gamma)=\Gamma_{24}(\one)$ and
$\Gamma_{13}(\gamma)=\Gamma_{13}(\one)$, which are of the form
$\gamma=\phi^{[3,4]}\> \psi^{[1,4]}\> \omega^{[1,2,3]}$.
Using similar arguments to those that lead to~\rf{MoreThanOne}, the
corresponding manifold of effective field configurations can be realised
as a coset manifold in terms of $G^{[3,4]}\times G^{[1,4]}\times
G^{[1,2,3]}$. 
Since $g^{[1,2,3,4]}$ is trivial, $G^{[1,2,3,4]}= H$ and one has to
take into account the invariance of $\gamma$ under $\phi^{[3,4]}\ra
\phi^{[3,4]}\rho$, $\psi^{[1,4]}\ra
\rho^{-1}\>\psi^{[1,4]}\beta$, and $\omega^{[1,2,3]}\ra
\beta^{-1}\omega^{[1,2,3]}$ for each
$\rho\in G^{[1,3,4]}$ and $\beta\in H$, which leads to the identification of
the fixed point as a coset conformal field theory associated with
\begin{equation}
\frac{SU(3)_k}{SU(2)_k\times U(1)}
\times \frac{SU(3)_k}{U(1)^{2}}\times
\frac{SU(2)_k}{U(1)}\>.
\end{equation}
In this particular case, the left and right cosets
coincide.

We can keep considering smaller and smaller energy scales until we reach the
deep IR limit $\overline\Lambda\lll {m}_{ii}$ for all $i=1\ldots4$, where
all the field configurations are decoupled. The resulting flow of effective
field theories is summarised by
\bea
\gamma\equiv \left(\frac{SO(8)_k}{U(1)^{4}}\right)^{\>\><{3}>}
&&\hskip-0.6cm\Buildrel {m}_{14}\over{\hbox to
35pt{\rightarrowfill}}\under{\Gamma_{14}}\> \phi^{[4]}\;
\psi^{[1]}\equiv
\left(\frac{SU(4)_k}{SU(3)_k\times
U(1)}\times\frac{SU(4)_k}{U(1)^3}\right)^{\>\>
<{14\over5}>}\allowdisplaybreaks\nn
\noalign{\vskip0.3truecm}
&&\hskip-2.5truecm\Buildrel {m}_{24}\over{\hbox to
35pt{\rightarrowfill}}\under{\Gamma_{24}}\>  \phi^{[4]}\;
\psi^{[1,2,3]}\equiv
\left(\frac{SU(4)_k}{U(1)^3}\times
\frac{SU(2)_k}{U(1)}\right)^{\>\> <\frac{5}{2}>}\allowdisplaybreaks\nn
\noalign{\vskip0.3truecm}
&&\hskip-2.5truecm\Buildrel {m}_{13}\over{\hbox to
35pt{\rightarrowfill}}\under{\Gamma_{13}}\> \phi^{[3,4]}\;
\psi^{[1,4]}\;
\omega^{[1,2,3]} \equiv \left(\frac{SU(3)_k}{SU(2)_k\times U(1)}
\times \frac{SU(3)_k}{U(1)^{2}}\times
\frac{SU(2)_k}{U(1)}\right)^{\>\> <\frac{12}{5}>}\allowdisplaybreaks\nn
\noalign{\vskip0.3truecm}
&&\hskip-2.5truecm\Buildrel {m}_{23}\over{\hbox to
35pt{\rightarrowfill}}\under{\Gamma_{23}}\> \phi^{[3,4]}\;
\psi^{[1,2,4]}\;
\omega^{[1,2,3]}\equiv \left(\frac{SU(3)_k}{U(1)^{2}}\times
\left[\frac{SU(2)_k}{U(1)}\right]^{\times2}\right)^{\>\>
<{11\over5}>}\allowdisplaybreaks\nn
\noalign{\vskip0.3truecm}
&&\hskip-2.5truecm\Buildrel {m}_{12}\over{\hbox to
35pt{\rightarrowfill}}\under{\Gamma_{12}}\> \phi^{[2,3,4]}\;
\psi^{[1,3,4]}\;
\omega^{[1,2,4]}\;
\chi^{[1,2,3]}\equiv \left(\left[\frac{SU(2)_k}{U(1)}
\right]^{\times4}\right)^{\>\> <{2}>}\allowdisplaybreaks\nn
\noalign{\vskip0.3truecm}
&&\hskip-2.5truecm
\Buildrel {m}_{ii}\over{\hbox to
35pt{\rightarrowfill}}\under{\Gamma_{ii}}\>
\phi^{[1,2,3,4]} \equiv {\rm Massive}^{\>\> <{0}>}\>.
\lab{SO8flow}
\ena
Strictly speaking, the Lagrangian action~\rf{Action} should only be expected
to
give a full description of the theory in the semiclassical (large~$k$)
limit.
However, the resulting values of the effective central charge reproduce the
approximate values calculated using the TBA equations for any value of $k$.
In
particular, in~\rf{SO8flow}, the superscripts $^{<\;>}$  provide the central
charges of the corresponding coset conformal field theories for level $k=2$.
They match the TBA results illustrated
in figure~\ref{fig2},
despite the fact that this value of the level is far from the
semiclassical
regime. The same occurs for the other examples discussed along the
paper.

\sect{Comments on form factor calculations}
\label{FFcomp}

\iindent
There is another context where crossover phenomena can be
seen, namely the behaviour of correlation functions.
It is natural to ask whether, and to what extent, the results
we have obtained from a study of finite-size effects can be reproduced.
In integrable theories, this can be addressed using the
form-factor approach, which provides an infrared series expansion for
correlation functions that typically has good convergence properties down
to short distances.  The method was first applied to the HSG
models associated with $SU(N)_2/U(1)^{N-1}$ for $N=3$
in~\cite{FFother,FFRG}, and for $N\geq4$ in~\cite{FF2}.
However, while crossover effects for unstable particles of height $2$
were observed in \cite{FFRG,FF2}, no transitions associated with
roots of height greater than $2$ were found.
This is as expected for the $SU(3)_2/U(1)^2$ case of \cite{FFRG} since
$SU(3)$ has no roots of height greater than $2$ anyway, but it is
more puzzling for the cases discussed in \cite{FF2}.
Indeed, the plot shown in~\cite{FF2} for $SU(4)_2/U(1)^3$
with $\sigma_{12}=50$ and $\sigma_{23}=20$
should qualitatively match the corresponding flow presented in
\cite{Ltalk,Dorey:2002sc} and in figure \ref{fig1} above,
while in fact even the number of steps is
different. Similarly, the plotted $SU(5)_2/U(1)^4$ flows in \cite{FF2} do
not match our predictions.
These discrepancies between form-factor and finite-size results were
first remarked in \cite{Ltalk} (see~\cite{Dorey:2002sc}),
and for these particular instances they were later traced
\cite{CDF} to a misattribution in the signs of the resonance parameters as
originally given in \cite{FF2} -- so, for example, the $SU(4)$
plot is actually for $\sigma_{12}=50$ and $\sigma_{23}=-20$.
Using our terminology, the scale $m_{13}$ is then shielded by
$m_{12}$ and this resolves the immediate
mismatch.
However, it remains the case that crossovers associated with roots of
heights greater than $2$ have yet
to seen using the form-factor approach.
In this section we shall argue that the reasons for this run, at least
in part, deeper than a simple question of the correct allocation
of the signs of the resonance parameters in the form factors, and that
they have a bearing on a claimed slow-down in the convergence of the
form factor approach for the HSG models.

The investigation of HSG RG flows
in~\cite{FFRG,FF2} was based on
the numerical evaluation of Zamolodchikov's $c$-function~\cite{ZAMc}
using Cardy's sum rule \cite{CARDYc}, expanding the relevant
two-point functions of
the trace of the energy-momentum tensor in terms of
$n$-particle form factors. In practice, such calculations must always
be truncated at some point. Usually, while this affects the accuracy of
results to some -- albeit small \cite{CM} -- extent, no important
information is lost provided one calculates at least as far as
the two-particle contributions. However, as we now show, in theories
with unstable particles there are good reasons to predict that the
effects of early truncation can be more dramatic, sometimes
obscuring physically-relevant crossovers, and sometimes leading to
misleading values for the ultraviolet central charge.

Form factors are matrix elements of some local operator ${\cal O}$
between
a multiparticle in-state and the vacuum. They can be written as
\begin{equation}
F_{n}^{{\cal O} \mid \mu_1\ldots \mu_n} (\theta_1,\ldots, \theta_n)
=\langle 0\mid {\cal O}(0,0) \mid V_{\mu_1}(\theta_1) \ldots
V_{\mu_n}(\theta_n)\rangle_{\rm in}\>,
\lab{FF}
\end{equation}
where the symbol $V_\mu(\theta)$ represents a particle of
species $\mu$ and rapidity $\theta$.
Two-point functions can then be obtained by inserting a complete set
of asymptotic states between the two operators involved,
resulting in a sum over (integrated)
one-, two- and higher- particle contributions.
Only stable particles appear directly in the asymptotic states, and
so this raises the question of how and in what way
the form-factor expansion can be influenced by the existence of any
unstable particles.
Taking the particle interpretation as a guiding principle, we conjecture
that two conditions must be met for the
$n^{\rm th}$-order term associated with the form
factor~\rf{FF} to be sensitive to an unstable
state $\tilde\mu_R$.
The
first condition is that it must be possible to form $\tilde\mu_R$
as a bound state of some subset $\{\mu_{i_1}\dots\mu_{i_m}\}$ of
the particles $\{\mu_1\ldots \mu_n\}$;
the second,
apparently trivial but as we
shall see important, is that
$F_{n}^{{\cal O} \mid \mu_1\ldots \mu_n}$
should not be identically zero.
Notice that the second condition will often require that
$n$ be strictly larger than $m$.

All HSG form factor calculations to date have been performed for level
$k=2$, and for simplicity we restrict to such cases here too. Then
the quantum number $a$ of
single-particle states $(i,a)$ is always equal to $1$ and can
be dropped, while $i$ labels
a simple root $\alv_i$ of $g$. The remaining, unstable, particles are
similarly indexed by the positive roots
$\bfm{\beta}=\sum n_i \bfm{\alpha}_i$ of height $\geq2$. The
condition that it be possible to form the unstable particle $\bev$ from
some subset of the set
$\{\alv_{i_1}\dots\alv_{i_n}\}$
of stable particles
is that it should contain $n_1$ times
$\bfm{\alpha}_1$,
$n_2$ times $\bfm{\alpha}_2$, and so on.

The second condition, the non-vanishing of the relevant form
factors, can be partially analysed using a spin-zero conserved charge.
As in the unperturbed theory of level-$k$
$G$-parafermions~[13], the HSG theory associated with the coset
$G_k/U(1)^{\rg}$ possesses a discrete conserved charge taking values in
$\Lambda_g$ modulo
$k\times\Lambda_g$, where $\Lambda_g$ is the root lattice of $g$. This
generalises the ${\fl Z}_k$ charge of the usual parafermions, which is
recovered
for $G=SU(2)$. Stable particles correspond to states of definite
charge, and the particle $(i,a)$ carries the charge
$a\alv_i$~[3].
Taking this into account, a necessary condition to ensure that a given
form
factor is non-vanishing is that
the total charge of the multiparticle state matches the charge of the
local operator ${\cal O}$. In the particular case of neutral
operators
such as the trace of the energy-momentum tensor, $\Theta$, this implies
that
the form factor associated with a multiparticle state
$(i_1,a_1)....(i_n,a_n)$ will vanish unless
\begin{equation}
\sum_{j=1}^n a_j \alv_{i_j} \in k\times\Lambda_g
\end{equation}
Therefore, for $k=2$, the only form factors of neutral operators
that can be non-vanishing are those corresponding to multiparticle
states
where each simple root appears an even number of times. For $\Theta$,
this
general rule is in agreement with the results of~[6,8].

Consider now the calculation of Zamolodchikov's
$c$-function for the HSG theories at $k=2$. This requires the two-point
function of $\Theta$, and hence the form
factors $F_n^{\Theta|\alv_{i_1}\ldots \alv_{i_n}}$.
Combining our two conditions, we conclude that the effects of
unstable resonance states associated with the roots $\bev$ of height
${\rm ht\/}(\bev)$ will only be seen in
the $n^{\rm th}$-order term in the form factor expansion if
$n\ge 2\>{\rm ht\/}(\bev)$.
Since the numerical calculations of~\cite{FF2} went up to
6-particle form factor contributions, this means that they are not expected
to
be sensitive to unstable particles associated with roots of height
larger that~$3$. 

All of this leads to the following predictions. For
$SU(4)_2/U(1)^3$, the calculation of Zamolodchikov
$c$-function for the values of the resonance parameters originally
quoted
in~\cite{FF2}, using the form factor expansion up to the $6^{\rm th}$-order
term, should qualitatively match the dotted line in
figure~\ref{fig1}
and detect the crossover transition at the scale fixed
by $m_{\bfm{\alpha}_1+\bfm{\alpha}_2+\bfm{\alpha}_3}$. In
contrast, for $SU(5)_2/U(1)^4$, it should only find the transitions
associated
with the scales $m_{\bfm{\alpha}_1+\bfm{\alpha}_2+\bfm{\alpha}_3}$ and
$m_{\bfm{\alpha}_2+\bfm{\alpha}_3+\bfm{\alpha}_4}$, but no crossover
transition associated with the maximal root alone, which in this case is
of height~4. In general, for $SU(N)_2/U(1)^{N-1}$ the
maximal number of crossovers that can be observed when truncating the form
factor series at the 6-particle contribution is $3(N{-}2)$, the
number
of roots of height $\leq 3$, while the maximal number
of separable scales quoted in table~\ref{MaxNumb} for $g=a_{N-1}$ is
$N(N{-}1)/2$. Hence for $N\ge 5$ there will always be choices of the
parameters for which some
crossovers are missed completely, if the form factor expansion is not
pushed further. 

On the other hand, if the parameters are chosen such that no new
physical crossover scales are associated with roots of height greater
than two, so that all crossovers are already noticed by the low-lying
roots, then another, more subtle, effect comes
into play -- while a crossover will be seen at the appropriate
energy scale, the change in the effective central charge will be wrong, as
not
all relevant states will have been taken into account. Recall that
the pattern
of crossovers observed in the HSG models can be viewed as a consequence of
the change in the number of field configurations that remain effectively
massless at the RG scale, both stable and unstable.
According to our analysis, a similar, but not identical,
effect is at work in the form factor expansion: truncating the
series at a certain number of particles will suppress the contribution
of some unstable particle states, which would otherwise have been seen
once the energy scale became large enough.
If we try to calculate the effective central charge in the far
ultraviolet, then this effect is sure to show up, no matter what
values are chosen for the S-matrix parameters. As will now be shown,
this idea can be checked quantatively against previously-published
data.

The central charge of the unperturbed theory can be recovered
by integrating the derivative of Zamolodchikov's
$c$-function not just up to some finite scale, but
all the way to the far ultraviolet.  Using the
form factor approach, this leads to a series
\begin{equation}
c_{UV} = \sum_{n=1}^{\infty} \Delta c^{(n)}
\lab{UVSeries}
\end{equation}
for the UV central charge,
where $\Delta c^{(n)}$ is the $n$-particle form factor contribution.
This typically has excellent convergence properties, for which there
are good analytic arguments~\cite{CM}.
However, in \cite{FF2} this general understanding was called into question:
a study of the truncations of the form factor series for the
$SU(N)_2/U(1)^{N-1}$ HSG models up to 6 particles led to the suggestion that
convergence was becoming slower and slower for
increasing values of $N$.
The proposed
decoupling of unstable particle states from form factors
involving a small number of particles allows an alternative
understanding of the data presented in \cite{FF2}: it is not that the
calculated
central charge is becoming more inaccurate, it is just that it is no
longer measuring
the UV central charge of the full theory, but rather that of a
subtheory
in which certain particle states are decoupled.
This phenomenon should be general to all theories with unstable particles,
and we predict that once sufficiently-many terms have been included in the
form factor expansion that no unstable particle contributions are
artificially
excluded, the good convergence properties argued for in \cite{CM} will be
restored.

For $SU(N)_2/U(1)^{N-1}$, this idea can be tested
against the numerical
results of~\cite{FF2} for the would-be UV central charge.
Let $c^{(m)}_{UV}\equiv\sum_{n=1}^{m}\Delta c^{(n)}$ denote the UV central
charge computed from the series \rf{UVSeries} truncated at the $m$-particle
contribution.
The prediction suggested by our analysis is that, for $m < 2(N{-}1)$,
$c^{(m)}_{UV}$ should not
approximate the UV central charge of the
full theory, but rather that of the effective theory obtained
by decoupling all
unstable particles associated with the roots of height larger that $m/2$.
In contrast, once $m\geq 2(N{-}1)$, the value of $c^{(m)}_{UV}$ should
converge in the usual manner
to the value of $c_{UV}$, which is $C_2(a_{N-1})$ in the
notation of \rf{cform}.

We start by quoting the individual
$n$-particle form factor contributions to the value of $c_{UV}$ for
$SU(N)_2/U(1)^{N-1}$, as given  in~\cite{FF2}:
\begin{eqnarray}
&&\Delta c^{(2)} =(N-1)\times 0.5\>, \nn
&&\Delta c^{(4)} =(N-2)\times 0.197\>, \nn
&&\Delta c^{(6)} =(N-2)\times 0.002 + (N-3)\times 0.0924\>.
\lab{FFnumbers}
\end{eqnarray}
Substituted into the expansion, these lead to
\begin{eqnarray}
&&c^{(2)}_{UV}= (N-1)\times 0.5\>, \nn
&&c^{(4)}_{UV}= N\times 0.697 - 0.894\>, \nn
&&c^{(6)}_{UV}= N\times 0.7914 -1.1752\>.
\lab{FFcalcs}
\end{eqnarray}
Our claim is that these numbers should approximate
the ultraviolet central charges of the effective
theories obtained by decoupling all the unstable particles associated with
the roots of height larger than $\chi$, where $\chi=1,2,3$
for $c^{(2)}_{UV}$, $c^{(4)}_{UV}$ and $c^{(6)}_{UV}$ respectively.
For the $SU(N)_2/U(1)^{N-1}$ theories, there is an alternative way to
realise the 
required effective theories, by
choosing the $S$-matrix parameters as follows
\begin{equation} m_i=  1 \quad
\text{and}\quad
\sigma_{i}=i\sigma\quad \forall\> i=1,\ldots, N-1 \quad
\Rightarrow
\;\; m_{ij}= \ex{|i-j|\sigma}\>,
\lab{decvals}
\end{equation}
with $\sigma\gg0$\,, and examining the theory at scales
$\ex{(\chi-1)\> \sigma/2}\lll 2r^{-1} \lll \ex{\chi\> \sigma/2}$,
where all the field configurations associated with the
roots of height larger than~$\chi$ have been decoupled.
Using the TBA analysis given earlier, for the parameters \rf{decvals}
the effective central charge $c(r)$ has
$N-1$ plateaux as $r$ varies from $0$ to $\infty$, matching
the successive decoupling of roots of
heights $N-1$, $N{-}2$, \dots, $1$:
\begin{equation}
c(r) \approx
\begin{cases}
c_{UV} \;, & \text{for $\;2r^{-1}\ggg \ex{(N-2)\>\sigma/2}\>,$}\\
\noalign{\vskip 0.2truecm}
c_\chi\;, & \text{for $\;\ex{(\chi-1)\> \sigma/2}\lll 2r^{-1} \lll
\ex{\chi\>
\sigma/2}\>, \quad \chi=1\ldots N-2\>,$}\\
\noalign{\vskip 0.2truecm}
0 \;, & \text{for $\;2r^{-1}\lll 1$.}
\end{cases}
\end{equation}
The plateau central charges $c_\chi$ can then
be calculated using the results of
section~\ref{TBAsec}. The result is
\begin{equation}
c_\chi= 
(N-\chi)\> C_2(a_\chi)-(N-\chi-1)\> C_2(a_{\chi-1})\>, \quad \chi=1\ldots
N-1\>,
\end{equation}
where $c_{N-1}=c_{UV}$, and we have used $C_2(a_0)=0$.
For
$\chi=1,2,3$,
this gives
\begin{equation}
c_1= (N-1)\times {1\over2}\>, \qquad
c_2= N\times {7\over10}\>-\>{9\over10}\>, \qquad
c_3= N\times {4\over5}\>-\>{6\over5}\>.
\end{equation}
Comparing with \rf{FFcalcs}, we see that
while the difference between
$c^{(6)}_{UV}$ and $c_{UV}=C_2(a_{N-1})$
becomes larger and larger with
increasing values of $N$, with a relative error that reaches up to
20\%\,,
$c^{(2\chi)}_{UV}$ approximates $c_\chi$ with a relative error of less
than $1\%$ for any value of $N$.
This offers strong support for our contention that the
problems previously observed in form-factor results for the HSG models
are not due to any general degrading
of the convergence properties of the series, but rather
to a controlled decoupling of states which could, in principle, be
remedied by adding a {\em finite}\/ number of further
terms.  The data is
illustrated in table \ref{tabcomp}.

\begin{table}[h]
\begin{center}
\begin{tabular}{|c||c||c|c||c|c|}
\hline 
$N$& $\qquad c_{UV}\qquad$ & $\qquad c_{UV}^{(6)}\qquad$ &  $\qquad
c_{3}\qquad $& $\qquad c_{UV}^{(4)}\qquad$ &  $\qquad
c_{2}\qquad $
\\
\hline \hline
4 &  2 &  1.9904 &  2 &1.894 & 1.9\\
\hline
5 &  2.85714 &  2.7818 &  2.8 & 2.591 & 2.6\\
\hline
6 &  3.75 &  3.5732 &  3.6 & 3.288 & 3.3\\
\hline
10 &  7.5 &  6.7388 &  6.8 & 6.076 & 6.1\\
\hline
100 &   97.0588 &   77.9648 &   78.8 & 68.806 & 69.1\\
\hline
\end{tabular}
\end{center}
\caption{\small
The UV central charges of the $SU(N)_2/U(1)^{N-1}$
HSG models, $c_{UV}$\,, compared with form factor
results, from \cite{FF2}, truncated at the $2\chi$-particle contribution,
$c_{UV}^{(2\chi)}$\,, and with the plateau central charges $c_{\chi}$
corresponding to the decoupling of all the unstable particles associated
with
roots of heights larger than $\chi$,
for $\chi=2,3$.}
\label{tabcomp}
\end{table}

\sect{Conclusions}
\iindent
In this paper, the patterns of crossover phenomena in the HSG models
have been analysed in detail, principally through a study of
finite-size effects using the thermodynamic Bethe ansatz.
We have restricted our attention to simply-laced~$G$, but we expect
that similar results will hold for the non simply-laced cases too.
For suitable values of the parameters, the finite-size scaling
function $c(r)$ undergoes a series of well-separated crossovers.
The positions of these crossovers allowed us to
identify scales associated with both stable and unstable
quantum particles, and to show that they match semiclassical data, even
far from that regime.  Although only well-separated scales can be seen in
this way, our results provide non-perturbative support for the idea that
{\em all}\/ of the semiclassical particles, both stable and unstable,
survive in the quantum theory, for any value of $k$.

The crossovers corresponding to unstable particles associated with
roots of height larger than two have yet to be observed in calculations
based on the form factor approach~\cite{FF2}. While this lack is
partially explained by the particular values of the parameters chosen
for the computations performed to date \cite{CDF}, we pointed out that
truncation effects in the
form factor series should be particularly important in the presence of
unstable particles, and proposed that this should lie behind
previously-observed losses in accuracy in such calculations.

Our results imply the existence of a great variety of `staircase'
renormalisation group flows which, starting from the UV, pass
close to a finite number of other fixed points before reaching their
ultimate destinations. The
multiparameter nature of the HSG models means that their flows in fact
sweep out whole manifolds
of integrability in the space of theories, and the staircases can
be understood as lying near the boundaries of these manifolds.  We have
provided general rules to
find the central charges of the fixed points visited by these
flows, by calculating the plateau values of $c(r)$.
We have also established the maximal number of steps for each model;
owing to the shielding phenomenon, for the $d$ and $e$ algebras this
number is less than might have been expected.  It would
be interesting to classify the different ways to realise the maximal
number of steps, and more generally to find a more group-theoretical
interpretation of the patterns of successive symmetry-breakings
revealed by the TBA analysis.

At the crossovers, the models are described by a set of effective 
TBA systems that extends the class of massless TBAs discussed
in~\cite{ZAMOcoset,RAVA,DYNK}, enabling the HSG models to unify 
these simpler flows within a common structure. This relationship
can be used to deduce a Lagrangian formulation for some of these massless
TBA systems from the formulation of the HSG models in terms of perturbed
gauged WZW actions. Another interesting aspect of the HSG TBA staircase
patterns is that the plateau values of the scaling function correspond to
certain exact multiple scaling limits.
The relationship between these limits and quantum group reduction
merits further study, perhaps making use of ideas discussed
in~\cite{Smirnov:ew}.

A final question, partially addressed in this paper,
is the precise
identification of the conformal field theories visited by the
RG flows. We have investigated this using the TBA
equations and the Lagrangian formulation, both of which lead
to natural candidates for the relevant fixed points.
In many cases these are
heterotic conformal field theories~\cite{VOLKER,BARS,GANNON}.
To the best of our knowledge, this is the first time that the issue of
RG flows to and from such conformal field theories has been
raised\footnote{But note,
{\em marginal}\/ heterotic deformations
are discussed in, for example, \cite{Israel:2004vv}, while
a perturbation with more extreme
left-right asymmetry, for which even
the perturbing dimensions on left and right differ, is
treated in~\cite{cdy}.},
and it clearly deserves further work. As suggested in more general
terms in \cite{Dorey:2002sc},
the full spectrum of excited states should provide important
information in this regard, and a study of this using the TBA
techniques developed in~\cite{Bazhanov:1996aq,Dorey:1996re}
would be worthwhile.

\vskip.8truecm

\noindent\centerline{\large\bf Acknowledgments}

\vskip0.15truecm
We would both like to thank
Pascal Baseilhac, Peter Bowcock, Aldo Delfino, Terry Gannon, Werner
Hoffmann, Angel Paredes, Volker Schomerus,
Roberto Tateo and Jean-Bernard Zuber for helpful discussions during
the course of this project, and SPhT Saclay for hospitality in its
final stages. PED and JLM also thank USC and SISSA, respectively,
for hospitality.
The work was partly supported by the EC
network ``EUCLID", contract number HPRN-CT-2002-00325, and partly by
a NATO grant PST.CLG.980424.
JLM also thanks MCyT (Spain) and FEDER (BFM2002-03881 and
FPA2002-01161), and Incentivos from Xunta de Galicia
for financial support.

\vskip1truecm

\appendix

\sect{Coset candidates and hints of heteroticity}
\label{Heterotic}

\iindent
The plateau values of the effective central
charge calculated in section~\ref{platsec} provide partial, but not
complete,
information about the fixed points visited by the RG flows. At various
points in this paper we have specified fixed points not by their central
charges, but rather by giving candidate coset conformal field theories.
In this appendix we review briefly how such candidates can be found directly
from the structure of the TBA equations, and explain why they should be
treated
with particular caution for the HSG models.

The basic procedure is simple.
In section \ref{platsec}, the plateau central charge was evaluated as a sum
over the left and right active energy terms of contributions
$c_p^{\pm}=\fract{1}{2}(C_k(g_p^{\pm})-C_k(\widehat g_p^{\pm}))$. The first
step is
to identify $2c_p^\pm$ with the central charge of
the coset conformal field theory
\begin{equation}
{\bigl( G_p^\pm\bigr)_k \over \bigl(
\widehat{G}_p^\pm\bigr)_k
\times U(1)}\>,
\lab{TensorBlock}
\end{equation}
where $G_p^\pm$ and $\widehat{G}_p^\pm$ are the compact subgroups of $G$
specified by the Dynkin diagrams $g_p^\pm$ and $\widehat{g}_p^\pm$,
respectively. 
By construction, $G_p^\pm$ is simple,
and $\widehat{G}_p^\pm$ semisimple. These cosets are
uniquely defined by the (regular) embeddings of
$\widehat{G}_p^\pm$ into $G_p^\pm$ provided by the inclusion $\widehat
g_p^{\pm} \subset g_p^{\pm}$, and the identification of the $U(1)$ factor
with the one-dimensional subgroup generated by the Cartan element
associated with the fundamental weight $\bfm{\lambda}_p$.
A candidate for the conformal field theory of the plateau is then
obtained by tensoring all these terms,
\begin{equation}
{\bigl( G_1^\pm\bigr)_k \over \bigl(
\widehat{G}_1^\pm\bigr)_k
\times U(1)} \times {\bigl( G_2^\pm\bigr)_k \over \bigl(
\widehat{G}_2^\pm\bigr)_k
\times U(1)} \times\;\cdots \;\times {\bigl( G_{\rg}^\pm\bigr)_k \over
\bigl(
\widehat{G}_{\rg}^\pm\bigr)_k
\times U(1)}\>,
\lab{Tensor}
\end{equation}
and reducing the resulting expression by means of the cancellation of
those simple factors that appear
associated with the same nodes of the Dynkin diagram of $g$
both in the numerator and the denominator.

To see the procedure in action, consider the plateau whose effective TBA
equations are summarised by figure~\ref{calc2}a. The $\{c_p^-\}$
contributions lead to the following `left' coset candidate:
\begin{equation}
{SU(2)^{\{1\}}_k\over U(1)^{\{1\}}}\times {SU(3)^{\{1,2\}}_k\over
SU(2)^{\{1\}}_k\times U(1)^{\{2\}}} \times {SU(4)^{\{1,2,3\}}_k\over
SU(3)^{\{1,2\}}_k\times U(1)^{\{3\}}},
\lab{SU4L}
\end{equation}
where we have introduced the notation $G^{\{i_1\ldots i_n\}}$ to indicate
that this group is associated with the nodes $i_1\ldots i_n$ of the Dynkin
diagram of $g$. After the cancellation of
the common terms of numerator and denominator, this coset simplifies to
\begin{equation}
{SU(4)^{\{1,2,3\}}_k\over
U(1)^{\{1\}}\times U(1)^{\{2\}}\times
U(1)^{\{3\}}} \>\equiv \> {SU(4)_k\over U(1)^3}\>.
\end{equation}
Similarly, the $\{c_p^+\}$ contributions lead
to the `right' coset candidate
\begin{equation}
{SU(2)^{\{1\}}_k\over U(1)^{\{1\}}}\times {SU(4)^{\{1,2,3\}}_k\over
SU(2)^{\{1\}}_k\times U(1)^{\{2\}} \times SU(2)^{\{3\}}_k} \times
{SU(2)^{\{3\}}_k\over U(1)^{\{3\}}}\>,
\lab{SU4R}
\end{equation}
which, after the cancellations, also leads to $SU(4)_k/ U(1)^3$. Notice that
the cancellation performed in~\rf{SU4L} and~\rf{SU4R} takes place only
between simple factors associated with the same nodes of the Dynkin
diagram of $g$.

In this first example, the left and
right candidates coincide, which should be expected
since the effective TBA equations represented by figure~\ref{calc2}a
correspond to the deep UV limit of the
$SU(4)_k/U(1)^3$ HSG model. This correspondence
provides an {\em a posteriori\/} justification for the cancellations
leading from~\rf{SU4L} and~\rf{SU4R} to the coset $SU(4)_k/U(1)^3$.

\begin{figure}[ht]
\begin{center}
\epsfig{file=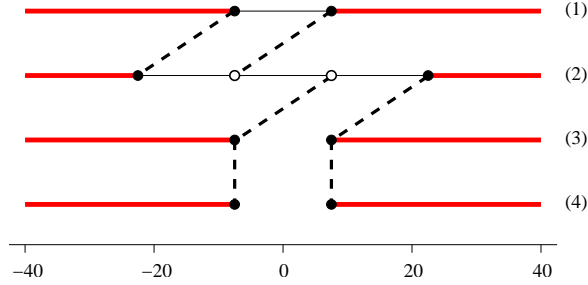,height=0.23\linewidth}
\caption{\small Effective TBA systems with asymmetric candidate
cosets.}
\label{calcasym2}
\end{center}
\end{figure}

However, since the individual contributions $c_p^+$ and $c_p^-$
are different in general, they often lead to different
left and right candidates. This will be illustrated by our second example.
Consider the effective TBA equations
summarised by
figure~\ref{calcasym2}. In this case, some of the Dynkin diagrams
$\{g_p^\pm\}$ coincide,
namely $g_2^+=g_3^+=g_4^+$, $g_1^-=g_2^-$,
$g_3^-=g_4^-$, and the general rules to calculate the plateau central
charge have to be modified as described in section~\ref{effsection},
so that the separate contributions
$\{c_1^+\ldots c_4^+\}$ and
$\{c_1^-\ldots c_4^-\}$ to the total effective central charge are
replaced by
$\{c_1^+,c_{2+3+4}^+\}$ and $\{c_{1+2}^-,c_{3+4}^-\}$, respectively
(see eq.~\rf{eeceff}). Then, the value of, say, $c_{p_1+\ldots +p_n}^\pm$ is
calculated as for the generic case, the only difference being that the
diagram $\widehat{g}_{p_1}^\pm$ is now found by deleting all the nodes
$p_1\ldots p_n$ from $g_{p_1}^\pm$, instead of just $p_1$. Correspondingly,
in our construction,
$2c_{p_1+\ldots +p_n}^\pm$ is identified with the central charge of the
coset conformal field theory
\begin{equation}
{\bigl( G_{p_1}^\pm\bigr)_k \over \bigl(
\widehat{G}_{p_1}^\pm\bigr)_k
\times U(1)^n}\>,
\lab{GenTensorBlock}
\end{equation}
which generalises~\rf{TensorBlock}.
The
$\{c_p^-\}$ contributions lead to the coset candidate
\begin{equation}
{SU(3)^{\{1,2\}}_k\over U(1)^{\{1\}} \times
U(1)^{\{2\}}}\times {SU(4)^{\{2,3,4\}}_k\over
SU(2)^{\{2\}}_k\times U(1)^{\{3\}}\times U(1)^{\{4\}}} \equiv
{SU(3)_k\over U(1)^2} \times {SU(4)_k\over SU(2)_k\times U(1)^2} \>,
\lab{SU5L}
\end{equation}
while the candidate suggested by $\{c_p^+\}$ is
\begin{equation}
{SU(3)^{\{1,2\}}_k\over SU(2)^{\{2\}}_k\times U(1)^{\{1\}}}\times
{SU(4)^{\{2,3,4\}}_k\over U(1)^{\{2\}}\times U(1)^{\{3\}}\times
U(1)^{\{4\}}}\equiv
{SU(3)_k\over SU(2)_k\times U(1)}\times {SU(4)_k\over
U(1)^3} \>.
\lab{SU5R}
\end{equation} 
Clearly,
for this case, our prescription indeed leads to different left and right
coset conformal field theory candidates, although with the same central
charge.

Since
the different left and right candidates are constructed by analysing the
left and right active energy term contributions, it is natural to identify
the resulting candidates with the holomorphic and anti-holomorphic sectors
of a conformal field theory. In cases where those candidates are different,
one might predict that
the corresponding fixed point visited by the RG flow should be of
heterotic type. In the particular case
of~\rf{SU5L} and~\rf{SU5R}, they would correspond to an asymmetric coset
conformal field theory~\cite{VOLKER,BARS} associated with
\begin{equation}
{SU(3)_k\times SU(4)_k \over SU(2)_k\times U(1)^4}\>,
\end{equation}
where~\rf{SU5L} and~\rf{SU5R} specify the different left and right actions
of $SU(2)\times U(1)^4$ on $SU(3)\times SU(4)$.

This matches the Lagrangian calculations of
section \ref{LAGsec}. The effective TBA
equations represented by figure~\ref{calcasym2} correspond to the
$SU(4)_k/U(1)^3$ HSG model in the regime $\mu_{pq}\lll 2r^{-1}\lll
\mu_{41}= \mu_{31}$, $\forall (p,q)\not= (4,1), (3,1)$. According
to~\rf{MoreThanOne},
the effective theory is specified by the field configurations of the form
$\gamma=\phi^{[1]}\psi^{[3,4]}$, which can always be
written as $\gamma=\widetilde{\phi}^{[1]}\widetilde{\psi}^{[3]}$.
The two candidates~\rf{SU5L} and~\rf{SU5R} then coincide with the cosets
in~\rf{LagCosetsB} for $i=3$ and $j=1$.

We finish this appendix by showing that,
despite the fact that the left and right candidates can be different, they
always have equal central charges. This follows
from the general indentity
\begin{equation}
\sum_{p=1}^{\rg} c_p^+(r) =\sum_{p=1}^{\rg} c_p^-(r)\>,
\end{equation}
which can be proved as follows. Eq.~\rf{LeftRight} and the TBA
equations~\rf{TBAGen} lead to
\begin{eqnarray}
\sum_{p=1}^{\rg} c_p^{+}(r) -\sum_{p=1}^{\rg} c_p^{-}(r)&=& \frac{3}{\pi^2}
\sum_{p=1}^{\rg} \sum_{a=1}^{k-1}
\int_{-\infty}^{+\infty} d\theta\> m_{i}\mu_a r\> \sinh\theta\>
L_a^p(\theta)\nn
&=&\frac{3}{\pi^2} \sum_{p=1}^{\rg} \sum_{a=1}^{k-1}
\int_{-\infty}^{+\infty} d\theta\> {\nu_a^{p\>}}'(\theta)
L_a^p(\theta)\nn[3pt]
&=& {\cal J}(r) +{\cal I}_0(r) + {\cal I}_\sigma(r)\>,
\end{eqnarray}
where $f'(\theta)=df(\theta)/d\theta$,
\begin{eqnarray}
&&{\cal J}(r) = \frac{3}{\pi^2} \sum_{p=1}^{\rg} \sum_{a=1}^{k-1}
\int_{-\infty}^{+\infty} d\theta \> {\varepsilon_a^{p\>}}'(\theta)
L_a^p(\theta) \>, \allowdisplaybreaks\nn
&&{\cal I}_0(r) =\frac{3}{\pi^2} \sum_{p=1}^{\rg} \sum_{a,b=1}^{k-1}
\int_{-\infty}^{+\infty} d\theta \> \phi_{ab}'\ast L_b^p(\theta) \>
L_a^p(\theta)\>,\allowdisplaybreaks\nn
&&{\cal I}_\sigma(r) =\frac{3}{\pi^2} \sum_{p,j=1}^{\rg}\>
\sum_{a,b=1}^{k-1}
\int_{-\infty}^{+\infty} d\theta \> I_{pj}^g\> \psi_{ab}'\ast
L_b^j(\theta-\sigma_{jp}) \> L_a^p(\theta)\>.\allowdisplaybreaks
\end{eqnarray}
All these contributions vanish. For instance, the first one is just
\begin{equation}
{\cal J}(r) = \frac{3}{\pi^2} \sum_{p=1}^{\rg} \sum_{a=1}^{k-1}
\int_{\varepsilon_a^{p\>}(-\infty)}^{\varepsilon_a^{p\>}(+\infty)}
d\varepsilon\> \ln(1+\ex{-\varepsilon})=0\>,
\end{equation}
where we have used~\rf{nudec}. The second is
\begin{eqnarray}
{\cal I}_0(r) &=&\frac{3}{\pi^2} \sum_{p=1}^{\rg} \sum_{a,b=1}^{k-1}
\int_{-\infty}^{+\infty} d\theta \> \int_{-\infty}^{+\infty}
\frac{d\tilde{\theta}}{2\pi}
\>\phi_{ab}'(\theta-\tilde{\theta}) L_b^p(\tilde{\theta}) \>
L_a^p(\theta)\nn
&=&-\frac{3}{\pi^2} \sum_{p=1}^{\rg} \sum_{a,b=1}^{k-1}
\int_{-\infty}^{+\infty} d\tilde{\theta} \> \int_{-\infty}^{+\infty}
\frac{d\theta}{2\pi}
\>\phi_{ba}'(\tilde{\theta}-\theta) L_a^p(\theta) \> L_b^p(\tilde{\theta})
\nn
&=&- {\cal
I}_0(r) =0\>,
\end{eqnarray}
where we have just swapped the integration variables\foot{The integral over
the entire $\theta,\tilde{\theta}$ plane is not absolutely convergent, which
might make a naive swap of integration variables lead to a wrong result. A
more careful analysis of integrals of this kind has been performed
in~\cite{PAT} (see eq.~(3.14)), which in our case leads to the same
results.}
and used that
$\phi_{ab}(\theta)=\phi_{ba}(-\theta)$. Similarly, it can be checked that
${\cal I}_\sigma(r)$ vanishes by using
$\psi_{ab}(\theta)=\psi_{ba}(-\theta)$,
$I_{pj}^g=I_{jp}^g$, and $\sigma_{jp}=-\sigma_{pj}$.

\sect{An
{\mathversion{bold}$E_6$} HSG flow with maximal number of steps.}
\label{AppFlows}

\iindent
As a confirmation of the results
quoted in table~\ref{MaxNumb} for $g=e_n$, we include, as a final example,
an $(E_6)_2/U(1)^6$ HSG model that generates a flow with
the predicted maximal
number of steps, {\em i.e.\/} $20=6(6{+}1)/2-1$.
\begin{figure}[ht]
\begin{center}
\epsfig{file=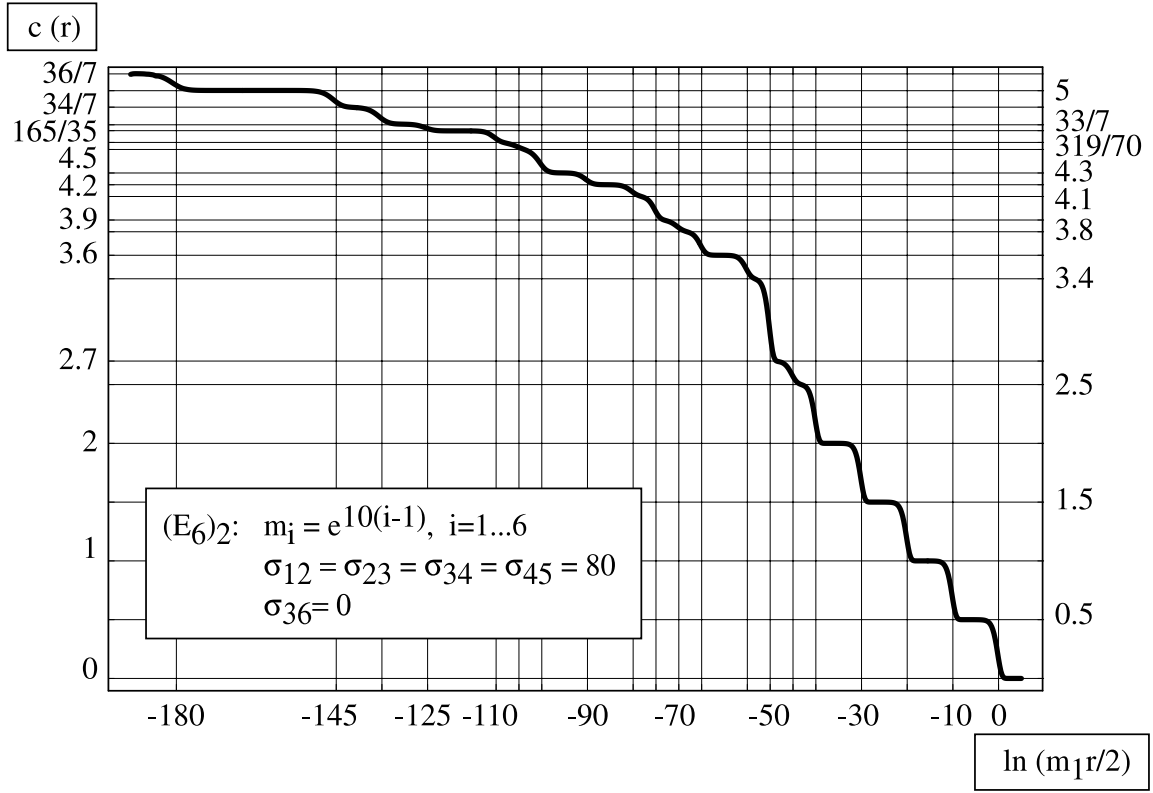,width=0.95\linewidth}
\caption{\small The TBA scaling function for the $(E_6)_2/U(1)^6$  HSG
model.}
\label{fige6}
\end{center}
\end{figure}
Figure \ref{fige6} shows the flow of the effective central charge for
the following choice of parameters:
\begin{equation}
m_i= \ex{(i-1)a}\>, \quad \sigma_i= ib\>, \quad \forall i=1\ldots 5\>,
\quad m_6= \ex{5a}\>, \quad \sigma_6=\sigma_3
\end{equation}
with $a=10$ and $b=80$. Using the results of section~\ref{platsec}, the
central
charges of the fixed points visited by the corresponding staircase flow are
predicted to be 
\begin{eqnarray}
&&\hskip-0.75cm\{e_6\}=\ArrFract{36}{7}
\ArrFl{15}{180}
\{2d_5\}-\{d_4\}={5}\;
\ArrFl{25}{145}
\{d_5,a_4\}-\{a_3\}=\ArrFract{34}{7}\;
\allowdisplaybreaks\nn[5pt]
&&
\ArrFl{14}{135}
\{d_4,2a_4\}-\{2a_3\}=\ArrFract{33}{7}\;
\ArrFl{56}{125}
\{d_4,a_4\}-\{a_2\}=\ArrFract{163}{35}\;
\allowdisplaybreaks\nn[5pt]
&&
\ArrFl{35}{110}
\{d_4,a_4,a_2\}-\{a_3,a_1\}=\ArrFract{319}{70}\;
\ArrFl{16}{105}
\{d_4,a_3\}-\{a_1\}=\ArrFract{9}{2}\;
\allowdisplaybreaks\nn[5pt]
&&
\ArrFl{24}{100}
\{3a_3\}-\{a_2,a_1\}=\ArrFract{43}{10}\;
\ArrFl{13}{90}
\{2a_3,a_2\}-\{2a_1\}=\ArrFract{21}{5}\;
\allowdisplaybreaks\nn[5pt]
&&
\ArrFl{46}{80}
\{a_3,3a_2\}-\{3a_1\}=\ArrFract{41}{10}\;
\ArrFl{45}{75}
\{a_3,2a_2\}-\{a_1\}=\ArrFract{39}{10}\;
\allowdisplaybreaks\nn[5pt]
&&
\ArrFl{26}{70}
\{4a_2\}-\{2a_1\}=\ArrFract{19}{5}
\ArrFl{34}{65}
\{3a_2\}=\ArrFract{18}{5}\;
\ArrFl{23}{55}
\{2a_2,2a_1\}=\ArrFract{17}{5}\;
\allowdisplaybreaks\nn[5pt]
&&
\ArrFl{6}{50}
\{a_2,3a_1\}=\ArrFract{27}{10}\;
\ArrFl{12}{45}
\{5a_1\}=\ArrFract{5}{2}
\ArrFl{5}{40}
\{4a_1\}={2}
\allowdisplaybreaks\nn[5pt]
&&
\ArrFl{4}{30}
\{3a_1\}=\ArrFract{3}{2}\;
\ArrFl{3}{20}
\{2a_1\}={1}\;
\ArrFl{2}{10}
\{a_1\}=\ArrFract{1}{2}\;
\Buildrel {m}_{1}\over{\hbox to
25pt{\rightarrowfill}}\under{0}
{\rm Massive}\>,
\lab{E6Flow}
\end{eqnarray}
where we have used the notation
\begin{equation}
\{p_1 g_1,\ldots, p_n g_n\}=p_1 C_2(g_1)+ \cdots +p_n C_2(g_n)\>.
\end{equation}
Here, $p_1\ldots p_n$ are positive integers, $g_1\ldots g_n$ are simple Lie
algebras, and
$C_k\bigl(g\bigr)$ is
the central charge of the
$G_k/U(1)^{\rg}$ coset conformal field theory,
given by~\rf{cform}.
Above and below the arrows, we have indicated the mass scale associated
with each crossover, say $m_{ij}$, and the value of $-\ln(m_1 r/2)$ at
$r=2/m_{ij}$\,.
These results are in complete agreement  with the numerical data
presented in fig.~\ref{fige6}.

The coset identifications for the fixed points visited by the
flow can be worked out as explained in
section~\ref{LAGsec} and appendix~\ref{Heterotic}, and provide
several examples where the fixed point is a true asymmetric coset
model. The first occurs in the regime $m_{14}\lll 2r^{-1} \lll
m_{25}$\,,
and is determined by the solutions to $\Gamma_{51}(\gamma)=
\Gamma_{51}(\one)$ and $\Gamma_{52}(\gamma)= \Gamma_{52}(\one)$.
Taking~\rf{LagShield} into account, the second condition implies
the first, so
the effective
theory is specified by left and right cosets of the form~\rf{LagCosetsB}
with $i=5$ and $j=2$, which in this case are
\begin{equation}
{SU(5)_k\over U(1)^4}
\otimes{SO(10)_k\over SU(4)\otimes U(1)^2} \quad\text{and}\quad
{SU(5)_k\over SU(4)_k \otimes U(1)}
\otimes{SO(10)_k\over U(1)^5} \>.
\end{equation}


\end{document}